\documentclass[english, onecolumn, 11pt]{IEEEtran}
\usepackage[centertags]{amsmath}
\usepackage{amsfonts}
\usepackage{amssymb}
\usepackage{epsfig}
\usepackage{amsmath,  amssymb}
\usepackage{graphicx}
\usepackage[centertags]{amsmath}
\usepackage{clrscode}
\usepackage{cite}
\usepackage{fullpage}
\usepackage{amsmath,amssymb,mathrsfs,pseudocode}
\usepackage{color}
\usepackage[normalem]{ulem}
\usepackage{psfrag}
\usepackage{url}
\usepackage{babel}
\usepackage{authblk}
\usepackage{makecell}
\usepackage{mathtools}
\usepackage{subfigure}
\usepackage{cases}
\usepackage[linesnumbered,ruled]{algorithm2e}

\usepackage[margin=22.5mm]{geometry}
\usepackage{xspace}
\usepackage{multicol}
\usepackage{booktabs} 
\usepackage{multirow}
\usepackage[x11names,dvipsnames,table]{xcolor}

\usepackage{lipsum,cuted}
\usepackage{stfloats}

\newcommand{\opertype}[1]{\begin{minipage}{105mm}\centering\vspace{1mm} #1\vspace{1mm}\end{minipage}}
\newcommand{\opertypesmall}[1]{\begin{minipage}{25mm}\centering\vspace{1mm} #1\vspace{1mm}\end{minipage}}
\newcommand{\opertypesmalltwo}[1]{\begin{minipage}{36mm}\centering\vspace{1mm} #1\vspace{1mm}\end{minipage}}
\newcommand{\opertypemed}[1]{\begin{minipage}{64mm}\centering\vspace{1mm} #1\vspace{1mm}\end{minipage}}
\newcommand{\opertypemedminus}[1]{\begin{minipage}{60mm}\centering\vspace{1mm} #1\vspace{1mm}\end{minipage}}

\newtheorem{theorem}{Theorem}
\newtheorem{lemma}{Lemma}

\newtheorem{proposition}{Proposition}
\newtheorem{example}{Example}
\newtheorem{definition}{Definition}
\newtheorem{corollary}{Corollary}
\newtheorem{remark}{Remark}

\usepackage{xspace}
\usepackage{bbm}
\input{mathlig}

\newcommand{\muspace}{\mspace{1mu}}

\DeclareRobustCommand{\scond}{\mathchoice{\muspace\vert\muspace}{\vert}{\vert}{\vert}}
\mathlig{|}{\scond}

\DeclareRobustCommand{\discint}{\mathchoice{\mspace{-1.5mu}:\mspace{-1.5mu}}{\mspace{-1.5mu}:\mspace{-1.5mu}}{:}{:}}
\mathlig{::}{\discint}

\newcommand{\Dc}{\mathcal{D}}

\newcommand{\Pc}{\mathcal{P}}

\newcommand{\Acal}{\mathcal{A}}

\newcommand{\Ecal}{\mathcal{E}}

\newcommand{\Gcal}{\mathcal{G}}

\newcommand{\Kcal}{\mathcal{K}}

\newcommand{\Pcal}{\mathcal{P}}
\newcommand{\Qcal}{\mathcal{Q}}

\newcommand{\Tcal}{\mathcal{T}}
\newcommand{\Ucal}{\mathcal{U}}
\newcommand{\Vcal}{\mathcal{V}}
\newcommand{\Wcal}{\mathcal{W}}

\newcommand{\Zcal}{\mathcal{Z}}

\newcommand{\Cr}{\mathscr{C}}

\newcommand{\Rr}{\mathscr{R}}

\newcommand{\Cv}{{\bf C}}
\newcommand{\Dv}{{\bf D}}
\newcommand{\Du}{\underline D}
\newcommand{\Duv}{{\bf \underline D}}

\newcommand{\Xv}{{\bf X}}
\newcommand{\Pv}{{\bf P}}
\newcommand{\Yv}{{\bf Y}}

\newcommand{\Vv}{{\bf V}}
\newcommand{\Rv}{{\bf R}}

\newcommand{\tv}{{\bf t}}
\newcommand{\xv}{{\bf x}}
\newcommand{\yv}{{\bf y}}

\newcommand{\rv}{{\bf r}}

\newcommand{\Xh}{{\hat{X}}}

\newcommand{\wh}{{\hat{w}}}
\newcommand{\xh}{{\hat{x}}}
\newcommand{\xvh}{{\hat{\xv}}}

\newcommand{\Gt}{{\tilde{G}}}

\def\d{\delta}
\def\e{\epsilon}

\let\P\relax
\DeclareMathOperator\P{\textsf{P}}

\def\textiid{i.i.d.\@\xspace}
\newcommand\iid{\ifmmode\text{ i.i.d. } \else \textiid \fi}

\def\clap#1{\hbox to 0pt{\hss#1\hss}}
\def\mathclap{\mathpalette\mathclapinternal}
\def\mathclapinternal#1#2{%
  \clap{$\mathsurround=0pt#1{#2}$}}

\let\oldstackrel\stackrel
\renewcommand{\stackrel}[2]{\oldstackrel{\mathclap{#1}}{#2}}

\newcommand{\ntouch}[1]{\overline{{#1}}}

\allowdisplaybreaks

\newcommand\blfootnote[1]{%
  \begingroup
  \renewcommand\thefootnote{}\footnote{#1}%
  \addtocounter{footnote}{-1}%
  \endgroup
}

\begin{document}
\title{Capacity Theorems for Distributed Index Coding}

\author{Yucheng Liu, Parastoo Sadeghi, Fatemeh Arbabjolfaei, and Young-Han Kim}

\maketitle


\begin{abstract}

In index coding, a server broadcasts multiple messages to their respective receivers, each with some side information that can be utilized to reduce the amount of communication from the server. Distributed index coding is an extension of index coding in which the messages are broadcast from multiple servers, each storing different subsets of the messages. In this paper, the optimal tradeoff among the message rates and the server broadcast rates, which is defined formally as the capacity region, is studied for a general distributed index coding problem. Inner and outer bounds on the capacity region are established that have matching sum-rates for all 218 non-isomorphic four-message problems with equal link capacities for all the links from servers to receivers. The proposed inner bound is built on a distributed composite coding scheme that outperforms the existing schemes by incorporating more flexible decoding configurations and enhanced fractional rate allocations into two-stage composite coding, a scheme that was originally introduced for centralized index coding. The proposed outer bound is built on the polymatroidal axioms of entropy, as well as functional dependences such as the  $\rm{fd}$-separation introduced by the multi-server nature of the problem. This outer bound utilizes general groupings of servers with different levels of granularity, which allows a natural tradeoff between computational complexity and tightness of the bound, and includes and improves upon all existing outer bounds for distributed index coding. Specific features of the proposed inner and outer bounds are demonstrated through concrete examples with four or five messages.
\end{abstract} 


\blfootnote{Preliminary results of this paper were presented, in part, at the IEEE International Symposium on Information Theory (ISIT), Aachen, Germany, June 2017 \cite{isit:2017} and at the IEEE Information Theory Workshop (ITW), Cambridge, United Kingdom, September 2016~\cite{Sadeghi--Arbabjolfaei--Kim2016}.}
\blfootnote{Yucheng Liu and Parastoo Sadeghi are with the Research School of Electrical, Energy and Materials Engineering, Australian National University, Canberra, ACT, 2601, Australia. Fatemeh Arbabjolfaei is currently with the Department of Electrical Engineering and Computer Science, University of Michigan, Ann Arbor, MI 48109 USA. Young-Han Kim is with the Department of Electrical and Computer Engineering, University of California, San Diego, CA 92093, USA. Emails:  \{yucheng.liu, parastoo.sadeghi\}@anu.edu.au, arbab@umich.edu,  yhk@ucsd.edu.}


\section{Introduction}\label{sec:intro}

Index coding has been recognized as one of the canonical problems in network information theory. In its classic setting, the index coding problem studies the broadcast rate of $n$ messages from a single \emph{centralized} server to multiple receivers with side information. In this paper, we study the \emph{distributed} index coding problem, whereby different subsets of the messages are stored over multiple servers. Such communication model has clear applications for practical scenarios, in which the information is geographically distributed and stored across many locations.

\subsection{Background}\label{sec:background}

Since its introduction by Birk and Kol \cite{Birk--Kol1998} in 1998, the centralized index coding problem has intrigued various research communities and has been extensively investigated from various perspectives such as algebraic coding theory, graph theory, network coding, Shannon theory, and interference alignment. See, for example, \cite{Birk--Kol2006, bar2011index,lubetzky2009nonlinear, tehrani2012bipartite, blasiak2011lexicographic, shanmugam2013local, arbabjolfaei2014local, Arbabjolfaei--Kim2015a, el2010index, effros2015equivalence, Arbabjolfaei--Bandemer--Kim--Sasoglu--Wang2013,jafar2013topological,maleki2014index, sun2015index}, and the references therein. However, the index coding problem remains open in general. 

Among all linear and nonlinear coding schemes proposed in the literature for the centralized index coding problem \cite{foundation}, we focus on the composite coding scheme from \cite{Arbabjolfaei--Bandemer--Kim--Sasoglu--Wang2013} for the following reasons. {Many coding schemes in the literature have been designed to achieve lower bounds on the symmetric capacity (upper bounds on the broadcast rate) only. These schemes include clique covering \cite{Birk--Kol1998}, maximum distance separable (MDS) codes \cite{Birk--Kol1998}, partial clique covering \cite{Birk--Kol1998}, minrank-based codes \cite{bar2011index,lubetzky2009nonlinear} and interference-alignment-based codes \cite{maleki2014index,jafar2013topological}. Fractional local partial clique covering and recursive coding proposed in \cite{arbabjolfaei2014local} generalize some of the aforementioned schemes and allow characterization of non-symmetric rates (inner bounds on the capacity region). However, it is shown in \cite{foundation} that an enhanced version of composite coding strictly subsumes the inner bound provided by both fractional local clique covering and recursive coding.\footnote{Also note that in general, searching the maximum symmetric rate in the convex hull of (asymmetric) inner bounds on the capacity region can result in a higher symmetric rate (compared to searching for the maximum symmetric rate among individual coding schemes).}} Furthermore, composite coding can give tight inner bounds on the capacity region for all centralized index coding problems with five or fewer messages \cite{Arbabjolfaei--Bandemer--Kim--Sasoglu--Wang2013}. {In plain terms, composite coding} is a two-stage nonlinear coding scheme based on random coding.
In the first stage of encoding, each nonempty subset of messages is mapped via random coding to a corresponding random codeword, referred to as \emph{composite index}. In the second stage, all composite indices are mapped together to a final codeword, again via random coding. Decoding takes place in the reverse order of stages. In the second stage of decoding, each receiver can choose which messages to decode, which dictates which composite indices are useful in this stage of decoding. 
Each such \emph{decoding configuration} results in an achievable rate region for the centralized index coding problem, which is an inner bound on its capacity region. 
Although composite coding is not practical due to random coding, {it achieves a computable achievable rate region that can be used as a benchmark when designing linear index codes}. Simplification methods for composite coding were recently proposed in \cite{isit:2018} for eliminating unnecessary composite indices and excluding unnecessary decoding configurations, leading to reduced computational complexity.

An outer bound on the capacity region of the centralized index coding problem was presented in \cite{blasiak2011lexicographic}, \cite{Arbabjolfaei--Bandemer--Kim--Sasoglu--Wang2013}, which is based on the polymatroidal (PM) axioms of the entropy function.\footnote{The resulting outer bound on the rate region itself is not polymatroidal. However, since the bound uses all the polymatroidal axioms for the entropy function, it is referred to as the PM outer bound.} The well-known maximum acyclic induced subgraph (MAIS) lower bound on achievable broadcast rates, proposed in \cite{bar2011index},  is implied by the PM outer bound and can be strictly looser \cite{Arbabjolfaei--Bandemer--Kim--Sasoglu--Wang2013}.
 
The distributed index coding problem was first introduced in \cite{Ong--Ho--Lim2016}, which derived lower and upper bounds on the broadcast rate in a special case in which each receiver has a distinct message as side information. Subsequently, the authors in \cite{Thapa--Ong--Johnson2016} considered another interesting special case with two servers and arbitrary message subsets. 

Previous work \cite{Sadeghi--Arbabjolfaei--Kim2016} is the first known work to consider the most general setting for distributed index coding, where there is a server for every subset of messages. These servers are connected to their subsets of messages and communicate the messages to all receivers via individual noiseless broadcast channels with fixed capacities. {The centralized index coding problem can be seen as a special case of this model, in which only one server with all messages and a nonzero broadcast channel capacity is present. The general distributed index coding problem shows non-trivial behavior compared to the centralized version, as first reported in \cite{Sadeghi--Arbabjolfaei--Kim2016}. See also Figure \ref{fig:toy} and Example \ref{exmp:toy} in Section \ref{sec:model}.} Paper \cite{Sadeghi--Arbabjolfaei--Kim2016} is also the first work to apply composite coding to distributed index coding and present a basic outer bound on its capacity region, which is based on the PM axioms of the entropy function and includes distributed MAIS bound as a special case. Papers \cite{Li--Ong--Johnson--ISIT--2017, isit:2017} both built on \cite{Sadeghi--Arbabjolfaei--Kim2016} and were independently developed. Two key differences between \cite{Li--Ong--Johnson--ISIT--2017} and \cite{isit:2017} are as follows. The authors in \cite{Li--Ong--Johnson--ISIT--2017} demonstrated the necessity of \emph{cooperative composite coding} among distributed servers for achieving tighter inner bounds (as opposed to independently splitting composite index rates across servers as used in \cite{Sadeghi--Arbabjolfaei--Kim2016}). This paper subsequently formed a basis for \cite{Li--Ong--Johnson2017}. Neither  \cite{Li--Ong--Johnson--ISIT--2017} nor \cite{Li--Ong--Johnson2017} develop new performance bounds and use a simplified version of the distributed MAIS bound in \cite{Sadeghi--Arbabjolfaei--Kim2016} for benchmarking their coding scheme. On the other hand, \cite{isit:2017} utilizes a more flexible \emph{enhanced fractional allocation} of server link capacities over different decoding configurations compared to \cite{Sadeghi--Arbabjolfaei--Kim2016, Li--Ong--Johnson--ISIT--2017, Li--Ong--Johnson2017}. In addition, it  nontrivially extended and strictly tightened the outer bound in \cite{Sadeghi--Arbabjolfaei--Kim2016}.

\subsection{Contributions}\label{sec:contribution}
In this paper, we build upon the accumulated knowledge in general distributed index coding \cite{Sadeghi--Arbabjolfaei--Kim2016, isit:2017, Li--Ong--Johnson--ISIT--2017, Li--Ong--Johnson2017} and provide further contributions as follows. 

Firstly in Section \ref{sec:dist:compcod}, we propose a more general distributed composite coding scheme that strictly subsumes those in \cite{Sadeghi--Arbabjolfaei--Kim2016, isit:2017, Li--Ong--Johnson--ISIT--2017, Li--Ong--Johnson2017}, thus establishing a tighter inner bound on the capacity region of the distributed index coding problem. 
Notably, the proposed coding scheme combines our previously devised enhanced fractional composite coding \cite{isit:2017} and the cooperative composite coding of \cite{Li--Ong--Johnson--ISIT--2017, Li--Ong--Johnson2017} and adds a new dimension of decoding flexibility. In our new scheme, each receiver can choose a different group of servers for decoding, independent of other receivers. Therefore, the decoding configuration has many more possibilities compared to \cite{Sadeghi--Arbabjolfaei--Kim2016, isit:2017, Li--Ong--Johnson--ISIT--2017, Li--Ong--Johnson2017}.

The second main contribution is a more general outer bound on the capacity region of the distributed index coding problem that strictly subsumes our previous outer bound in \cite{isit:2017}. It still uses the PM axioms of the entropy function. The novelty, however, is twofold. First, \cite{isit:2017} only considered server groups based on the \emph{touch} (intersecting) structure of the servers with messages. In this paper, we incorporate the most flexible use of server groups to derive the necessary conditions for the achievable rates. Second, due to the specific touch structure in server groups, some functional structures of the problem were not reflected in the formulation of the outer bound in \cite{isit:2017}. In this paper, we incorporate conditional independence relations among messages, as identified according to $\mathrm{fd}$-separation \cite{kramer,satyajit}. We refer to this general bound as \emph{the grouping polymatroidal (PM) outer bound}.

For gentler presentation, in Section \ref{sec:fixed}, we first describe the basic form of our distributed composite coding scheme with fixed decoding configuration and provide the achievable rate region and detailed error analysis (Theorem \ref{thm:compcod}). We present the resulting inner bound in a series of equivalent or simplified forms that can help better understand the coding scheme and reduce the complexity of computation (Proposition \ref{propo:compcod:equiv} and Corollaries \ref{corr:allservers}--\ref{corr:simplify}). In Section \ref{sec:var}, we present our general composite coding scheme (Theorem \ref{thm:compcod:fractional} and Corollary~\ref{corr:simplify:fractional}). We provide detailed numerical examples and discussions to showcase the use and novelty of our results.

In Section \ref{subsec:outer:grouping}, we first present the general grouping PM outer bound (Theorem \ref{thm:pm:group}). Then in Sections \ref{sec:touch:grouping} and \ref{sec:outer:bounds:fd}, we specify a number of construction techniques for grouping servers and present the corresponding specialized grouping PM outer bounds (Corollaries \ref{cor:pm:touch:aggregate}--\ref{cor:pm:fd:group} and Proposition \ref{propo:uv}). We present the examples for which these simplified server group constructions can provide tight sum capacity results. In Section \ref{sec:hierarchy}, we formalize the hierarchy of server groupings in terms of tightness and computational complexity of the outer bound (Corollaries \ref{cor:pm:finest}--\ref{cor:pm:coarsest} and Propositions \ref{propo:refine}-\ref{propo:allserver:itw}).

In Section~\ref{sec:numerical}, utilizing the inner and outer bounds derived thus far, we establish the sum-capacity of all $218$ four-message distributed index coding problems with equal link capacities. Finally, we summarize our  insights and present an outlook for future research directions in Section \ref{sec:conclusion}.
 
For a positive integer $n$, $[n]$ denotes the set $\{1, 2, \ldots, n\}$. 

For a finite set $A$, $2^A$ denotes the set of all subsets of $A$. In particular, $N = 2^{[n]}$ denotes the set of all subsets of $[n]$. For a subset $A$ of a ground set, $A^c$ denotes its complement with respect to the ground set. For two sets $A$ and $B$, $A \setminus B$ denotes $A \cap B^c$.


\section{The System Model and the Formal Definition of the Problem}\label{sec:model}

Consider the distributed index coding problem 
with $n$ messages, $x_i \in \{0,1\}^{t_i}, i \in [n]$. For brevity, when we say message $i$, we mean message $x_i$. Let $X_i$ be the random variable corresponding to $x_i$. For any $K\subseteq [n]$, we use the shorthand notation $\xv_K$ and $\Xv_K$ to denote the collection of messages and the collection of message random variables, whose index is in $K$, respectively. We use the convention $\xv_{\emptyset} = \Xv_\emptyset = \emptyset$. We assume that $X_1, \ldots, X_n$ are uniformly distributed and independent of each other. 

There are $2^n-1$ servers, one per each nonempty subset of the messages. The server indexed by $J \in N$ has access to messages $\xv_J$. For brevity, when we say server $J$, we mean the server indexed by $J$. Server $J$ is connected to all receivers via a noiseless broadcast link of finite capacity $C_J\geq 0$.
{Note that this model allows for all possible servers that contain any subset of messages to be present in the system.}
If $C_J = 1$ for $J= [n]$ and is zero otherwise, we recover the centralized index coding problem. Let $y_J$ be the output of server $J$, which is a function of $\xv_J$, and $Y_J$ be the random variable corresponding to $y_J$. For simplicity of notation, we also assume that there exists a dummy server indexed by $J = \emptyset$, which does nothing ($C_\emptyset = 0$ and $Y_\emptyset = \emptyset$).
For a collection $P\subseteq N$ of servers, we use the shorthand notation $\yv_P$ ($\Yv_P$) to denote the corresponding collection of outputs (output random variables) from servers in $P$. In particular, for $P = N$, $\Yv_N$ denotes the collection of output random variables from all the servers.

There are $n$ receivers, where receiver $i \in [n]$ wishes to obtain $x_i$ and knows $\xv_{A_i}$ as side information for some $A_i \subseteq [n]\setminus \{i\}$. 
The set of messages which receiver $i$ does not know or want is denoted by $B_i = [n] \setminus (A_i \cup\{i\})$. 

The central question of distributed index coding is to find the maximum amount of information that can be communicated to the receivers and the optimal coding scheme that achieves this maximum.
To answer this question formally, we define 
a $(\tv,\rv) = ((t_i, i \in [n]), (r_J, J \in N))$ {\em distributed index code} by
\begin{itemize}
\item $2^n$ encoders, one for each server $J \in N$, such that $\phi_J: \prod_{j \in J} \{0,1\}^{t_j} \to \{0,1\}^{r_J}$ maps the messages $\xv_J$ in server $J$ to an $r_J$-bit sequence $y_J$, and
\item $n$ decoders, one for each receiver $i \in [n]$, such that $\psi_i: \prod_{J \in N} \{0,1\}^{r_J} \times \prod_{k \in A_i} \{0,1\}^{t_k} \to \{0,1\}^{t_i}$ maps the sequences $\yv_N$ and the side information $\xv_{A_i}$ to $\hat{x}_i$.
\end{itemize}

We say that a rate--capacity tuple $(\mathbf{R}, \mathbf{C}) = ((R_i, i \in [n]),(C_J, J \in N))$ is achievable if for every $\epsilon > 0$, there exists a $(\tv,\rv)$ code and a {common normalization positive integer} $r$ such that the message rates $\frac{t_i}{r}$, $i \in [n]$ and broadcast rates $\frac{r_J}{r}$, $J \in N$ satisfy
\begin{equation}
R_i \le \frac{t_i}{r}, \quad i \in [n], \qquad \qquad
C_J \geq \frac{r_J}{r}, \quad J \in N,  \label{eq:model:code}
\end{equation}
and the probability of error satisfies
\begin{equation}
\P\{ (\Xh_1,\ldots, \Xh_n) \ne (X_1, \ldots, X_n)\} \le \epsilon.
\end{equation}
For a given link capacity tuple $\Cv$, the capacity region $\Cr = \Cr(\Cv)$ of this index coding problem is the closure of the set of all rate tuples $\Rv$ such that $(\Rv,\Cv)$ is achievable.
Unlike the centralized case in which the capacity region is equal to 
the zero-error capacity region \cite{Langberg--Effros2011}, it is not known whether these two capacity regions are equal for the distributed index coding problem.

We will compactly represent a distributed index coding instance by a sequence  $(i|j \in A_i)$, $i \in [n]$. For example, for $A_1 = \emptyset$, $A_2 = \{3\}$, and $A_3=\{2\}$, we write
$(1|-),\, (2|3),\, (3|2)$. The list of main symbols introduced so far and to be introduced later in the paper is summarized in Table \ref{tab:notations}.

\begin{figure}[t]
\begin{center}
\subfigure[][]{
\label{fig:toy:a}
\includegraphics[scale=0.18]{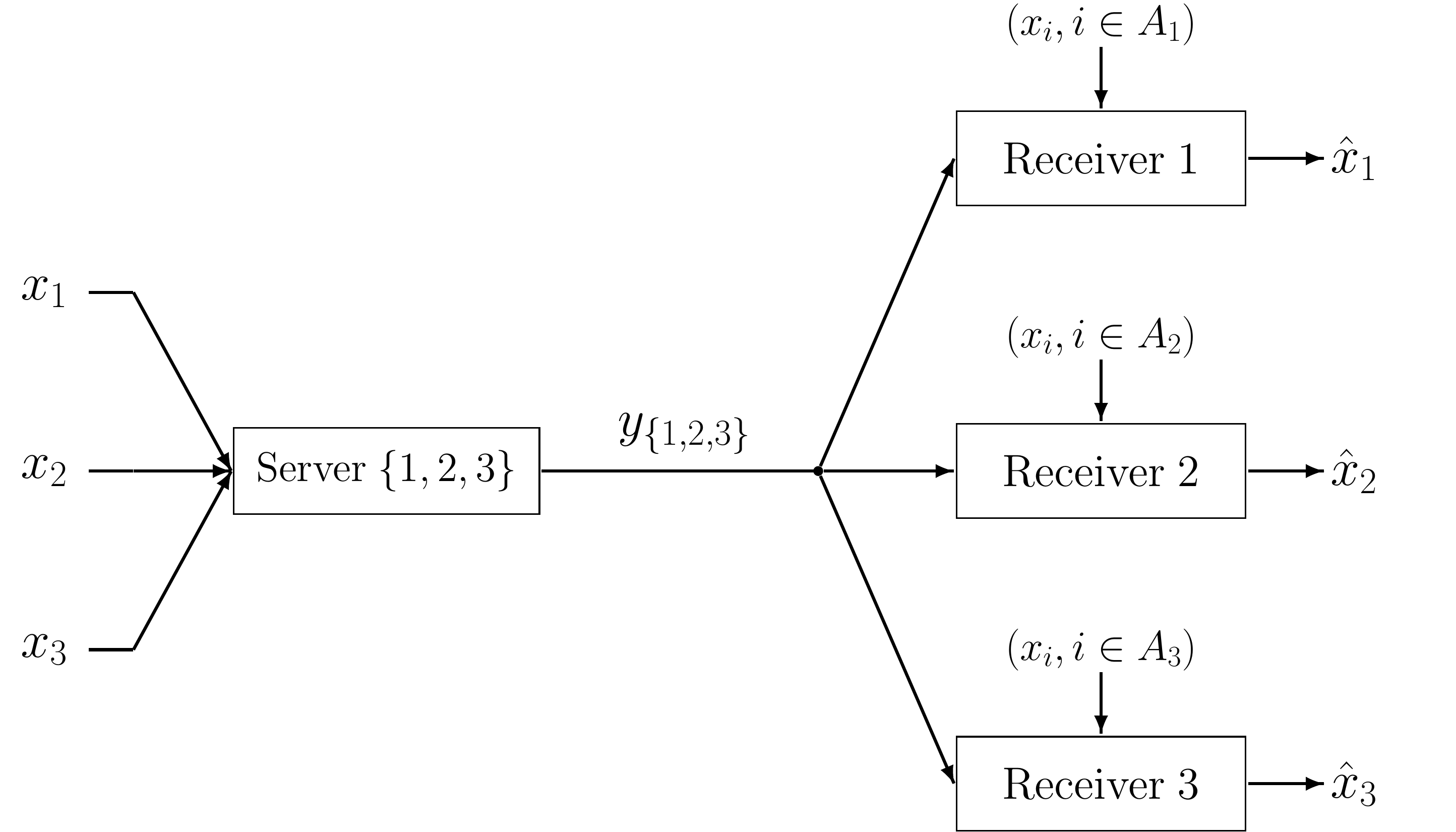}}\\
\subfigure[][]{
\label{fig:toy:b}
\includegraphics[scale=0.16]{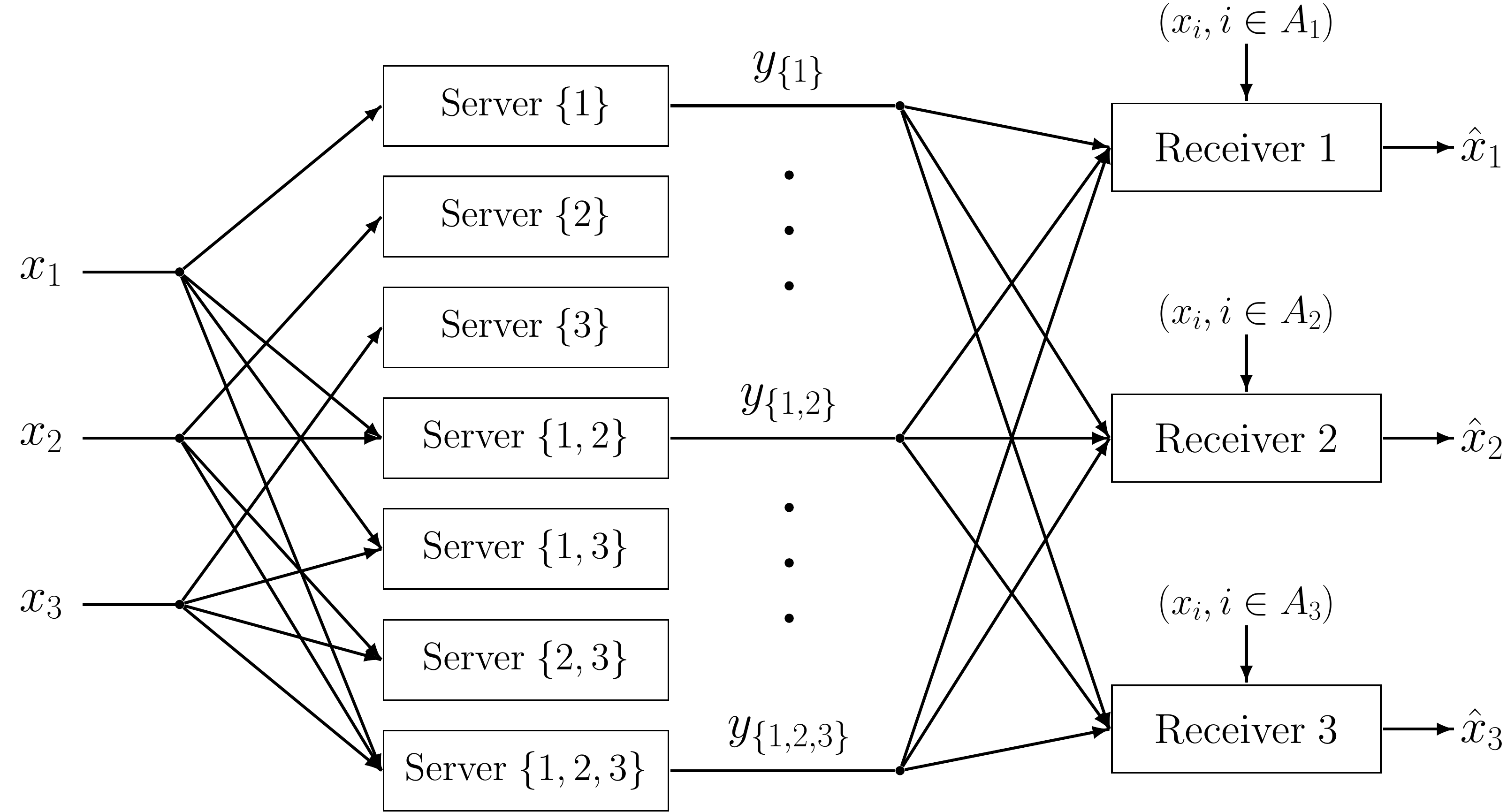}}
\caption{Centralized and distributed index coding problems with $n=3$ messages.}
\label{fig:toy}
\end{center}
\end{figure}

\begin{example}\label{exmp:toy}
Figures \ref{fig:toy:a} and \ref{fig:toy:b} respectively show the system models for the centralized index coding problem and the distributed index coding problem  with $n=3$ messages. Recall that the server indexed by $J \in N$ contains messages $\xv_J=(x_j, j \in J)$. The multi-server nature of the distributed index coding problem can lead to fundamentally different properties compared with the centralized index coding problem. For example, consider the centralized and distributed index coding problems, both with the same receiver side information $(1|3), (2|3), (3|2)$. For the centralized problem, the capacity region remains unchanged if the side information at receiver $1$ is removed \cite{tahmasbi2015critical}, i.e., if $A_1$ becomes $\emptyset$. However, in its distributed counterpart, as long as the broadcast channel capacities from the servers $J_1 = \{1,2\}$ and $J_2 = \{1,3\}$ are positive, removing the side information at receiver $1$ does result in a strictly smaller capacity region \cite{Sadeghi--Arbabjolfaei--Kim2016}.
\end{example}

\begin{table*}[t]
\begin{center}
\caption{List of Symbols}
\begin{tabular}{|c|c|}
\hline

Symbol&Meaning\\
\hline
$n \geq 1$&\opertype{Number of messages (and receivers) in the system}\\
\hline
$[n] = \{1,2,\cdots, n\}$&\opertype{Index set of messages (and receivers)}\\
\hline
$i,j,k,\ell$&\opertype{General-purpose running indices}\\
\hline
$x_i\in \{0,1\}^{t_i}$&\opertype{Binary message $i$ of length $t_i$}\\
\hline
$R_i$&\opertype{Rate of message $i$}\\
\hline
$X_i$&\opertype{Random variable corresponding to message $x_i$}\\
\hline
$\hat{x}_i, \hat{X}_i$&Decoded message or decoded message random variable $i$\\
\hline
$A_i \subseteq [n]\setminus \{i\}$&\opertype{Index of side information for receiver $i$}\\
\hline
$B_i= [n]\setminus (A_i \cup \{i\})$&\opertype{Index of interfering messages that receiver $i$ neither wants nor has}\\
\hline
$D_i \subseteq [n]\setminus A_i, i \in D_i$&\opertype{Index of messages receiver $i$ wishes to decode}\\
\hline
$L,L',K,K' \subseteq [n]$&\opertype{Subset of message indices}\\
\hline
$\xv_K$, $\Xv_K$&\opertype{Collection of messages (message random variables)  indexed by $K$}\\
\hline
$w_K(\xv_K)$&\opertype{Composite message that is randomly generated from messages $\xv_K$\\ Shorthand: $w_K$, composite index $K$}\\
\hline
$S_K$&\opertype{Rate of composite index $K$}\\
\hline
$N = 2^{[n]}$&\opertype{The set of all possible message subsets}\\
\hline
$J \in N$&\opertype{Server $J$ that contains messages $\xv_J$}\\
\hline
$C_J \geq 0$&\opertype{Link capacity of server $J$}\\
\hline
$y_J =\{0,1\}^{r_J}$&\opertype{Output of server $J$ of length $r_J$}\\
\hline
$Y_J$&\opertype{Random variable corresponding to $y_J$}\\
\hline
$P,Q \subseteq N$&\opertype{Arbitrary collection of servers}\\
\hline
$\yv_P$, $\Yv_P$&\opertype{Collection of outputs (output random variables) for servers in $P$}\\
\hline
$T_K \subseteq N$&\opertype{Group of servers that contain at least one message from message set indexed by $K$;\\ $J \in T_K \Leftrightarrow J \cap K \neq \emptyset$}\\
\hline
$m$&Number of server groups in a grouping\\
\hline
$[m] = \{1,2,\cdots, m\},[\ell]=\{1,2,\cdots,\ell \}$&\opertype{Index set of server groups}\\
\hline
$\Pcal=\{P_1,P_2,\cdots,P_m\}$&\opertype{A valid server grouping of $m$ server groups}\\
\hline
$P_G = \bigcup_{i\in G}P_i, \quad G\subseteq [m]$& \opertype{Collection of servers in the server groups identified by $G$}\\
\hline
$M \subseteq N$&\opertype{A collection of composite indices (corresponding to all $w_{K}: K \in M$)}\\
\hline
$\Gamma_*(P) = \bigcup_{J \in P}\{K: K \subseteq J\}$& \opertype{The subset completion of $P$. For a group of servers $P\subseteq N$, $\Gamma_*(P)$ signifies the set of all composite indices that the servers in $P$ can collectively access.}\\
\hline
$\Gamma^*(M) = \bigcup_{K \in M}\{J\in N: K \subseteq J\}$& \opertype{The superset completion of $M$ with respect to the ground set $N$. For a collection of composite indices $M$, $\Gamma^*(M)$ signifies the set of servers $P \subseteq N$ that can access at least one composite index in $M$.}\\
\hline
 \opertypemed{$f(G,K)$, $G\subseteq [m]$, $K \subseteq [n]$, for a valid server grouping $\Pcal=\{P_1,P_2,\cdots,P_m\}$}& \opertype{A general double set function that is used for establishing outer bounds on the capacity region of the distributed index coding problem.}\\
\hline
$g(K)$, $K \subseteq [n]$& \opertype{A single set function that is used for establishing a simplified outer bound  on the capacity region of the distributed index coding problem.}\\
\hline
\end{tabular}\label{tab:notations}
\end{center}
\end{table*}


\section{Composite Coding for the Distributed Index Coding Problem}\label{sec:dist:compcod}
\subsection{Composite Coding for a Fixed Decoding Configuration}\label{sec:fixed}

Let $r > 0$, $t_i = \lceil{rR_i\rceil}$, $i \in [n]$, where $t_i$ is the length and $R_i$ is the rate of message $i$, respectively. Composite coding scheme is a two-stage nonlinear coding scheme based on random coding. In the first stage of encoding, each nonempty subset of messages $K \subseteq [n]$ is mapped via random coding to a corresponding random codeword, referred to as \emph{composite index},  denoted by $w_K\in \{0,1\}^{s_K}$, with rate $S_K$ and length $s_K = \lceil{rS_K\rceil}$. We set $s_\emptyset = S_\emptyset = 0$ by convention. In the second stage, composite indices $K \subseteq J$ that are available at server $J$ are mapped together to a codeword $y_J\in \{0,1\}^{r_J}$, again via random coding, where $r_J = \lfloor{r{C_J}\rfloor}$, $J \in N$.

Decoding takes place in the reverse order of stages. 
For each receiver $i\in [n]$, fix a set $P_i \subseteq N$ called the decoding server group and a set $D_i \subseteq [n]\setminus A_i$ called the decoding message set such that $i \in D_i$. The tuples $\Pv = (P_i, i\in [n])$ and $\Dv = (D_i, i\in[n])$ are collectively referred to as the decoding configuration. In this subsection, we provide an achievable rate region for composite coding for a fixed $(\Pv, \Dv)$. Note that $\bigcup_{J \in P_i} J$ is the set of message indices that is collectively available to servers in $P_i$. Thus, receiver $i$ can effectively recover messages whose indices are in $\Delta_i = (\bigcup_{J \in P_i} J)\cap D_i$.  For notational brevity, the dependence of $\Delta_i$ on $(\Pv, \Dv)$ is implicit.

For any $P \subseteq N$, $\Gamma_*(P) = \bigcup_{J \in P}\{K: K \subseteq J\} = \bigcup_{J \in P}2^J$ is the subset completion of $P$. Similarly, for any $M \subseteq N$, $\Gamma^*(M) = \bigcup_{K \in M}\{J\in N: K \subseteq J\}$ is the superset completion of $M$ (with respect to $N$). One can think of $\Gamma_*(P)$ as the set of all composite indices 
that the servers in $P$ can collectively access. One can think of $\Gamma^*(M)$ as
the set of all servers that have access to at least one composite index in $M$.

\begin{theorem}\label{thm:compcod}
A rate-capacity tuple $(\Rv, \Cv)$ 
is achievable for the distributed index coding problem $(i| A_i)$, $i \in [n]$, 
under a given decoding configuration $(\Pv, \Dv)$ if $R_i = 0$ for all $i\in[n]$ such that $i \notin \Delta_i$ and\footnote{As a degenerate case, if $i \notin \Delta_i$, then
this ``poor'' choice of $P_i$ will simply result in $R_i = 0$. This is reflected in the conditions of Theorems \ref{thm:compcod} and \ref{thm:compcod:fractional}. }
\begin{align}\label{eq:second:decode:fixed}
\sum_{j \in L} R_j &< \sum_{\substack{K \subseteq \Delta_i \cup A_i,\\ K \in \Gamma_*(P_i),\\ K \cap L \neq \emptyset}} S_K, \quad \quad \forall L \subseteq \Delta_i,\\\label{eq:first:decode:fixed}
\sum_{K \in M} S_K &< \sum_{J \in \Gamma^*(M)\,\cap \, P_i} C_J, \qquad \forall M \subseteq \Gamma_*(P_i) \setminus2^{A_i},
\end{align}
 for some $S_K \ge 0$, $K \subseteq [n]$, for all other $i\in[n]$ such that $i \in \Delta_i$.
\end{theorem}

One can use Fourier--Motzkin elimination (FME) \cite[Appendix D]{elgamal_yhk} to express the achievable
rate--capacity region in Theorem~\ref{thm:compcod} without the intermediate variables $S_K$, $K \subseteq [n]$. A linear program (LP) can also find a (weighted) achievable sum-rate of composite coding, which typically has a lower computational complexity. Before outlining the  coding scheme corresponding to Theorem \ref{thm:compcod}, we showcase its use via the following example.

\begin{example}\label{exmp:simple3}
Consider the distributed index coding problem $(1|-),(2|3),(3|2)$ with $n=3$ messages and non-negative, but otherwise arbitrary link capacities $C_J \ge 0, J \in N\setminus \{\emptyset\}$ and $C_{\emptyset} = 0$. Note that the set $N=\{\emptyset, \{1\},  \{2\},  \{1,2\},  \{3\}, \{1,3\}, \{2,3\}, \{1,2,3\}\}$ is subset complete and hence, $\Gamma_*(N) = N$.

We fix $P_1 = \{\{1\}\}$, $P_2 = \{\{2\},\{1,2\}, \{2,3\}\}$, and $P_3 = \{\{3\}, \{1,3\},\{1,2,3\}\}$. We fix $D_1 = \{1\}$, $D_2 = \{1,2\}$, and $D_3 = \{1,3\}$. Hence, $D_1\cup A_1 = \{1\}$, and $D_i\cup A_i = \{1,2,3\}$,  $i = 2, 3$. Note that $\Delta_i = D_i$, $i \in [n]$. Note also that $\Gamma_*(P_1) =  \{\emptyset,\{1\}\}$, $\Gamma_*(P_2) = \{\emptyset,\{1\},\{2\},\{1,2\},\{3\},\{2,3\}\}$ and $\Gamma_*(P_3) = N$. Inequality \eqref{eq:second:decode:fixed} (including active and inactive inequalities) gives \eqref{eq:exmp:simple3:second:decode:fixed}. Inequality \eqref{eq:first:decode:fixed} (excluding inactive inequalities) yields \eqref{eq:exmp:simple3:first:decode:fixed}.

\begin{figure*}[h]
\begin{align}
\begin{matrix*}[l]
R_1 < S_{\{1\}}, & i = 1, L = \Delta_1,\\
R_1 < S_{\{1\}}+S_{\{1,2\}}, & i = 2, L = \{1\}\subset \Delta_2,\\
R_2 < S_{\{2\}}+S_{\{1,2\}}+S_{\{2,3\}}, & i = 2, L = \{2\}\subset \Delta_2,\\
R_1+R_2 < S_{\{1\}}+S_{\{2\}}+S_{\{1,2\}}+S_{\{2,3\}}, & i = 2, L = \Delta_2,\\
R_1 < S_{\{1\}}+S_{\{1,2\}}+S_{\{1,3\}}+S_{\{1,2,3\}}, & i = 3, L = \{1\}\subset \Delta_3,\\
R_3 < S_{\{3\}}+S_{\{1,3\}}+S_{\{2,3\}}+S_{\{1,2,3\}}, & i = 3, L = \{3\}\subset \Delta_3,\\
R_1+R_3 < S_{\{1\}}+S_{\{1,2\}}+S_{\{3\}}+S_{\{1,3\}}+S_{\{2,3\}}+S_{\{1,2,3\}}, & i = 3, L = \Delta_3.
\end{matrix*} \label{eq:exmp:simple3:second:decode:fixed}
\end{align}
\begin{align}
\begin{matrix*}[l]
 S_{\{1\}} < C_{\{1\}}, &\hspace{-1mm} i = 1, M_1 = \Gamma_*(P_1),\Gamma^*(M_1)\cap P_1=  P_1,\\
 S_{\{1\}}+S_{\{1,2\}} < C_{\{1,2\}}, &\hspace{-1mm} i = 2, M_2 =\{\{1\},\{1,2\}\}, \Gamma^*(M_2)\cap P_2=  \{\{1,2\}\},\\
S_{\{2,3\}} < C_{\{2,3\}}, &\hspace{-1mm} i = 2, M_3 =\{\{2,3\}\}, \Gamma^*(M_3)\cap P_2=  \{\{2,3\}\},\\
 S_{\{1\}}+S_{\{1,2\}} +S_{\{2,3\}}< C_{\{1,2\}}+C_{\{2,3\}}, &\hspace{-1mm} i = 2, M_4 =\{\{1\},\{1,2\}, \{2,3\}\},\\
 S_{\{1\}}+S_{\{2\}}+S_{\{1,2\}} +S_{\{2,3\}}< C_{\{2\}}+C_{\{1,2\}}+C_{\{2,3\}}, &\hspace{-1mm} i = 2, M_5 =\{\{1\},\{2\},\{1,2\}, \{2,3\}\},\\
 S_{\{1,2\}}+S_{\{2,3\}} +S_{\{1,2,3\}}< C_{\{1,2,3\}}, &\hspace{-1mm} i = 3, M_6 =\{\{1,2\}, \{2,3\}\, \{1,2,3\}\},\\
 S_{\{1\}}+S_{\{1,2\}}+S_{\{1,3\}}+S_{\{2,3\}} +S_{\{1,2,3\}}< C_{\{1,3\}}+C_{\{1,2,3\}}, &\hspace{-1mm} i = 3, M_7 =\{\{1\},\{1,2\}, \{1,3\}, \{2,3\}\, \{1,2,3\}\},\\
\sum_{K \in M_8} S_{K}< C_{\{3\}}+C_{\{1,3\}}+C_{\{1,2,3\}}, &\hspace{-1mm} i = 3, M_8 = \Gamma_*(P_3)\setminus \{\{2\}\}, \Gamma^*(M_8)\cap P_3 = P_3.
\end{matrix*} \label{eq:exmp:simple3:first:decode:fixed}
\end{align}
\rule[0.6ex]{\textwidth}{0.3mm}
\end{figure*}

We apply FME to eliminate all present $S_K$ and find that the achievable rate-capacity tuple $(\Rv, \Cv)$ satisfies
\begin{align*}
 R_1 &< \min\{C_{\{1\}}, C_{\{1,2\}}, C_{\{1,3\}}+C_{\{1,2,3\}}\},\\
 R_1+R_2 &<C_{\{2\}}+C_{\{1,2\}}+C_{\{2,3\}},\\
R_1+R_3 &< C_{\{3\}}+C_{\{1,3\}}+C_{\{1,2,3\}}.
\end{align*}
 
Note that we are not claiming that the choice of $(\Pv,\Dv)$ is optimal for this toy example. The chosen $\Dv$ is indeed optimal, but for a more compact exposition, we chose different and smaller suboptimal sets $P_i \subset N$, instead of the optimal $P_i = N$, $i \in [n]$.
\end{example}

\begin{IEEEproof}[Outline of Coding Scheme for Theorem \ref{thm:compcod}]
\textbf{Codebook generation.}  
\emph{Step 1.} For each $K \subseteq [n]$ and each value of $\xv_K$, generate a composite index $w_K(\xv_K)$ drawn independently and uniformly at random from $[2^{s_K}]$. That is, $w_K$ is a random mapping as
\begin{align*}
w_K: \prod_{j \in K} [2^{t_j}] \to [2^{s_K}].
\end{align*}
{For brevity, when we say composite index $K$, or $w_K$, we mean composite index $w_K(\xv_K)$.}
\emph{Step 2.} For each $J \in N$ and each value of composite index tuple $(w_K, K \in 2^J)$, generate a codeword $y_J((w_{K}, K\in 2^J))$ drawn independently and uniformly at random from $[2^{r_J}]$. That is, $y_J$ is a random mapping as
\begin{align*}
y_J: \prod_{K \in 2^J} [2^{s_K}] \to [2^{r_J}].
\end{align*}
The codebook $\{(w_K(\xv_K), K \subseteq [n]), (y_J((w_K, K \in 2^J)), J \in N)\}$ is revealed to all \emph{corresponding} parties.\footnote{{It is worth contrasting single-layer \emph{flat} coding of messages into random codewords versus two-layer \emph{composite} coding of messages into random composite indices and then composite indices into random codewords. In other words, composite coding can be viewed as random construction of an approximate MDS code for composite indices, rather than messages themselves.  This adds flexibility in decoding conditions and can enhance the achievable rate region. Please also see \cite{Arbabjolfaei--Bandemer--Kim--Sasoglu--Wang2013,foundation}.}}$^,$\footnote{One of the servers must act as a representative or a central processing unit to generate the codebook and reveal the codebook to \emph{all corresponding} servers and all users. This is because the random mapping of $\xv_K$ to composite index $w_K$ should be identical among  all servers that can generate $w_K$. For composite index $K$, the corresponding servers are indexed by the superset of $K$ with respect to the set of all servers $N$, $\Gamma^*(K)$. \label{footnote:parties}
}

\textbf{Encoding.} To communicate messages $\xv_{[n]}$, 
each server $J \in N$ computes $w_K(\xv_K)$ for each $K \in 2^J$ and
transmits $y_J((w_{K}, K\in 2^J))$.

\textbf{Decoding.} 
\emph{Step 1.} For $i\in [n]$ such that $i \in \Delta_i$, receiver $i$ finds the unique composite index tuple $(\wh_K, K \in \Gamma_*(P_i))$ such that $y_J = y_J((\wh_K, K \in 2^J))$ for every $J \in P_i$. If there is more than one such tuple, it declares an error. \emph{Step 2.} Assuming that $(\wh_K, K \in \Gamma_*(P_i))$ is correct, receiver $i$ 
finds the unique message tuple $\xvh_{\Delta_i}$ 
such that
$\wh_K = w_K(\xvh_K)$ for every $K \in \Gamma_*(P_i)$ with $K\subseteq \Delta_i \cup A_i$. If there is more than one such tuple, it declares an error.

The inequalities in \eqref{eq:second:decode:fixed} signify 
the second-step decoding constraints for the messages in $\Delta_i$ to be
recovered with vanishingly small probability of error from all composite indices $K$ in $\Gamma_*(P_i)$, with the help of side information $A_i$. The inequalities in \eqref{eq:first:decode:fixed}
signify the first-step decoding constraints for the composite indices that the servers in $P_i$
have access to (except those that can be generated from side information) to be recovered with vanishingly small probability of error  from
the outputs $y_J$ from the servers $J$ in $P_i$.
The details of error analysis of Theorem~\ref{thm:compcod} is provided in Appendix \ref{app:error:proof}.  
\end{IEEEproof}

To help with understanding of Theorem \ref{thm:compcod}, we also present the error analysis for a specific example as follows. 
\begin{example}\label{exmp:simple3:error:analysis}
Let us revisit Example \ref{exmp:simple3} and consider decoding for receiver $2$ with $P_2=\{ \{2\},\{1,2\},\{2,3\} \}$, $\Delta_2=D_2=\{1,2\}$.

In the first step of decoding, receiver $2$ tries to decode composite indices $\wh_K,K\in \Gamma_*(P_2)$ from the received codewords $y_J,J\in P_2$ and its side information $\xv_{A_2}$. 
The decoding error probability $P_e$ is the probability that there exists some composite index tuple $(\wh_K,K\in \Gamma_*(P_2))=(\wh_{\{1\}},\wh_{\{2\}},\wh_{\{1,2\}},\wh_{\{3\}},\wh_{\{2,3\}})$ other than the correct (actually transmitted) tuple $(w_{\{1\}},w_{\{2\}},w_{\{1,2\}},w_{\{3\}},w_{\{2,3\}})$ such that they are mapped to the same codeword $y_J$ for every $J\in P_2=\{ \{2\},\{1,2\},\{2,3\} \}$. 
{To utilize the union bound for upper bounding $P_e$, we partition the error event according to the erroneous composite index set $M \subseteq \Gamma_*(P_2) \setminus 2^{A_2}=\{ \{1\},\{2\},\{1,2\},\{2,3\} \}$. That is, $\wh_K \neq w_K$ iff $K \in M$. 
Note that $2^{A_2}$ is always excluded from $M$ for the reason that $\wh_{\{3\}}$ can be generated by receiver $2$ from $\xv_{A_2}=\xv_{\{3\}}$ and thus will never be erroneous. }
Therefore, by the union bound, we have $P_e\leq \sum_{M \subseteq \{ \{1\},\{2\},\{1,2\},\{2,3\} \}}P_{e}^M$, where 
\begin{align}
P_e^M
&\doteq \sum_{\substack{(\wh_{\{1\}},\wh_{\{2\}},\wh_{\{1,2\}},w_{\{3\}},\wh_{\{2,3\}}):\\\wh_K \neq w_K, K \in M,\\\wh_K = w_K, K \notin M} }  \nonumber  \\
&\qquad \mathrm{P}\left\{\bigcap_{\substack{J \in \{ \{2\},\{1,2\},\{2,3\} \}, \\J \in \Gamma^*(M)}}\left\{y_J = y_J(\wh_K, K \in 2^J)\right\}\right\}.  \nonumber  
\end{align}
To ensure vanishingly small decoding error probability $P_e$, each $P_e^M$ has to be vanishingly small. In particular, we present detailed analysis for $P_e^M$ with $M=M_4=\{ \{1\},\{1,2\},\{2,3\} \}$ as specified in Example \ref{exmp:simple3}, while other $P_e^M$ can be analyzed similarly. Since $P_2\cap \Gamma^*(M)=\{ \{2\},\{1,2\},\{2,3\} \} \cap \Gamma^*(\{ \{1\},\{1,2\},\{2,3\} \})=\{ \{1,2\},\{2,3\} \}$, we have
\begin{align}
&P_e^{M_4} =\sum_{\substack{(\wh_{\{1\}},w_{\{2\}},\wh_{\{1,2\}},w_{\{3\}},\wh_{\{2,3\}}):\\\wh_K \neq w_K, K \in \{ \{1\},\{1,2\},\{2,3\} \} }}  \nonumber  \\
&\qquad \mathrm{P}\left\{ \bigcap_{\substack{J \in \{ \{1,2\},\{2,3\} \}}}\left\{y_J = y_J(\wh_K, K \in 2^J)\right\} \right\}  \nonumber  \\
&=\frac{(2^{s_{\{1\}}}-1)\cdot(2^{s_{\{1,2\}}}-1)\cdot(2^{s_{\{2,3\}}}-1)}{2^{r_{\{1,2\}}}\cdot 2^{r_{\{2,3\}}}}    \label{eq:exmp:simple3:error:analysis:step:one:11}   \\
&< 2^{s_{\{1\}}} \cdot 2^{s_{\{1,2\}}} \cdot 2^{s_{\{2,3\}}} \cdot 2^{-r_{\{1,2\}}} \cdot 2^{-r_{\{2,3\}}}       \nonumber   \\ 
&< 2^{rS_{\{1\}}+1+rS_{\{1,2\}}+1+rS_{\{2,3\}}+1-(rC_{\{1,2\}}-1)-(rC_{\{2,3\}}-1)}       \nonumber   \\ 
&=2^5\cdot 2^{r(S_{\{1\}}+S_{\{1,2\}}+S_{\{2,3\}}-C_{\{1,2\}}-C_{\{2,3\}})},       \label{eq:exmp:simple3:error:analysis:step:one:22}   
\end{align}
where \eqref{eq:exmp:simple3:error:analysis:step:one:11} holds since there are $(2^{s_{\{1\}}}-1)\cdot(2^{s_{\{1,2\}}}-1)\cdot(2^{s_{\{2,3\}}}-1)$ erroneous tuples $(\wh_{\{1\}},w_{\{2\}},\wh_{\{1,2\}},w_{\{3\}},\wh_{\{2,3\}})$ where $\wh_K \neq w_K, K \in M_4$, 
and for each erroneous composite index tuple, it is mapped to the same codeword $y_J$ as the correct tuple for all $J\in \{ \{1,2\},\{2,3\} \}$ with probability $1/(2^{r_{\{1,2\}}}\cdot 2^{r_{\{2,3\}}})$ due to the uniform random codebook generation. According to \eqref{eq:exmp:simple3:error:analysis:step:one:22}, $P_e^{M_4}$ tends to $0$ as $r\to \infty$, provided that
\begin{align*}
S_{\{1\}}+S_{\{1,2\}}+S_{\{2,3\}}<C_{\{1,2\}}+C_{\{2,3\}}. 
\end{align*}
Note that the above constraint has appeared in the system of inequalities given by \eqref{eq:first:decode:fixed} in Example \ref{exmp:simple3}. 
Other inequalities given by \eqref{eq:first:decode:fixed} with $i=2$ in Example \ref{exmp:simple3} are required to ensure vanishingly small $P_e^M$ for other $M\subseteq \{ \{1\},\{2\},\{1,2\},\{2,3\} \}$. 
All these inequalities are to be satisfied to ensure a vanishingly small first-step decoding error probability $P_e$ for receiver $2$.  

The error analysis for the second step of decoding can be done in a similar way. Assume that all the composite indices $\wh_K,K\in \Gamma_*(P_2)$ have been correctly decoded. In the second step of decoding, receiver $2$ tries to decode messages $\xvh_{\Delta_2}=\xvh_{\{1,2\}}$ from the decoded composite indices $\wh_K,K\in \Gamma_*(P_2)$ and its side information $\xv_{A_2}=\xv_{\{3\}}$. 
The decoding error probability $P_e$ is the probability that there exists some message tuple $(\xh_1,\xh_2,x_3)$ other than the correct tuple $(x_1,x_2,x_3)$ such that they are mapped to the same composite index $w_K$ for every $K\in 2^{\Delta_2\cup A_2} \cap \Gamma_*(P_2)=2^{\{1,2,3\}}\cap \{ \{1\},\{2\},\{1,2\},\{3\},\{2,3\} \}=\{ \{1\},\{2\},\{1,2\},\{3\},\{2,3\} \}$. 
We partition this error event according to the erroneous message set $L \subseteq \Delta_2=\{ 1,2 \}$. That is, $\xh_j \neq x_j$ iff $j \in L$.  
Therefore, by the union bound, we have $P_e\leq \sum_{L \subseteq \{ 1,2 \}}P_{e}^L$, where 
\begin{align}
P_e^L
&\doteq \sum_{\substack{(\xh_1,\xh_2):\\ \xh_j \neq x_j, j \in L, \\ \xh_j = x_j, j \notin L} }\mathrm{P}\left\{\bigcap_{\substack{K\in \{ \{1\},\{2\},\\\{1,2\},\{3\},\{2,3\} \}, \\K \cap L\neq \emptyset}}\left\{\wh_K = w_K(\xvh_K)\right\}\right\}.  \nonumber  
\end{align}
To ensure vanishingly small $P_e$, each $P_e^L$ has to be vanishingly small. In particular, we present detailed analysis for $P_e^L$ with $L=\Delta_2=\{1,2\}$, while other $P_e^L$ can be analyzed similarly. We have
\begin{align}
P_e^L
&=\sum_{\substack{(\xh_1,\xh_2):\\ \xh_1 \neq x_1, \xh_2 \neq x_2} }\mathrm{P}\left\{\bigcap_{\substack{K\in \{ \{1\},\{2\},\\\{1,2\},\{2,3\} \} }}\left\{\wh_K = w_K(\xvh_K)\right\}\right\}   \nonumber    \\
&=\frac{ (2^{t_1}-1) \cdot (2^{t_2}-1) }{ 2^{s_{\{1\}}} \cdot 2^{s_{\{2\}}} \cdot 2^{s_{\{1,2\}}} \cdot 2^{s_{\{2,3\}}} }  \label{eq:exmp:simple3:error:analysis:step:two:12} \\
&<2^{t_1} \cdot 2^{t_2} \cdot 2^{-s_{\{1\}}} \cdot 2^{-s_{\{2\}}} \cdot 2^{-s_{\{1,2\}}} \cdot 2^{-s_{\{2,3\}}}  \nonumber  \\
&<2^{rR_1+1+rR_2+1-rS_{\{1\}}-rS_{\{2\}}-rS_{\{1,2\}}-rS_{\{2,3\}}}  \nonumber  \\
&=2^2 \cdot 2^{r(R_1+R_2-S_{\{1\}}-S_{\{2\}}-S_{\{1,2\}}-S_{\{2,3\}})},       \label{eq:exmp:simple3:error:analysis:step:two:22}
\end{align}
where \eqref{eq:exmp:simple3:error:analysis:step:two:12} holds since there are $(2^{t_1}-1) \cdot (2^{t_2}-1)$ erroneous message tuples,
and for each erroneous tuple, it is mapped to the same composite index $w_K$ as the correct tuple for all $K\in \{ \{1\},\{2\},\{1,2\},\{2,3\} \}$ with probability $1/(2^{s_{\{1\}}} \cdot 2^{s_{\{2\}}} \cdot 2^{s_{\{1,2\}}}\cdot 2^{s_{\{2,3\}}})$ due to the uniform random codebook generation. According to \eqref{eq:exmp:simple3:error:analysis:step:two:22}, $P_e^{L}$ tends to $0$ as $r\to \infty$, provided that
\begin{align*}
R_1+R_2<S_{\{1\}}+S_{\{2\}}+S_{\{1,2\}}+S_{\{2,3\}}.
\end{align*}
Note that the above constraint  appears in the system of inequalities given by \eqref{eq:second:decode:fixed} in Example \ref{exmp:simple3}. 
Other inequalities given by \eqref{eq:second:decode:fixed} with $i=2$ in Example \ref{exmp:simple3} are required to ensure vanishingly small $P_e^L$ for other $L\subseteq \{ 1,2 \}$. 
All these inequalities are to be satisfied to have a vanishingly small second-step decoding error probability $P_e$ for receiver $2$. 
\end{example}

The rate region in Theorem~\ref{thm:compcod} can be represented equivalently as follows.
\begin{proposition} \label{propo:compcod:equiv}
A rate-capacity tuple $(\Rv, \Cv)$ 
is achievable for the distributed index coding problem $(i| A_i)$, $i \in [n]$, 
under a given decoding configuration $(\Pv, \Dv)$ if $R_i = 0$ for all $i\in[n]$ such that $i \notin \Delta_i$. And for $i\in[n]$ such that $i \in \Delta_i$, \eqref{eq:second:decode:fixed} holds and
\begin{align}\label{eq:first:decode:fixed:minli}
\sum_{K \in \Gamma_*(Q)\setminus \Gamma_*(P_i \setminus Q)\setminus 2^{A_i}} S_K < \sum_{J \in Q} C_J, \qquad \forall Q \subseteq P_i.
\end{align}
\end{proposition}
Here, the summand on the LHS of \eqref{eq:first:decode:fixed:minli} signifies the set of composite indices that can be accessed only by the servers in $Q$ (and not by the servers in $P_i \setminus Q$), and that are not generated freely from the side information $A_i$. A formal proof of Proposition~\ref{propo:compcod:equiv} is provided in Appendix \ref{app:compcod:equiv}. 

\begin{remark}\label{rem:own:earlier}
In previous work~\cite{isit:2017, Sadeghi--Arbabjolfaei--Kim2016}, composite index rates had been split across servers, where server-specific rates, $S_{K,J}$, $K\subseteq J$, were limited by the corresponding server capacity, $C_J$. However, as demonstrated by \cite{Li--Ong--Johnson--ISIT--2017, Li--Ong--Johnson2017} such rate splitting can be suboptimal, and cooperative composite coding (CCC) can generally achieve tighter inner bounds on the capacity region of distributed index coding problems, where the same subset of messages are mapped to the same composite index at different servers. Subsequently, composite indices $w_K$ and their corresponding rates $S_K$ are not server-specific. In the current work, we have adopted cooperative compression of composite indices as baseline. 
\end{remark}

\begin{remark}\label{rem:own:ournovelty}
Compared to all previous work~\cite{isit:2017, Sadeghi--Arbabjolfaei--Kim2016,Li--Ong--Johnson--ISIT--2017, Li--Ong--Johnson2017}, we have introduced user-specific decoding server groups, $P_i\subseteq N$, $i\in [n]$. Compared to \cite{Li--Ong--Johnson--ISIT--2017, Li--Ong--Johnson2017} we use a more flexible enhanced fractional allocation of link capacities over decoding configurations (see Section \ref{sec:var}), which was also reported in earlier work \cite{isit:2017}. See Remarks~\ref{rem:better:than:minli:pi} and \ref{rem:better:than:minli:convex}, as well as Examples~\ref{ex:tighter:pi} and \ref{ex:tighter} for more details on how these improvements can lead to generally tighter inner bounds on the capacity region.
\end{remark}

\begin{remark}\label{rem:better:than:minli:pi}
If $C_J = 0$ for some $J \in N$, we can limit our attention to the set of \emph{active} servers $J$ with positive capacity, denoted by  $N_A = \{ J \in N: C_J > 0 \}$. 
Our results in Theorem~\ref{thm:compcod} and Proposition~\ref{propo:compcod:equiv} can easily incorporate the set of active servers $N_A$, which can reduce the computational complexity of characterizing the rate region. 
Example \ref{exmp:simple4} in Appendix \ref{app:example} illustrates how the rate region is easily specialized when active servers $N_A$ is a strict subset of $N$. 
Example \ref{exmp:simple4} also shows an instance of equivalence of Proposition \ref{propo:compcod:equiv} and Theorem \ref{thm:compcod}.
Note that the results in \cite{Li--Ong--Johnson--ISIT--2017,Li--Ong--Johnson2017} are presented based on the set of active servers $N_A$.
\end{remark}

We now present a few simplifications of Theorem~\ref{thm:compcod}. First, setting $P_i = N$, $i \in [n]$ yields  
the following.
\begin{corollary}\label{corr:allservers}
A rate-capacity tuple $(\Rv, \Cv)$ 
is achievable for the distributed index coding problem $(i| A_i)$, $i \in [n]$,
under given decoding message sets $\Dv$ if
\begin{align}
\sum_{j \in L} R_j &< \sum_{\substack{K \subseteq D_i \cup A_i,\\ K \cap L \neq \emptyset}} S_K, \qquad \forall L \subseteq D_i, \, i\in[n],
\label{eq:corr:allservers:second:decode}
\intertext{and}
\sum_{K \in M} S_K &< \sum_{J \in \Gamma^*(M)} C_J, \quad \enskip \forall M \subseteq N \setminus2^{A_i}, \, i\in[n].
\label{eq:corr:allservers:first:decode}
\end{align}
\end{corollary}
The simplification in Corollary~\ref{corr:allservers} can still result in a tight sum-rate.
\begin{example}\label{exmp:case14}
Consider the distributed index coding problem $(1|-),(2|4),(3|4),(4|3)$ with equal unit link capacities $C_J = 1$ for all $J \in N \setminus \{\emptyset\}$. Choose $P_i = N$, $i \in [n]$. Choose $D_1 = \{1\}$ and $D_i = [n]\setminus A_i$ for $i = 2, 3, 4$. Maximizing the sum-rate under
the constraints \eqref{eq:corr:allservers:second:decode} and~\eqref{eq:corr:allservers:first:decode} 
results in $R_1+R_2+R_3+R_4 < 21$, which is the sum-capacity of this index coding problem under equal link capacities; see Example~\ref{exm:uv} for the matching upper bound.
\end{example}

We now further simplify Corollary~\ref{corr:allservers} by choosing the composite rates explicitly (and potentially suboptimally) as $S_K = C_K$, $K \in N$, which essentially prevents cooperation among the servers and forces server~$J$ to transmit the composite index $w_J$ by a one-to-one mapping $y_J(w_J(\xv_J))$.
\begin{corollary}\label{corr:simplify}
A rate-capacity tuple $(\Rv, \Cv)$ 
is achievable for the distributed index coding problem $(i| A_i)$, $i \in [n]$,
under given decoding message sets $\Dv$ if
\begin{align}\label{eq:second:decode:corollary}
\sum_{j \in L} R_j &< \sum_{\substack{J \subseteq D_i \cup A_i,\\ J \cap L \neq \emptyset}} C_J, \qquad \forall L \subseteq D_i, \, i\in[n].
\end{align}
\end{corollary}
With no need for Fourier--Motzkin elimination of the composite index rates, the rate region in Corollary~\ref{corr:simplify} can be easily evaluated.
\begin{example}\label{exm:no:sK}
Consider the distributed index coding problem $(1|4),(2|4),(3|2),(4|3)$ with equal unit link capacities $C_J = 1$ for all $J \in N \setminus \{\emptyset\}$. Choose $D_i = [n]\setminus A_i$, $i\in [n]$. Corollary~\ref{corr:simplify} then simplifies to the set of inequalities
$R_i < 8$, $i\in [n]$, $R_i+R_j < 12$, $i \ne j\in [n]$, $R_1+R_2+R_3 < 14$, $R_1+R_2+R_4 < 14$, 
and $R_1+R_3+R_4 < 14$. For this problem, Theorem~\ref{thm:compcod} (with no pre-specified restriction on $S_K$) yields the same rate region under equal link capacities.
\end{example}

\subsection{Enhanced Fractional Composite Coding}\label{sec:var}

The main idea behind enhanced fractional composite coding is to allow message rates $R_i$, $i \in [n]$, and composite index rates $S_K$, $K \subseteq [n]$, to be a function of the decoding configuration $(\Pv, \Dv)$ at the receivers. More formally, let $\Pc_i = 2^N \setminus \{\emptyset\}$ be the set of all possible nonempty decoding server groups
and $\Dc_i = \{D_i \subseteq [n]\setminus A_i: i\in D_i\}$ be the set of all possible decoding message sets
at receiver $i$. Whenever we refer to a decoding configuration $(\Pv, \Dv)$, we refer to a decoding server group tuple
$\Pv = (P_i, i \in [n]) \in \prod_{i=1}^{n}\Pc_i$ and a decoding message set tuple $\Dv = (D_i, i \in [n]) \in \prod_{i=1}^{n}\Dc_i$. Recall that $\Delta_i = (\bigcup_{J \in P_i} J)\cap D_i$, for each $i\in [n]$ and $(\Pv, \Dv)$.

{Let $r \in \mathbb{N}$. Let for each $(\Pv, \Dv)$ and $i \in [n]$, $t_i(\Pv,\Dv) = \lceil{rR_i(\Pv,\Dv)\rceil}$, where $R_i(\Pv,\Dv)$ is the rate of message $i$ communicated via decoding configuration $(\Pv,\Dv)$. Let $x_i(\Pv,\Dv)\in [2^{t_i(\Pv,\Dv)}]$ be the part of message $i$ communicated via decoding configuration $(\Pv, \Dv)$. Denote $s_K(\Pv,\Dv) = \lceil{rS_K(\Pv,\Dv)\rceil}$, $K \subseteq [n]$, where $S_K(\Pv,\Dv)$ is the rate of composite index $K$ and configuration $(\Pv, \Dv)$. Denote $r_J(\Pv) = \lfloor{r{C_J(\Pv)}\rfloor}$, where $C_J(\Pv)$ is the fractional capacity of server $J$ for decoding server group $\Pv$. By convention, $s_\emptyset(\Pv,\Dv) = S_\emptyset(\Pv,\Dv) = 0$ for each $(\Pv, \Dv)$.}

\begin{theorem}\label{thm:compcod:fractional}
{A rate-capacity tuple $(\Rv, \Cv)$ 
is achievable for the distributed index coding problem $(i| A_i)$, $i \in [n]$, if
\begin{align}
R_i &= \sum_{\Pv,\Dv} R_i(\Pv,\Dv), \qquad i \in [n],\\
C_J &= \sum_{\Pv} C_J(\Pv), \qquad J \in N,\label{enhanced:fractional}
\end{align}
for some $R_i(\Pv,\Dv)$, $C_J(\Pv)$, and $S_K(\Pv,\Dv) \geq 0$ that satisfy
\begin{align}\label{eq:second:decode:frac}
\sum_{j \in L} R_j(\Pv, \Dv) &< \sum_{\substack{K \subseteq \Delta_i \cup A_i,\\ K \in \Gamma_*(P_i),\\ K \cap L \neq \emptyset}} S_K (\Pv, \Dv), \qquad \forall L \subseteq \Delta_i,
\end{align}
and

\begin{align}
\sum_{\Dv}\sum_{K \in M} S_K(\Pv, \Dv)  
<&\sum_{J \in \Gamma^*(M)\,\cap \, P_i} C_J(\Pv), \nonumber  \\&  \qquad\forall M \subseteq \Gamma_*(P_i) \setminus2^{A_i},\label{eq:first:decode:frac}
\end{align}
for $i\in [n]$ such that $i \in \Delta_i$ and for $J \in N$ such that $J \in \bigcup_{i\in [n]}P_i $. Otherwise, set $R_i(\Pv,\Dv) = 0$ for $i\in [n]$ such that $i \notin \Delta_i$ and set $C_J(\Pv) = 0$ for $J \in N$ such that $J \notin \bigcup_{i\in [n]}P_i $.}
\end{theorem}

We now outline the  coding scheme corresponding to the achievable rate region of Theorem \ref{thm:compcod:fractional}. The details of error analysis is omitted for brevity and follows similar steps as in the proof of Theorem \ref{thm:compcod}. 

{\textbf{Codebook generation:} \emph{Step 1.} For each $K \subseteq [n]$, $(\Pv, \Dv)$, and each value of $\xv_K(\Pv,\Dv)$, a corresponding composite index $w_{K,\Pv,\Dv}(\xv_K(\Pv,\Dv))$ is drawn independently and uniformly at random from $[2^{s_K(\Pv,\Dv)}]$. That is,
\begin{align*}
w_{K,\Pv,\Dv}: \prod_{j \in K} [2^{t_j(\Pv,\Dv)}] \to [2^{s_{K}(\Pv,\Dv)}].
\end{align*}
\emph{Step 2.} For each server $J \in N$, decoding server group tuple $\Pv\in \prod_{i=1}^{n}\Pc_i $, and each value of composite index tuple $(w_{K,\Pv,\Dv}, (K,\Dv)\in 2^{J}\times\prod_{i=1}^{n}\Dc_i)$, a fractional server index $y_{J,\Pv}((w_{K, \Pv,\Dv}, (K,\Dv)\in 2^{J}\times\prod_{i=1}^{n}\Dc_i))$ is drawn independently and uniformly at random from $[2^{r_J(\Pv)}]$.  That is,
\begin{align*}
y_{J, \Pv}: \prod_{K \in 2^J}  \prod_{\Dv\in \prod_{i=1}^{n}\Dc_i}[2^{s_K(\Pv,\Dv)}] \to [2^{r_J(\Pv)}].
\end{align*}
For each $J \in N$, the final codeword $y_J$ is the deterministic concatenation of the fractional server index tuples, $(y_{J,\Pv}, \Pv\in \prod_{i=1}^{n}\Pc_i)$. The random codebook 
\begin{align*}
&\{(w_{K,\Pv,\Dv}(\xv_K(\Pv,\Dv)), K \subseteq [n], \Pv\in \prod_{i=1}^{n}\Pc_i, \Dv\in \prod_{i=1}^{n}\Dc_i), \\
&(y_{J,\Pv}((w_{K, \Pv,\Dv}, \!(K,\Dv)\in 2^{J}\!\times\!\prod_{i=1}^{n}\Dc_i)),\! J \in N, \Pv \!\in \!\prod_{i=1}^{n}\Pc_i)\}
\end{align*}
is revealed to all corresponding parties. See Footnote \ref{footnote:parties}.}

{\textbf{Encoding:} To communicate messages $\xv_{[n]}$, 
each server $J \in N$ computes $w_{K,\Pv,\Dv}(\xv_K(\Pv,\Dv))$ for each $K \in 2^J$ and $(\Pv, \Dv)$, as well as $y_{J,\Pv}((w_{K, \Pv,\Dv}, (K,\Dv)\in 2^{J}\times\prod_{i=1}^{n}\Dc_i))$ for each $\Pv$ and then transmits the codeword $y_J = (y_{J,\Pv}, \Pv\in \prod_{i=1}^{n}\Pc_i)$.}

\textbf{Decoding:} \emph{Step 1.} For each $i \in [n]$ and each $\Pv$, receiver $i$ finds the unique tuple $(\hat w_{K, \Pv,\Dv}, (K,\Dv)\in\Gamma_*(P_i) \times\prod_{i=1}^{n}\Dc_i)$ such that $y_{J,\Pv} = y_{J,\Pv}((\hat w_{K, \Pv,\Dv}, (K,\Dv)\in 2^{J}\times\prod_{i=1}^{n}\Dc_i))$ for every $J \in P_i$. If there is more than one such tuple, it declares an error. \emph{Step 2.} Assuming Step 1 is correctly executed and for each $i\in [n]$ and each $(\Pv, \Dv)$ such that $i \in \Delta_i$, receiver $i$ finds the unique message tuple $\xvh_{\Delta_i}(\Pv,\Dv)$
such that $\wh_{K,\Pv,\Dv} = w_{K,\Pv,\Dv}(\xvh_K(\Pv,\Dv))$ for every $K \in \Gamma_*(P_i)$ with $K\subseteq \Delta_i \cup A_i$. If there is more than one such tuple, it declares an error.

A few remarks are in order.

\begin{remark}\label{rem:simplification}
Computing the rate region in Theorem~\ref{thm:compcod:fractional} over all decoding configurations $(\Pv, \Dv)$ is quite expensive, but a few simplifications are possible. First, as mentioned earlier, if $i \notin \Delta_i = D_i \cap (\bigcup_{J \in P_i} J)$, then the corresponding $R_i(\Pv,\Dv) = 0$. {Also if $J \notin \cup_{i\in [n]}P_i $, then $C_J(\Pv) = 0$}. Second, as mentioned in Remark~\ref{rem:better:than:minli:pi},
we can focus on active servers and consider $P_i \subseteq N_A = \{ J \in N: C_J > 0 \}$. {Third, it suffices to consider subset complete decoding server groups, such that  $P_i = \Gamma_*(P_i)$. This is because, all subsets of indices that can be generated by the servers in $P_i$ (except those that are already known) appear in the LHS of \eqref{eq:first:decode:frac}. That is, composite indices $K$ in $M \subseteq\Gamma_*(P_i)\setminus 2^{A_i}$ appear on the LHS of \eqref{eq:first:decode:frac}. However, in the RHS  of \eqref{eq:first:decode:frac}, the contributing server capacities are ``cut" by $P_i$. Therefore, the region becomes no smaller if we use subset completion of $P_i$, $\Gamma_*(P_i)$, instead of $P_i$.}
\end{remark}

\begin{remark}\label{rem:better:than:minli:convex}
A less general version of the fractional coding over different decoding configurations was proposed in the cooperative composite coding (CCC) scheme \cite{Li--Ong--Johnson2017}. In our notation, CCC uses
the same decoding server group $P_i = P \subseteq N$ for all the receivers and {for a fixed decoding message set tuple $\Dv$, the corresponding achievable rate region $\Rr_\mathrm{CCC}(\Dv)$ can be written as}
\begin{align}\label{eq:average:capacity:constraint:ccc:rem3}
R_i &= \sum_{P} R_i(P), \qquad i \in [n],\\
C_J &= \sum_{P} C_J(P), \qquad J \in N, \label{eq:average:capacity:constraint:ccc}
\end{align}
for some $R_i(P)$, $S_K(P)$, and $C_J(P)$ such that
\begin{align}
\sum_{j \in L} R_j(P) &< \sum_{\substack{K \subseteq \Delta_i \cup A_i,\\ K \in \Gamma_*(P),\\ K \cap L \neq \emptyset}} S_K (P), \qquad \forall L \subseteq \Delta_i,\\
\sum_{K \in M} S_K(P) &< \sum_{J \in \Gamma^*(M)\,\cap \, P} C_J(P), \quad \enskip \forall M \subseteq \Gamma_*(P) \setminus2^{A_i},\label{eq:average:capacity:constraint:ccc:rem3:end}
\end{align}
 {for $i \in [n]$ such that $i \in \Delta_i$ and for $J \in N$ such that $J \in P$. Otherwise, set $R_i(P) = 0$ for $i\in [n]$ such that $i \notin \Delta_i$  and set $C_J(P) = 0$ for $J \in N$ such that $J \notin P$. Here $\Delta_i = (\bigcup_{J \in P} J)\cap D_i$, for each $i\in [n]$ and $(P, \Dv)$.}
By taking time sharing over different $\Rr_\mathrm{CCC}(\Dv)$ regions, the convex hull of $\bigcup_\Dv \Rr_\mathrm{CCC}(\Dv)$ is then achievable.
The fractional composite coding scheme in Theorem~\ref{thm:compcod:fractional} is more general in two aspects. First, as mentioned above, our coding scheme allows more degrees of freedom in choosing the decoding server groups $P_i\subseteq N$ that are receiver-dependent.\footnote{This is also an improvement over our preliminary work \cite{isit:2017}.} Second, more subtly, our coding scheme requires the {fractional} link capacity constraints to be satisfied on average over {$\Dv$ (cf.~\eqref{eq:first:decode:frac})}, whereas CCC in~\cite{Li--Ong--Johnson2017} requires the link capacity constraints to be satisfied for each $\Dv$ (cf.~\eqref{eq:average:capacity:constraint:ccc:rem3:end}). As illustrated by Examples~\ref{ex:tighter:pi} and~\ref{ex:tighter}, respectively, flexibility in choosing different decoding server groups or averaging {fractional} link capacities over different decoding configurations can strictly increase the achievable rates.
\end{remark}

\begin{example} \label{ex:tighter:pi}
Consider the distributed index coding problem $(1|2,5),(2|3,4),(3|-),(4|2,5), (5|1,2,4)$ {with $C_J = 1$ for $J = J_1 = \{1,2,3\}$ and $J=J_2 = \{1,4\}$, $C_J = 2$ for $J=J_3 = \{1,3,4,5\}$, and  $C_J = 0$ otherwise.} Hence, the set of active servers is $N_A = \{J_1, J_2, J_3\}$. We use a single decoding message set tuple $\Dv$ with
$D_2 = \{2\} $, $D_3 = \{3\}$, and $D_i = [n]\setminus A_i$ for $i = 1, 4, 5$. The sum-rate achievable by CCC, which is computed using \eqref{eq:average:capacity:constraint:ccc:rem3}-\eqref{eq:average:capacity:constraint:ccc:rem3:end} in Remark~\ref{rem:better:than:minli:convex} across all 7 decoding server group tuples $\Pv$, $P_i = P \subseteq N_A$, $i \in [n]$, satisfies $R_1+R_2+R_3+R_4+R_5< 6$. The sum-rate achievable by Theorem \ref{thm:compcod:fractional}, which is computed with variable $P_i$ as a function of $i\in[n]$ using the following 7 randomly found decoding server group tuples, satisfies $R_1+R_2+R_3+R_4+R_5< 7$. Therefore, there can be a benefit in allowing $P_i$ to vary across $i$.

In $\Pv_1$, we set $P_1 = \{J_1, J_2\}$, $P_2 = \{J_3\}$, $P_3 = \{J_2\}$, $P_4 = \{J_2,J_3\}$, and $P_5 = N_A$. 

In $\Pv_2$, we set $P_1 = P_2 = P_4 = \{J_1, J_2\}$,  $P_3 = \{J_2\}$, and $P_5 = \{J_2,J_3\}$.

In $\Pv_3$, we set $P_1 = \{J_2,J_3\}$,  $P_2 = P_5 = \{J_3\}$, $P_3 = \{J_1,J_2\}$, 
and $P_4 = \{J_2\}$.

In $\Pv_4$, we set $P_1 = \{J_1,J_3\}$,  $P_2 = \{J_1,J_2\}$, $P_3 = \{J_2\}$, 
$P_4 = \{J_3\}$, and $P_5 = \{J_1\}$.

In $\Pv_5$, we set $P_1 = \{J_2,J_3\}$,  $P_2 =N_A$, $P_3 = \{J_2\}$, 
$P_4 = \{J_1,J_3\}$, and $P_5 = \{J_2\}$.

In $\Pv_6$, we set $P_1 = \{J_2\}$,  $P_2 = \{J_1,J_2\}$, $P_3 = \{J_2,J_3\}$, 
$P_4 = \{J_1, J_3\}$, and $P_5 =N_A$.

In $\Pv_7$, we set $P_1 = \{J_1, J_3\}$,  $P_2 = \{J_2\}$, $P_3 = P_4 = \{J_1,J_3\}$, 
and $P_5 =N_A$. 

We note that even with slight variations in the above 7 decoding server group tuples, one can still obtain $R_1+R_2+R_3+R_4+R_5< 7$. For example, if we keep $\Pv_2$ to $\Pv_7$ unchanged and in $\Pv_1$ we set $P_1 = \{J_3\}$, $P_2 = N_A$, $P_3 = \{J_1\}$, $P_4 = \{J_2,J_3\}$, and $P_5 = \{J_1, J_2\}$, we still obtain the same sum-rate. Applying Corollary~\ref{cor:pm:touch:aggregate} (see Section~\ref{sec:outer:bounds} for details) with the touch grouping $\Pcal_{\mathrm{t}} =\{T_{\{2,5\}}\cap N_A,T_{\{1,3,4\}}\cap N_A\}$ will also give a matching outer bound on the sum-capacity, thus establishing the  sum-capacity to be 7. 
\end{example}

\begin{example} \label{ex:tighter:pi:region}
In Example \ref{ex:tighter:pi}, we can compute the whole capacity region using FME. In our programs, there were 181 variables to eliminate, which was completed in a few minutes on an Apple iMac 4GHz Intel Core i7 with 16 GB memory and using {Matlab\textsuperscript{\tiny\textregistered}} R2017b and the FME software \cite{fme:permuter}. The achievable rate region is 
\begin{align*}
R_2&< 1, \qquad R_4 <3, \qquad R_5<2,\\
R_1+R_2+R_3&<4, \qquad R_1+R_3+R_4<4,\\
R_2+R_3+R_4&<4, \qquad R_2+R_3+R_5<3.
\end{align*}
Comparison of this region with that obtained using Corollary \ref{cor:pm:finest} (see Section~\ref{sec:outer:bounds} for details) shows the region is tight, thus establishing the capacity region for this problem.
\end{example}

\begin{example} \label{ex:tighter}
Consider the distributed index coding problem $(1|4),(2|1,3,4),(3|1,2,4),(4|1,3), (5|3)$ with $C_J = 1$ for $J = \{1,2,5\},\,
\{1,2,3,5\}$, and $\{2,4,5\}$, and  $C_J = 0$ otherwise. 
The sum-rate achievable by CCC in Remark~\ref{rem:better:than:minli:convex}, after taking the convex hull
over all possible decoding servers $P\subseteq N_A = \{\{1,2,5\},\{1,2,3,5\},\{2,4,5\}\}$ (7 possibilities) 
and $\Dv \in \prod_{i=1}^{n}\Dc_i$ (1024 possibilities according to $D_i \subseteq [n]\setminus A_i, i\in D_i$) 
satisfies $R_1+R_2+R_3+R_4+R_5 < 4$. In contrast, if we set 
$R_i(\Pv,\Dv)$, $C_J(\Pv,\Dv)$, and $S_K(\Pv, \Dv)$, to zero in  Theorem~\ref{thm:compcod:fractional} except
one decoding server group tuple $\Pv$ with $P_i =  N_A$, $i \in [n]$ 
and two decoding message set tuples $\Dv$ with 
$D_1 = \{1\}$, $D_5 = \{5\}$, and $D_i = [n]\setminus A_i$ for $i = 2, 3, 4$,
and $\Dv'$ with $D_1 = \{1,2\}$, $D_5 = \{5\}$, and $D_i = [n]\setminus A_i$ for $i = 2, 3, 4$,
then the sum-rate satisfies $R_1+R_2+R_3+R_4+R_5< 5$, which is strictly larger than
that of CCC. Applying Corollary~\ref{cor:pm:touch:aggregate} (see Section~\ref{sec:outer:bounds} for details) with the touch grouping $\Pcal_{\mathrm{t}} =\{T_{\{1,3\}}\cap N_A,T_{\{2,4,5\}}\cap N_A\}$ will also give a matching outer bound on the sum-capacity, thus establishing the  sum-capacity to be 5.
\end{example}

Just like Theorem~\ref{thm:compcod},
Proposition~\ref{propo:compcod:equiv} and Corollaries~\ref{corr:allservers} and~\ref{corr:simplify} can be extended by fractional allocation of rates over decoding configurations. In the following, we present the extension of 
Corollary~\ref{corr:simplify}  as an illustration. {Fix a single decoding server group tuple $\Pv$ with $P_i = N$ for all $i\in [n]$. Let $S_K(\Pv, \Dv) = S_K(\Dv) = C_K(\Dv)$, $K \in N$, where $C_K(\Dv)$ is to be determined. This essentially prevents cooperation among the servers and turns the coding scheme to
\begin{align*}
w_{J,\Dv}: \prod_{j \in J} [2^{t_j(\Dv)}] \to [2^{C_{J}(\Dv)}], \qquad J \in N,
\end{align*}
where the final codeword $y_J$ is the deterministic concatenation of composite index tuples, $(w_{J,\Dv}, \Dv\in \prod_{i=1}^{n}\Dc_i)$. }

\begin{corollary}\label{corr:simplify:fractional}
A rate-capacity tuple $(\Rv, \Cv)$ 
is achievable for the distributed index coding problem $(i| A_i)$, $i \in [n]$, if
\begin{align}
R_i &= \sum_{\Dv} R_i(\Dv), \qquad i \in [n],\\
C_J &= \sum_{\Dv} C_J(\Dv), \qquad J \in N,
\end{align}
for some $R_j(\Dv)$ and $C_J(\Dv)$ such that
\begin{align}\label{eq:second:decode:frac:corollary}
\sum_{j \in L} R_j(\Dv) < \sum_{\substack{J \subseteq D_i \cup A_i,\\ J \cap L \neq \emptyset}} C_J (\Dv), \qquad \forall L \subseteq D_i,\, i\in[n],
\end{align}
for every $\Dv\in \prod_{i=1}^{n}\Dc_i$.
\end{corollary}


\section{Outer Bounds on the Capacity Region}\label{sec:outer:bounds}

\subsection{Preliminaries}\label{sec:touch}
We start by introducing two definitions that will play crucial roles in the general outer bound and its special cases to be developed henceforth.
\begin{definition}[Touch structure]\label{def:structure:J}
For any set of messages $K\subseteq [n]$, we say that server $J$ \emph{touches} $K$ if $J \cap K \ne \emptyset$ and that $J$ \emph{does not touch} $K$ if $J \cap K = \emptyset$. We denote by $T_K$ the collection of servers that touch $K$ and denote by $T_{\ntouch{K}}$ the collection of servers that do not touch $K$, that is,
\begin{align*}
T_{K} &= \{J \in N: J \cap K \not = \emptyset\},\\
T_{\ntouch{K}} &= \{J \in N: J \cap K=\emptyset\} \\
&= \{J \in N: J \subseteq [n]\setminus K\}  = N \setminus T_K. 
\end{align*}
And furthermore, for two sets of messages $K$ and $L$, 
\begin{align*}
T_{K,L} = T_{L,K} &= \{J \in N: J \cap K \not = \emptyset, J \cap L \not = \emptyset\},\\
T_{K,\ntouch{L}} = T_{\ntouch{L},K} &=  \{J \in N: J \cap K \not = \emptyset, J \cap L = \emptyset\}.
\end{align*}
\end{definition}

\begin{example}
If $n=4$, then 
\begin{align*}
T_{\{1\}}&=\{\{1\},\{1,2\},\{1,3\},\{1,4\},\{1,2,3\},\{1,2,4\},\\
&\qquad \{1,3,4\},\{1,2,3,4\} \},\\
T_{\{1\}, \ntouch{\{2\}}}&=\{\{1\},\{1,3\},\{1,4\},\{1,3,4\}\}.
\end{align*}
\end{example}

\begin{remark} \label{remark:JT:basic}
Any set $T_K$ can be broken into two disjoint subsets $T_{K,L}$ and $T_{K,\ntouch{L}}$.
Thus, $T_{K,L} \subseteq T_K, T_{K,\ntouch{L}} \subseteq T_K$ in general and $T_{K,L} = T_K$ if $K \subseteq L$. 
It is also easy to verify that $T_K \cup T_L = T_{K\cup L}$ and $T_{K\cap L} \subseteq T_{K,L} = T_K \cap T_L$.
Definition~\ref{def:structure:J} can be naturally extended to three or more message sets. 
\end{remark}

In our usual notation, $\Yv_{T_K}$ thus denotes the output random variables
from servers that have at least one message from the message set $K$, e.g., 
$\Yv_{T_{[n]}} = \Yv_N$.
When the context is clear, we shall use the shorthand notation $T_K$ for $\Yv_{T_K}$, e.g., $H(T_K)$ means $H(\Yv_{T_K})$.

\begin{definition}[Valid server grouping]\label{def:grouping}
A \emph{server grouping} $\Pcal=\{P_1,P_2,\cdots,P_m\}$, consisting of $m$ server groups $P_i\subseteq N$, $i \in [m]$, is
said to be \emph{valid} if $\bigcup_{i\in [m]}P_i=N$. Given a server grouping $\Pcal$, we denote by 
$P_G = \bigcup_{i\in G}P_i$ the collection of servers in the server groups identified by 
$G\subseteq [m]$. By convention, $P_\emptyset = \emptyset$.
\end{definition}

Note that the significance and use of server groups $P_i$ in this section is completely different than that in composite coding in Section \ref{sec:dist:compcod}. Also note that we allow overlaps between different server groups in a grouping. This is why $\Pcal$ is called a server grouping rather than a \emph{server partition} that consists of disjoint server groups. Also note that $P_{[m]}=\bigcup_{i\in [m]}P_i=N$. As usual, $\Yv_{P_G}$ denotes the output random variables from the server collection $P_G$, e.g., 
$\Yv_{P_{[m]}}=\Yv_N$. When the context is clear, we shall use the shorthand notation $P_G$ for $\Yv_{P_G}$, e.g., $H(P_G|\Xv_{K^c})$ means $H(\Yv_{P_G}|\Xv_{K^c})$. 

Whenever convenient, we do not include the inactive dummy server, $J = \emptyset$, with $C_\emptyset = 0$ in the server groupings.

\subsection{The Grouping Polymatroidal Outer Bound}  \label{subsec:outer:grouping}

Before establishing the outer bound on the capacity region, 
we review the standing assumptions and conditions of achievable rate--capacity tuples $(\Rv, \Cv)$ and $(\tv, \rv)$ distributed index codes that will be used in the derivations. 
Since the messages are assumed to be independent and uniformly distributed, for any two disjoint sets $K,K' \subseteq [n]$ we have 
\begin{equation} \label{eq:message:independence}
H(\Xv_K|\Xv_{K'}) = H(\Xv_K) = \sum_{i\in K} t_i.
\end{equation}
The \emph{encoding condition} at server $J\in N$ is 
\begin{equation} \label{eq:encoding:condition}
H(Y_J|\Xv_J)=0.
\end{equation}
The \emph{decoding condition} at receiver $i$ stipulates 
\begin{equation} \label{eq:decoding:condition}
H(X_i|\Yv_N,\Xv_{A_i})\le t_i\cdot \d(\e)
\end{equation}
with $\lim_{\e\to0}\d(\e) = 0$ by Fano's inequality. 

We are ready to state the main result of this section, namely, the \emph{grouping polymatroidal (PM) outer bound}.

\begin{theorem}
\label{thm:pm:group} 
If a rate--capacity tuple $(\Rv, \Cv)$ is achievable for the distributed
index coding problem $(i|A_i)$, $i \in [n]$,  then
for any valid server grouping $\Pcal=\{P_1,P_2,\cdots,P_m\}$, 
\begin{align}
R_i \leq f([m],B_i \cup \{i\}) - f([m],B_i), \qquad i\in[n], \label{eq:group:rate}
\end{align}
for some $f(G,K)$, for all $G, G'\subseteq [m]$, $K, K' \subseteq [n]$, such that 
\begin{align}
&f(G,K) = f(G',K),    \nonumber  \\
& \qquad \qquad \qquad \text{if $(P_G\cup P_{G'})\setminus (P_G \cap P_{G'})\subseteq T_{\ntouch{K}}$},   \label{eq:axiom1}
\\
&f(\emptyset,K)= f(G,\emptyset) =0,      \label{eq:axiom2}                                                                               \\
&f(G,K)\leq \sum_{J: J \in P_G, J \in T_K} C_J,    \label{eq:axiom3}                                                                      \\
&f(G,K)\leq f(G',K'),      \qquad   \qquad  \text{if $K\subseteq K', G\subseteq G'$},   \label{eq:axiom4}                      \\
&f(G\cup G',K\cap K') + f(G\cap G',K\cup K')   \nonumber  \\
&\qquad \qquad \leq f(G,K) + f(G',K'),    \label{eq:axiom5}\\
&f([m],B_i\cup \{i\})-f([m],B_i)=f([m],\{i\}), \enskip i\in [n], \hspace{-1mm} \label{eq:axiom:addi:de}\\
&f(G,K)+f(G,K')=f(G,K\cup K'),      \nonumber  \\
&\qquad \qquad \qquad \text{if $K\cap K'=\emptyset, P_G\subseteq (N\setminus T_{K,K'})$}.  \label{eq:axiom6}
\end{align}
\end{theorem}

\begin{IEEEproof}
If rate--capacity tuple $(\Rv,\Cv)$ is achievable, then for every $\e>0$ there exists a $(\tv,\rv)$ distributed index code
satisfying \eqref{eq:model:code}. For any $i\in [n]$, we have
\begin{align}
t_i&= H(X_i|\Xv_{A_i})    \label{eq:group:pm:proof:revision4:0}    \\
&\le H(X_i|\Xv_{A_i}) -  H(X_i|\Yv_N, \Xv_{A_i})+t_i\cdot {\d(\e)}  \label{eq:group:pm:proof:revision4:1} \\
&= I(X_i;\Yv_N | \Xv_{A_i})+t_i\cdot {\d(\e)} \\
&= H(\Yv_N|\Xv_{A_i})-H(\Yv_N|\Xv_{A_i\cup \{i\}})+t_i\cdot {\d(\e)},  \label{eq:group:pm:proof:revision4:2}
\end{align}
where \eqref{eq:group:pm:proof:revision4:0} follows from the fact that the messages are independent and uniformly distributed as specified in \eqref{eq:message:independence}, and \eqref{eq:group:pm:proof:revision4:1} is due to the decoding condition in \eqref{eq:decoding:condition}. 
Now, given the server grouping $\Pcal=\{P_1,P_2,\cdots,P_m\}$, define
\begin{align}
\label{eq:f-def2}
f_{\e}(G,K) \doteq \frac{1}{r}H(P_G|\Xv_{K^c})=\frac{1}{r}H(\Yv_{\{J: J\in \bigcup_{i\in G}P_i\}}|\Xv_{K^c}),  
\end{align}
for $G\subseteq [m]$ and $K\subseteq [n]$.\footnote{Recall that the $(\tv,\rv)$ distributed index code depends on $\e$ and thus so does the set function $f_{\e}(G,K)$. }
Then,
\begin{align}
R_i \leq \frac{t_i}{r} &\le \frac{H(\Yv_N|\Xv_{A_i})-H(\Yv_N|\Xv_{A_i\cup \{i\}})}{r\cdot (1-\d(\e))} \nonumber  \\
&= \frac{f_{\e}([m],B_i \cup \{i\}) - f_{\e}([m],B_i)}{1-\d(\e)},   \label{eq:group:pm:proof:revision4:rate:1}
\end{align}
where the second inequality follows from \eqref{eq:group:pm:proof:revision4:2} and the equality from the definition of $f_{\e}(G,K)$. 

We now show that the set function $f_{\e}(G,K)$ is bounded from above for any $G\subseteq [m]$, $K\subseteq [n]$ and any $\e>0$.
We have
\begin{align}
f_{\e}(G,K)&=\frac{1}{r}H(P_G|\Xv_{K^c})  \nonumber  \\
&\le \frac{1}{r}H(P_G)   \nonumber  \\
&\le \sum_{J\in P_G}\frac{1}{r}H(Y_J)\le \sum_{J\in P_G}\frac{r_J}{r}\le \sum_{J\in P_G}C_J.
\end{align}
Also, $f_{\e}(G,K)$ is bounded from below as $f_{\e}(G,K)\ge 0$ due to the nonnegativity of the entropy function. 

Given the boundedness of $f_\e(G,K)$, its limit infimum
\begin{align}
f(G,K)\doteq \liminf_{\e \to 0}f_{\e}(G,K)   \label{eq:group:pm:proof:def:f}
\end{align}
is real and bounded. Now taking the limit infimum as $\e$ approaches zero on both sides of \eqref{eq:group:pm:proof:revision4:rate:1} yields
\begin{align}
R_i&\le \liminf_{\e \to 0} \frac{f_{\e}([m],B_i \cup \{i\}) - f_{\e}([m],B_i)}{1-\d(\e)}  \\
&=\liminf_{\e \to 0}( f_{\e}([m],B_i \cup \{i\}) - f_{\e}([m],B_i) )       \\
&=\liminf_{\e \to 0} f_{\e}([m],B_i \cup \{i\}) - \limsup_{\e \to 0} f_{\e}([m],B_i )   \\
&\le \liminf_{\e \to 0} f_{\e}([m],B_i \cup \{i\}) - \liminf_{\e \to 0} f_{\e}([m],B_i )   \\
&=f([m],B_i\cup \{i\})-f([m],B_i).\label{eq:rate:leading}
\end{align}
We have thus far established \eqref{eq:group:rate}. It is also checked in Appendix \ref{app:outer:group} that $f(G,K)$ satisfies the conditions in \eqref{eq:axiom1}--\eqref{eq:axiom6}, which establishes Theorem \ref{thm:pm:group}. 
\end{IEEEproof}

\begin{remark} \label{rmk:pm:group}
The conditions \eqref{eq:axiom1}--\eqref{eq:axiom6} in Theorem~\ref{thm:pm:group} will be referred to as the \emph{Axioms} satisfied by the function $f(G,K)$. 
Axioms~\eqref{eq:axiom2}, \eqref{eq:axiom4}, and \eqref{eq:axiom5} capture standard polymatroidal properties of the entropy function. 
Axioms~\eqref{eq:axiom1} and~\eqref{eq:axiom3} capture 
the encoding conditions at servers, as well as the link capacity constraints. 
Axiom~\eqref{eq:axiom6} captures the conditional message independence given by the $\mathrm{fd}$-separation. Refer to Appendix~\ref{sec:fd} for a brief treatment on how $\mathrm{fd}$-separation applies to the distributed index coding problem. 
The inequality \eqref{eq:group:rate} will be referred to as the \emph{rate constraint inequality} jointly satisfied by $\Rv$ and $f(G,K)$, which is based on the message independence, as well as the decoding conditions at receivers. 
\end{remark}

\begin{remark} \label{rmk:additional}
Axiom~\eqref{eq:axiom:addi:de} captures the additional decoding conditions at the receivers, which also applies to the centralized index coding problem \cite{liu:vellambi:kim:sadeghi:itw18}. See Appendix \ref{app:outer:group} for more details. It was found in \cite{liu:vellambi:kim:sadeghi:itw18} that this axiom is strictly needed to obtain tight outer bounds on the capacity region for the \emph{secure} centralized index coding problem. Currently, we are not aware of any instance of \emph{non-secure} centralized or distributed index coding problem for which Axiom~\eqref{eq:axiom:addi:de} can tighten the outer bound. However, we include this axiom for the following reason. Through including the additional decoding conditions in the centralized index coding problem, it was shown in \cite{liu:vellambi:kim:sadeghi:itw18} that the PM outer bound is as tight as the apparently stronger bound in which all Shannon-type inequalities are used. See Remark \ref{rem:shannon} where we discuss a similar relation between the most refined PM outer bound for the distributed index coding problem and the one obtained based on all Shannon-type inequalities of the entropy function. 
\end{remark}

{The tightness and computational complexity of the grouping PM outer bound 
in Theorem~\ref{thm:pm:group} depends pivotally on the specific server grouping $\Pcal$. 
For a fixed problem size $n$, the number of variables in the theorem is exponential in $m$, the size of $\Pcal$. 
To fully compute $(\Rv, \Cv)$ satisfying \eqref{eq:group:rate}--\eqref{eq:axiom6} for a given $\Pcal$, one should use FME to remove all the $2^{m+n}$ intermediate variables $f(G,K)$, $G\subseteq [m]$, $K\subseteq [n]$. In general, this operation is prohibitively complex even for small $m$ and $n$. For a given $\Cv$, however, it is typically tractable to establish an upper bound on the  (weighted) sum-capacity using LP subject to \eqref{eq:group:rate}--\eqref{eq:axiom6}. 
In the next few subsections, we specialize the outer bound using a number of explicit constructions for server grouping. In general, the optimal tightness-complexity tradeoff in choosing a server grouping remains open.}

{Unless otherwise stated, we denote the outer bound given by the grouping PM outer bound in Theorem \ref{thm:pm:group} with a specific server grouping $\Pcal$ as $\Rr_{\Pcal}$. 
For brevity, whenever we say that one server grouping $\Pcal$ is tighter (looser) than another grouping $\Pcal'$, we mean that the outer bound $\Rr_{\Pcal}$ is tighter (looser) than the outer bound $\Rr_{\Pcal'}$.}

\subsection{{Outer Bounds Based on Server Groupings Utilizing the Touch Structure}\label{sec:touch:grouping}}

In this section, we explicitly construct server groupings based on the touch structure of Definition \ref{def:structure:J}. Let us motivate the construction through a series of examples and definitions. Together, they will lead to a closed-form upper bound on the total sum-capacity, which is implied by Theorem \ref{thm:pm:group} with server groupings based on a specific touch structure.

\begin{example} \label{exm:uv:u}
Consider the problem $(1|-),(2|4),(3|4),(4|3)$ discussed in Example \ref{exmp:case14} in Section~\ref{sec:dist:compcod}. Consider two sets $L = \{4,2\}$ and $K = \{1,3\}$, where set $L$ is ordered as $L = \{i_1 = 4, i_2 = 2\}$. For this problem, we can verify that
\begin{align*}
&A_{i_1} = A_4=\{3\} \subset K = \{1,3\},    \\ 
&A_{i_2} = A_2=\{4\} \subset K \cup \{i_1\} = \{1,3,4\}.
\end{align*}
We say that $L$ is an augmentation set of $K$. Similarly, $L = \{2,3\}$ is an augmentation set of $K= \{1,4\}$. When $K= \emptyset$, we find a unique \emph{maximal} augmentation set, called the peripheral. The set $U=\{1\}$ is the peripheral for this problem as $A_1=\emptyset \subseteq \emptyset$, and $A_i\not \subseteq \{1\}$, $i\in \{2,3,4\}$.
\end{example}

Generally, the idea is to find two disjoint subsets $L,K\subseteq [n]$, such that any valid index code \emph{augments} the singular message decoding condition \eqref{eq:decoding:condition} to $H(\Xv_L|\Yv_N,\Xv_K)\le \sum_{i\in L}t_i\cdot\d(\e)$. For the peripheral set $U$, the peripheral decoding condition gives $H(\Xv_U|\Yv_{N})\le \sum_{i\in U}t_i\cdot \d(\e)$. We formalize this through the following definitions.

\begin{definition}[Augmentation set] \label{def:augmentation:set}
For the distributed index coding  problem $(i|A_i),i\in [n]$ and any two disjoint sets $L,K\subseteq [n]$, we say $L$ is an \emph{augmentation set} of $K$ if there exists an ordering $i_1,i_2,\cdots,i_{|L|}$ of the elements in $L$ such that $A_{i_j}\subseteq \{i_1,\cdots,i_{j-1}\}\cup K, j\in [|L|]$. The empty set $\emptyset$ is an augmentation set of any set $K \subseteq [n]$. 
\end{definition}

\begin{definition}[Peripheral] \label{def:uv:u}
For the distributed index coding  problem $(i|A_i),i\in [n]$, we say that set $U\subseteq [n]$ is a \emph{peripheral} if $U$ is an augmentation set of the empty set $\emptyset$, and that for any $i\in U^c$, we have $A_i\not \subseteq U$. 
\end{definition}

For a given problem $(i|A_i),i\in [n]$, peripheral $U$ is unique. This can be verified by contradiction as follows. 
Assume that there exist two different peripherals $U,U'$, and $U\setminus U'\neq \emptyset$. Define $u\doteq |U|$, $U_1\doteq U\setminus U'$ and $U_0\doteq U\cap U'$. Then, $U=U_0\cup U_1$ and $U_1\cap U_0 = \emptyset$. By Definition \ref{def:uv:u}, there exists an ordering $i_1,i_2,\cdots,i_{u}$ of the elements in $U$ such that $A_{i_j} \subseteq \{ i_1,\cdots,i_{j-1} \}$, $j\in [u]$. 
There always exists some $s\in [u]$ such that $i_s\in U_1$ and $\{ i_1,\cdots,i_{s-1} \} \subseteq U_0$. 
Hence, we have $A_{i_s}\subseteq \{ i_1,\cdots,i_{s-1} \} \subseteq U_0 \subseteq U'$. We also have $i_s\in U'^c$ since $i_s\in U_1$ and $U_1\cap U'=\emptyset$. Combining that $A_{i_s}\subseteq U'$ and that $i_s\in U'^c$ leads to a contradiction against Definition \ref{def:uv:u}, and thus completes the proof.

\begin{remark}   \label{remark:decoding:chain}
In \cite{isit:2018,tangliu2018itw}, a decoding chain is established based on the idea that receiver $j$ can mimic another receiver $i$ and decode message $i$ at no cost to the achievable rates if receiver $j$ knows everything receiver $i$ does. This is similar to the notion of augmentation set in Definition \ref{def:augmentation:set}. 
A chaining procedure, starting from a receiver with an empty side information set,  is used in \cite{neely2013dynamic} to prove a lower bound on the broadcast rate of index coding. This is similar to the procedure of building the peripheral in Definition \ref{def:uv:u}.
\end{remark}

Based on Definitions \ref{def:augmentation:set} and \ref{def:uv:u}, we define the augmentation group as follows.

\begin{definition}[Augmentation group] \label{def:uv:v}
For the distributed index coding  problem $(i|A_i),i\in [n]$ with peripheral $U$, we use $\Vv=(V_1,V_2,\cdots,V_k)$, referred to as an \emph{augmentation group}, to denote the tuple of $k$ disjoint nonempty sets $V_1, \cdots,V_k\subseteq U^c$ for some $k\ge 1$ such that the following conditions are satisfied:
        \begin{enumerate}
        \item for any $j\in [k]$, $V_j$ is an augmentation set of its complement set $V^c_j$;
        \item set $W\doteq [n]\setminus U\setminus (\bigcup_{j\in [k]}V_j)$ is an augmentation set of its complement set $W^c$;
         \item there does not exist another tuple of disjoint nonempty sets $\Vv'=(V_1',V_2',\cdots,V_{k'}')$ such that it satisfies the first two conditions, and that $\bigcup_{j'\in [k']}V_{j'}'\subset \bigcup_{j\in [k]}V_j$.
        \item there does not exist another tuple of disjoint nonempty sets $\Vv'=(V_1',V_2',\cdots,V_{k'}')$ such that it satisfies the first two conditions, $\bigcup_{j'\in [k']}V_{j'}'= \bigcup_{j\in [k]}V_j$, and that $k'<k$. 
        \end{enumerate}
\end{definition}

Note that there can be multiple augmentation groups for a given problem $(i|A_i),i\in [n]$. 

\begin{example}\label{exm:v:number}
Consider the problem $(1|-),(2|4),(3|4),(4|3)$ discussed in Examples \ref{exmp:case14} and \ref{exm:uv:u}. We have $U=\{1\}$ and there are in total 2 augmentation groups $\Vv=(\{3\})$ and $\Vv'=(\{4\})$. Note that $\Vv''=(\{3\},\{4\})$ does satisfy the first two conditions of Definition \ref{def:uv:v}, yet given the existence of $\Vv$ and $\Vv'$, according to the third condition, $\Vv''$ is not a valid augmentation group. 
For another example, consider the six-message problem $(1|4),(2|3),(3|2),(4|1),(5|-),(6|5)$. We have $U=\{5,6\}$ and in total 4 augmentation groups shown as follows,
\begin{align*}
&\Vv^1=(\{1,2\}), \qquad \Vv^2=(\{1,3\}), \\ 
&\Vv^3=(\{2,4\}), \qquad \Vv^4=(\{ 3,4 \}).
\end{align*}
Note that $\Vv'=(\{1\},\{2\})$ does satisfy the first two conditions of Definition \ref{def:uv:v}, yet given the existence of $\Vv^1$, according to the fourth condition, $\Vv'$ is not a valid augmentation group. 
\end{example}

A closed-form upper bound on the sum-capacity is given by the following proposition. Note that set $[k+1:k]$ simply means the empty set $\emptyset$. 

\begin{proposition} \label{propo:uv}
For the distributed index coding  problem $(i|A_i),i\in [n]$ with peripheral $U$  and an augmentation group $\Vv=(V_1,V_2,\cdots,V_k)$, we have
\begin{align}
\sum_{i\in [n]}R_i \le \sum_{J\in N}C_J+\sum_{\ell \in [k]} \sum_{J\in \Tcal_\ell}C_J,   \label{eq:uv:propo}
\end{align}
where $\Tcal_\ell \doteq T_{V_{\ell}, (\bigcup_{j\in [\ell+1:k]}V_j)\cup W}$ and $W=[n]\setminus U\setminus (\bigcup_{j\in [k]}V_j)$.
\end{proposition}

We prove the proposition by showing that \eqref{eq:uv:propo} is implied by Theorem \ref{thm:pm:group} with a specific server grouping $\Pcal_{\rm \Vv}$ defined below. The proof details are given in Appendix \ref{app:uv:proof}.  

\begin{definition}[$\Pcal_{\rm \Vv}$ grouping] \label{def:pcal:uv}
For the distributed index coding problem $(i|A_i),i\in [n]$ with an augmentation group $\Vv=(V_1,V_2,\cdots,V_k)$, the server grouping $\Pcal_{\rm \Vv}$ is defined as follows
\begin{align}
\Pcal_{\rm \Vv}=\{ T_{V_1},T_{V_2},\cdots,T_{V_k},T_{(\bigcup_{j\in [k]}V_j)^c} \}. \label{eq:pcal:uv}
\end{align}
\end{definition}

In some cases, Proposition \ref{propo:uv} gives a tight sum-rate result, as illustrated below. 

\begin{example}\label{exm:uv}
Recall the four-message problem $(1|-),(2|4),(3|4),(4|3)$ discussed in Examples~\ref{exmp:case14}, \ref{exm:uv:u} and \ref{exm:v:number}.
The outer bound presented earlier in \cite{Sadeghi--Arbabjolfaei--Kim2016} yields that $R_1+R_2+R_3+R_4 \le 22$.  
In comparison, given the peripheral $U=\{1\}$ and the augmentation group $\Vv=\{ \{3\} \}$, we have $W=[n]\setminus \{1\} \setminus \{3\}=\{2,4\}$, and hence Proposition \ref{propo:uv} tightens the sum-capacity upper bound to 
\begin{align}
R_1+R_2+R_3+R_4  &\le  \sum_{J\in N}C_J+\sum_{J\in T_{\{3\},\{2,4\}}}C_J\nonumber\\&=15+6=21,
\end{align}
which matches the lower bound presented in Example~\ref{exmp:case14}. Note that with another augmentation group $\Vv'=\{ \{4\} \}$, Proposition \ref{propo:uv} yields the same tight upper bound of $21$ on the sum-rate. 
\end{example}

As one can see, $\Pcal_{\rm \Vv}$ is a server grouping whose server groups are in the form of touch structure. To generalize this further, we introduce the touch grouping and its resulting outer bound as follows. 

\begin{definition}[Touch grouping]\label{def:touch:group}
 For a given $m \leq n$ and disjoint nonempty sets $L_i \subseteq [n]$, $i \in [m]$, such that 
$\bigcup_{i\in [m]} L_i = [n]$, the \emph{touch grouping} $\Pcal_{\rm t}$ is defined as
\begin{align}
\Pcal_{\mathrm{t}} =\{T_{L_1},T_{L_2},\cdots,T_{L_m}\}.
\end{align}
\end{definition}

For the special case $m=n$, $L_i = \{i\}$, and $P_i=T_{\{i\}}$, the touch grouping is called the \emph{individual touch grouping} and is denoted by
\begin{align}
\Pcal_{\rm t}^*=\{T_{\{1\}},T_{\{2\}},\ldots,T_{\{n\}}\}.
\end{align}

Note that we have $P_G=\bigcup_{i\in G}T_{L_i}=T_{L_G}$, where $L_G\doteq \bigcup_{i\in G}L_i$. With the touch grouping $\Pcal_{\mathrm{t}}$,  
the grouping PM outer bound in Theorem~\ref{thm:pm:group} simplifies to the following outer bound, simply denoted as $\Rr_{\rm t}$ and referred to as the \emph{touch grouping outer bound}. 

\begin{corollary}\label{cor:pm:touch:aggregate}
The capacity region of the distributed
index coding problem $(i|A_i)$, $i \in [n]$, with link capacity tuple $\Cv$ satisfies $\Cr(\Cv) \subseteq \Rr_{\rm t}$, where $\Pcal_{\rm t} =\{T_{L_1},T_{L_2},\cdots,T_{L_m}\}$ is a valid touch grouping and $\Rr_{\rm t}$ consists of all rate tuples $\Rv$ such that
\begin{align}
R_i \le f([m],B_i\cup \{i\})-f([m],B_i), \qquad i\in[n],       \label{eq:touch:rate}
\end{align}
for some $f(G,K)$ for all $G,G'\subseteq [m],K,K'\subseteq [n]$ satisfying
\begin{align}
&f(G,K) = f(G',K),   \qquad \qquad \text{if $K\subseteq (L_G\cap L_{G'})$},   \label{eq:touch:axiom1}
\\
&f(\emptyset,K)= f(G,\emptyset) =0,      \label{eq:touch:axiom2}                                                                               \\
&f(G,K)\leq \sum_{J: J \in T_{L_{G},K}} C_J,    \label{eq:touch:axiom3}                                                                      \\
&f(G,K)\leq f(G',K'),      \qquad   \qquad  \text{if $K\subseteq K', G\subseteq G'$},   \label{eq:touch:axiom4}                      \\
&f(G\cup G',K\cap K') + f(G\cap G',K\cup K')   \nonumber  \\
&\qquad \qquad \leq f(G,K) + f(G',K'),    \label{eq:touch:axiom5}\\
&f([m],B_i\cup \{i\})-f([m],B_i)=f([m],\{i\}), \enskip i\in [n]. \hspace{-1mm} \label{eq:touch:axiom:addi:de}
\end{align}
\end{corollary}

\begin{IEEEproof}
It is obvious that Axioms \eqref{eq:axiom2}, \eqref{eq:axiom4}, \eqref{eq:axiom5}, and \eqref{eq:axiom:addi:de} in Theorem \ref{thm:pm:group} and Axioms \eqref{eq:touch:axiom2}, \eqref{eq:touch:axiom4}, \eqref{eq:touch:axiom5}, and \eqref{eq:touch:axiom:addi:de} above do not depend on the underlying server grouping, and thus remain unchanged. Since $P_G=\bigcup_{j\in G}T_{L_j}=T_{L_G}$, Axioms \eqref{eq:axiom3} and \eqref{eq:touch:axiom3} are the same. 

Note that $[n]\in T_{L}$ for any nonempty $L\subseteq [n]$, which indicates that the server $J = [n]$ containing all messages is common among all server groups in the touch grouping $\Pcal_{\rm t}$. 
Hence, with $\Pcal_{\rm t}$, there never exists two disjoint nonempty sets $K,K'\subseteq [n]$ such that $T_{L_G}\subseteq (N\setminus T_{K,K'})$ for any nonempty $G\subseteq [m]$. 
This implies that Axiom \eqref{eq:axiom6} in Theorem \ref{thm:pm:group}, namely, the $\mathrm{fd}$-separation axiom, can only give trivial inequalities (e.g., $f(\emptyset,K)+f(\emptyset,K')=f(\emptyset,K\cup K')$), and hence, in Corollary \ref{cor:pm:touch:aggregate}, there is no axiom corresponding to the $\mathrm{fd}$-separation axiom. 

It remains to prove that Axiom \eqref{eq:axiom1} in Theorem \ref{thm:pm:group} simplifies to Axiom \eqref{eq:touch:axiom1}, which is shown in Appendix \ref{app:pm:touch:aggregate}.
\end{IEEEproof}

Within the general class of touch grouping, it is unclear which touch grouping can give the tightest capacity outer bound with the lowest possible computational cost. Since $\Pcal_{\rm \Vv}$ grouping has an explicit construction, our proposed approach is to first try this grouping and compare the obtained performance bound with an achievable coding scheme. If the results match, no further action is required. Otherwise, the finest touch grouping $\Pcal_{\rm t}^*=\{T_{\{1\}},T_{\{2\}},\ldots,T_{\{n\}}\}$ can be tried, which results in the tightest outer bound on the capacity region among all possible touch  groupings. See Section \ref{sec:hierarchy} for the hierarchy of server groupings in terms of their tightness. 
Note that the outer bound in \cite{isit:2017} is also established based on the individual touch grouping and is identical to the touch grouping outer bound, Corollary \ref{cor:pm:touch:aggregate}, but without Axiom \eqref{eq:touch:axiom:addi:de}.

\begin{remark}  \label{rem:simplified:axiom:na}
The grouping PM outer bound of Theorem \ref{thm:pm:group} can easily incorporate the set of active servers $N_A = \{ J \in N: C_J > 0 \}$. 
One can simply replace $N$ with $N_A$ and $\Pcal=\{P_1,\cdots,P_m\}$ with $\Pcal_A=\{P_1\cap N_A,\cdots,P_m\cap N_A\}$. 
However, notice that the axioms of Corollary \ref{cor:pm:touch:aggregate} (and those of Corollary \ref{cor:pm:coarsest} to be introduced in Section \ref{sec:hierarchy}) are expressed in simplified forms based on the assumption that all the servers $J\in N$ are active. When $N_A\subset N$, using these simplified axioms might result in looser outer bounds. One can avoid this issue by using the axioms in their original unsimplified forms of Theorem \ref{thm:pm:group} with the desired server grouping. 
\end{remark}

\subsection{{Outer Bounds Based on Server Groupings Utilizing $\mathrm{fd}$-separation}}  \label{sec:outer:bounds:fd}
As discussed in the proof of Corollary \ref{cor:pm:touch:aggregate}, one limitation of the touch grouping outer bound is the missing $\rm fd$-separation axiom. 
To show the usefulness of the $\rm fd$-separation axiom, we present a list of problems with discussion, leading to a construction of server grouping based on $\mathrm{fd}$-separation. Consider the directed side information graph  $\Gcal=(\Vcal,\Ecal)$ of the index coding problem \cite{foundation} with $n$ vertices, where vertex $v_i$ represents message $i$. There exists a directed edge from vertex $v_i$ to vertex $v_j$ if and only if $i \in A_j$. A set of vertices $\{ v_{i_1},v_{i_2},\cdots,v_{i_k} \} \subseteq \Vcal$ form a \emph{cycle} if there is a directed edge from vertex $v_{i_j}$ to $v_{i_{j+1}}$ for any $j\in [k-1]$, and there is a directed edge from $v_{i_k}$ to $v_{i_1}$.

\begin{definition}[Isolated vertex and disjoint cycles]\label{def:isolated}
For the distributed index coding problem $(i|A_i),i\in [n]$ with side information graph $\Gcal=(\Vcal,\Ecal)$, a vertex $v\in \Vcal$ is said to be \emph{isolated} if it has no incoming edges. That is, there does not exist any edge $e =(v', v)\in \Ecal$ for some $v' \in [n]$.
Two cycles $K,K'\subseteq \Vcal$ in $\Gcal$ are said to be \emph{disjoint} if $K\cap K'=\emptyset$.
\end{definition}

\begin{example}\label{exm:fd:1}
Consider the following 6 distributed index coding problems with $n=4$ and equal link capacities $C_J = 1, J \in N\setminus \{ \emptyset \}$,
\begin{align}  \label{eq:exm:fd:1}
\begin{array}{c}
(1|-),(2|4),(3|2),(4|3),  \\  (1|-),(2|4),(3|2),(4|1,3),    \\
(1|-),(2|1,4),(3|1,2),(4|1,3),  \\  (1|4),(2|3),(3|2),(4|1,3),   \\
(1|4),(2|3),(3|2),(4|1,2,3),  \\  (1|4),(2|3),(3|1,2),(4|1,2),  
\end{array}
\end{align}
whose side information graphs are shown in Figure \ref{fig:fd:1}. 
For the problems shown in Figures \ref{fig:t1-a}, \ref{fig:t1-b}, and \ref{fig:t1-c}, the touch grouping outer bound $\Rr_{\rm t}$ yields $\sum_{i\in [4]}R_i\leq 19.5$. With $\Pcal=\{ P_{1},N\setminus P_1 \}$, where $P_{1}=\{ J\in N:|J\setminus \{1\}|\leq 1 \}$, a tighter upper bound of $19$ can be obtained by the grouping PM outer bound for these three problems, matching the sum-capacity. For the problems shown in Figures \ref{fig:t1-d}, \ref{fig:t1-e}, and \ref{fig:t1-f}, the touch grouping outer bound $\Rr_{\rm t}$ yields $\sum_{i\in [4]}R_i\leq 24$. With $\Pcal'=\{P'_1, P'_2,N\setminus P'_1\setminus P'_2 \}$, where $P'_1=\{ \{1\},\{2\},\{3\},\{4\} \},P'_2=\{ J\in N: |J|=2,J\in T_{\{1,4\},\{2,3\}} \}$, a tighter upper bound of $23.5$ can be obtained by the grouping PM outer bound for these three problems, matching the sum-capacity. 
For all these six problems in Figure \ref{fig:fd:1}, the achievable rates are obtained through Theorem~\ref{thm:compcod} with $(\Pv,\Duv)$, where $P_i=N$, $i\in [n]$, and $\Duv$ is generated by Algorithm \ref{alg:natural:D} presented in Section \ref{sec:numerical}. 
\begin{figure}[ht]
\begin{center}
\subfigure[][]{
\label{fig:t1-a}
\includegraphics[scale=0.25]{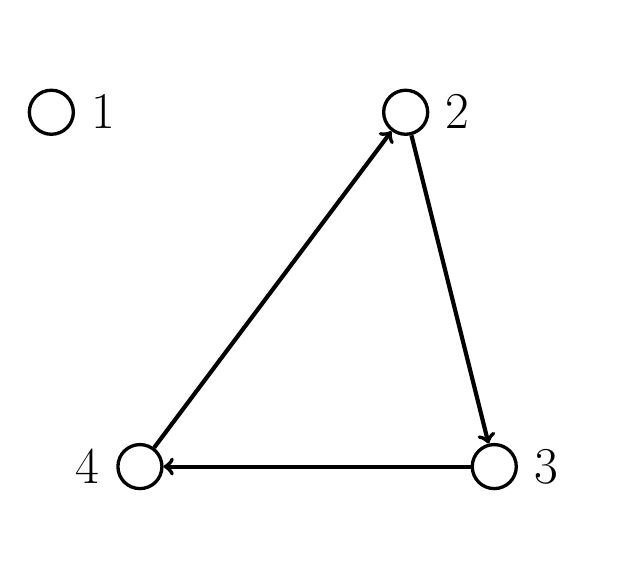}}
\subfigure[][]{
\label{fig:t1-b}
\includegraphics[scale=0.25]{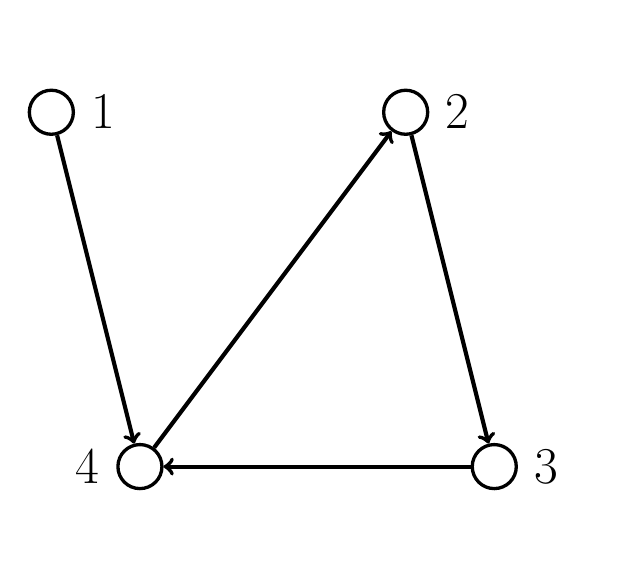}}
\subfigure[][]{
\label{fig:t1-c}
\includegraphics[scale=0.25]{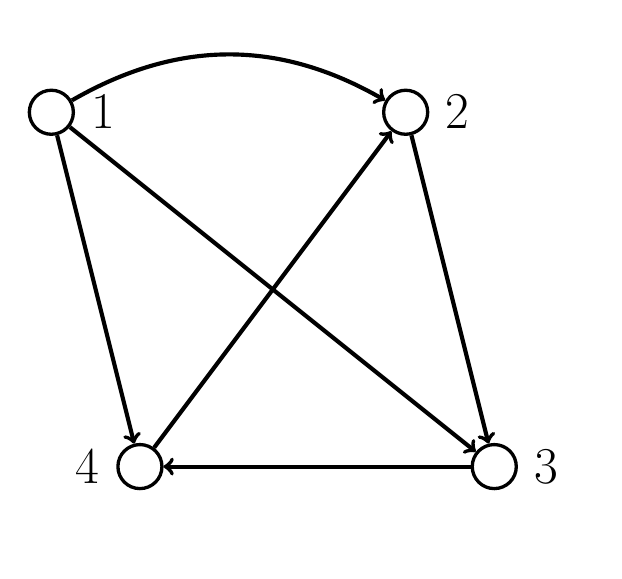}}
\subfigure[][]{
\label{fig:t1-d}
\includegraphics[scale=0.25]{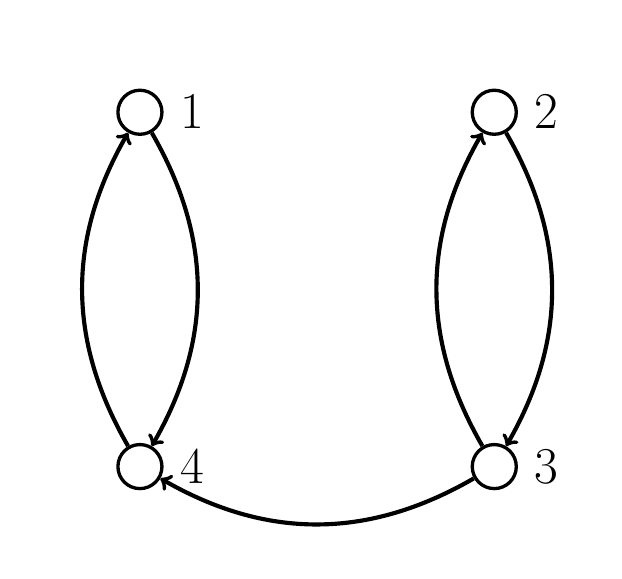}}
\subfigure[][]{
\label{fig:t1-e}
\includegraphics[scale=0.25]{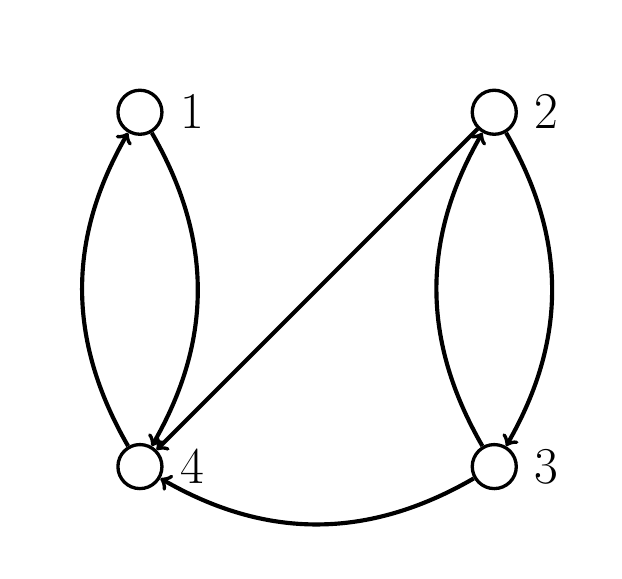}}
\subfigure[][]{
\label{fig:t1-f}
\includegraphics[scale=0.25]{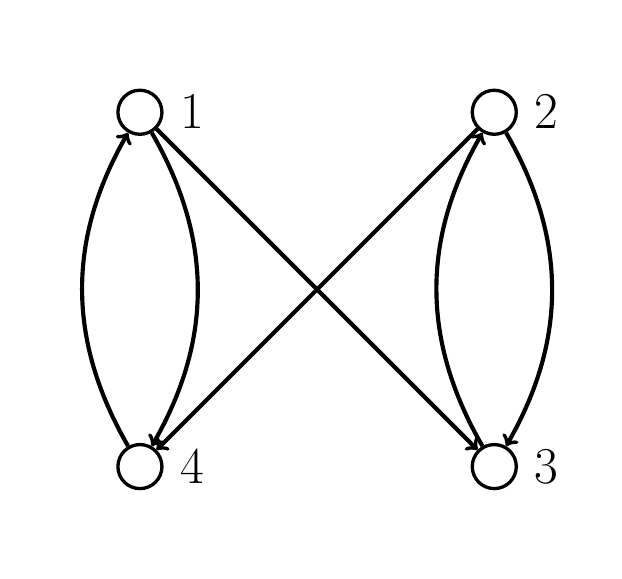}}
\caption{The side information graphs for the six $4$-message problems given in \eqref{eq:exm:fd:1}. In Figures (a), (b) and (c), there is one isolated vertex, vertex $1$, which has no incoming edges. In Figures (d), (e), and (f), there is a pair of disjoint cycles, $\{1,4\}$ and $\{2,3\}$.}
\label{fig:fd:1}
\end{center}
\end{figure}
As we can see, there is a common pattern among the problems of Figures \ref{fig:t1-a}, \ref{fig:t1-b}, and \ref{fig:t1-c}. That is, there is an isolated vertex, vertex $1$. Also, in their capacity-achieving server grouping $\Pcal$, $P_1$ only contains servers that have no more than one message apart from message $1$. 
There also exists a common pattern among the problems of Figures \ref{fig:t1-d}, \ref{fig:t1-e}, and \ref{fig:t1-f}. That is, there are two disjoint cycles, cycle $\{1,4\}$ and cycle $\{2,3\}$. Also, in their capacity-achieving server grouping $\Pcal'$, $P'_2$ only contains servers that have one message from each disjoint cycle.  
\end{example}

\begin{remark}
For the problems in Figures \ref{fig:t1-a}, \ref{fig:t1-b}, and \ref{fig:t1-c} with the capacity-achieving $\Pcal$, the following constraints are given by Axiom \eqref{eq:axiom6} 
\begin{align}
&f(\{1\},\{j\})+f(\{1\},\{2,3,4\} \setminus \{j\})    \nonumber  \\
&\qquad \qquad =f(\{1\},\{2,3,4\}),  \qquad  j\in \{2,3,4\}.
\end{align}
If the constraints above were to be removed from Theorem \ref{thm:pm:group}, then the upper bounds would become looser than $19$. Similarly, for the problems in Figures \ref{fig:t1-d}, \ref{fig:t1-e}, and \ref{fig:t1-f} with the capacity-achieving $\Pcal'$, the following two constraints are given by Axiom \eqref{eq:axiom6}
\begin{align*}
&f(\{1,2\},\{1\})+f(\{1,2\},\{4\})=f(\{1,2\},\{1,4\}),   \\
&f(\{1,2\},\{2\})+f(\{1,2\},\{3\})=f(\{1,2\},\{2,3\}).
\end{align*}
If the constraints above were to be removed from Theorem \ref{thm:pm:group}, then the upper bounds would become looser than $23.5$.
\end{remark}

The following definition generalizes the server grouping construction based on the $\mathrm{fd}$-separation.

\begin{definition}[$\mathrm{fd}$ grouping]\label{def:fd:group}
Consider the distributed index coding problem $(i|A_i),i\in [n]$, whose side information graph $\Gcal$ contains $k\geq 0$ mutually disjoint cycles, denoted by $K_j\subseteq [n],j\in [k]$, $K_j\cap K_{j'}=\emptyset$ for any $j,j'\in [k], j\neq j'$, and $|K_0| \geq 0$ isolated vertices, $v \in K_0\subseteq [n]$. The $\mathrm{fd}$  grouping, denoted by $\Pcal_{\rm fd}$, with $m=(2^k-1-k)+2=2^k-k+1$ groups is defined as follows. The first server group is given by
\begin{align}
P_1&=\{ J\in N: |J\setminus K_0|\le 1 \}.  \label{eq:def:pcalfd:1}
\end{align}
For $\ell = 2, \cdots, m-1$, the server groups are 
\begin{align}
P_{\Gt_\ell}=\{ J\in N: |J\setminus K_0|=|\Gt_\ell|,J\in \bigcap_{j\in \Gt_\ell}T_{K_{j}} \}, \label{eq:def:pcalfd:2}
\end{align}
where $\Gt_\ell \subseteq [k], |\Gt_\ell|\ge 2$. Note that there are $(2^k-1-k)$ such groups. Finally, the last server group $P_m$ is defined as
\begin{align}
P_m&=N\setminus (\bigcup_{\Gt \subseteq [k]:|\Gt|\ge 2}P_{\Gt})\setminus P_1.   \label{eq:def:pcalfd:3}
\end{align}

 \end{definition}

With $\Pcal_{\rm fd}$, we have the following corollary, namely, the \emph{$\rm fd$ grouping outer bound}, from Theorem \ref{thm:pm:group}. As the $\rm fd$ grouping does not result in any simplified expression for the grouping PM outer bound compared to Theorem \ref{thm:pm:group}, we do not repeat the rate constraint inequality \eqref{eq:group:rate} and Axioms \eqref{eq:axiom1}-\eqref{eq:axiom6} below.

\begin{corollary}\label{cor:pm:fd:group}
The capacity region of the distributed
index coding problem $(i|A_i)$, $i \in [n]$, with link capacity tuple $\Cv$ satisfies
\[
\Cr(\Cv) \subseteq \Rr_{\Pcal_{\rm fd}},
\]
where $\Rr_{\Pcal_{\rm fd}}$ denotes the outer bound given by the grouping PM outer bound with any valid $\rm{fd}$  grouping $\Pcal_{\mathrm{fd}}$.
\end{corollary}

The proof is trivial and thus omitted.

We give an example showing the efficacy of the $\rm fd$ grouping outer bound $\Rr_{\rm fd}$ when there are both disjoint cycles and isolated vertices in the side information graph. 

\begin{example}\label{example:3fd}
Consider the distributed index coding problem $(1|-),(2|3),(3|2),(4|5),(5|4)$ with equal link capacities $C_J=1,J\in N\setminus \{ \emptyset \}$. 
The side information graph of the problem consists of two disjoint cycles $K_1=\{2,3\}$ and $K_2=\{4,5\}$, as well as one isolated vertex $1$, and thus $K_0=\{1\}$. 
For easier notation, set 
\begin{align*}
Q_1&=\{ \{1\},\{2\},\{3\},\{4\},\{5\},\{1,2\},\{1,3\},\{1,4\},\{1,5\} \},      \\
Q_2&=\{ \{2,4\},\{2,5\},\{3,4\},\{3,5\},\{1,2,4\},  \\
&\qquad \qquad \qquad \qquad \{1,2,5\},\{1,3,4\},\{1,3,5\} \},
\end{align*}
and $Q_3=N\setminus Q_1 \setminus Q_2$. We have $m=2^2-2+1=3$, and the $\rm fd$ grouping as $\Pcal_{\rm fd}=\{ P_1,P_{\{1,2\}},P_3 \}$, where $P_1=Q_1$, $P_{\{1,2\}}=Q_2$, and $P_3=Q_3$. 
The $\rm fd$ grouping outer bound with $\Pcal_{\mathrm{fd}}$ yields $R_1+R_2+R_3+R_4+R_5\leq 47\frac{2}{3}$, which is tight and matches the lower bound on the sum-capacity.
For the latter, we use in Theorem~\ref{thm:compcod:fractional} 
seven decoding message set tuples $\Dv_1,\Dv_2,\cdots,\Dv_7$ as follows. 

In $\Dv_{1}$, we set $D_1=\{1\}$ and $D_{i}=\{ 1,i \}$, $i\in \{2,3,4,5\}$. 

In $\Dv_{2}$, we set $D_1=\{1\}$, $D_{i}=[n]\setminus A_i$, $i\in \{2,3\}$, and $D_{i}=\{ 1,i \}$, $i\in \{4,5\}$. 

In $\Dv_{3}$, we set $D_1=\{1,4,5\}$, $D_{i}=[n]\setminus A_i$, $i\in \{2,3\}$, and $D_{i}=\{ 1,i \}$, $i\in \{4,5\}$. 

In $\Dv_{4}$, we set $D_1=\{1\}$, $D_{i}=\{ 1,i \}$, $i\in \{2,3\}$, and $D_{i}=[n]\setminus A_i$, $i\in \{4,5\}$.

In $\Dv_{5}$, we set $D_1=\{1,2,3\}$, $D_{i}=\{ 1,i \}$, $i\in \{2,3\}$, and $D_{i}=[n]\setminus A_i$, $i\in \{4,5\}$.

In $\Dv_{6}$, we set $D_1=\{1\}$, $D_{i}=[n]\setminus A_i$, $i\in \{2,3,4,5\}$.

In $\Dv_{7}$, we set $D_{i}=[n]\setminus A_i$, $i\in [n]$.

We also use the following three decoding server groups $\Pv_1,\Pv_2,\Pv_3$. 
In $\Pv_1$, we set $P_i = Q_1$, $i \in [n]$. 
In $\Pv_2$, we set $P_i = Q_2$, $i \in [n]$, and in $\Pv_3$, we set $P_i=Q_3$, $i\in [n]$. 
Hence, there are in total $7*3=21$ decoding configurations, $(\Pv_j,\Dv_k)$, $j\in [3],k\in [7]$. We set $R_i(\Pv,\Dv), i \in [n]$, $C_J(\Pv,\Dv), J \in N$, and $S_K(\Pv, \Dv)$, $K \subseteq [n]$, to zero for all other $(\Pv,\Dv)$ configurations. 
Notice that there is an interesting correspondence between the server groups in $\Pcal_{\rm fd}$ used in the outer bound and the decoding server groups used in the inner bound. Whether such correspondence has its roots in some deeper structural properties of the problem remains to be studied in future. 
\end{example}

\subsection{A Hierarchy of Server Groupings} \label{sec:hierarchy}

In some distributed index coding problems, it is more advantageous to use ``finer'' server groupings than we have introduced so far. As hinted before, there is a natural hierarchy of server groupings in terms of tightness of the resulting outer bound. We need the following definition to formalize this. 
\begin{definition}[Grouping refinement and aggregation]\label{def:refined}
For any two valid server groupings $\Qcal= \{Q_1, Q_2, \cdots, Q_{\ell}\}$ and $\Pcal=\{P_1,\cdots,P_m\}$. We say that $\Pcal$ is a \emph{refinement} of $\Qcal$ and that $\Qcal$ is an \emph{aggregation} of $\Pcal$, if for every $i\in [\ell]$, $Q_i = P_G$ for some $G\subseteq [m]$.
\end{definition}

In words, every server group in $\Qcal$ is the union of some server groups in $\Pcal$. 
{We have the following relationship between the outer bounds $\Rr_{\Pcal}$ and $\Rr_{\Qcal}$. }
\begin{proposition}\label{propo:refine}
If $\Pcal$ is a refinement of $\Qcal$, {or equivalently, $\Qcal$ is an aggregation of $\Pcal$,} then $\Rr_{\Pcal}$ is no looser than $\Rr_{\Qcal}$, i.e., $\Rr_{\Pcal} \subseteq  \Rr_{\Qcal}$.
\end{proposition}

The proof is presented in Appendix \ref{app:propo:refine}. Note that Proposition~\ref{propo:refine} clarifies the relationship between the individual touch grouping and any other touch grouping.

\begin{definition}[Intersecting refinement of groupings]\label{def:refine:intersecting}
For two valid server groupings $\Pcal$ and $\Qcal$,
\begin{align}
\Pcal \wedge \Qcal = \{P\cap Q: P \in \Pcal, Q \in \Qcal\}
\end{align}
is the \emph{intersecting} refinement of both groupings.
\end{definition}

\begin{example}\label{exm:refine:touch:fd}
Consider the distributed index coding problem $(1|2),(2|1),(3|5),(4|3),(5|4)$ with equal link capacities $C_J = 1, J  \in N_A$ where 
\begin{align*} 
N_A&=\{ \{1,3\},  \{1,4\},  \{1,5\}, \{2,3\},\{2,4\}, \{2,5\}, \\
&\qquad \qquad \{3,4,5\}, \{1,3, 4, 5\}, \{2,3, 4, 5\}, \{1,2,3, 4, 5\}\}
\end{align*}
and $C_J = 0$ otherwise. The grouping PM outer bounds with the touch grouping 
\begin{align}
\Pcal_{\rm t}=\{ T_{\{1\}}\cap N_A,T_{\{2,3,4,5\}} \cap N_A\},
\end{align}
and the $\mathrm{fd}$ grouping $\Pcal_{\mathrm{fd}}=\{ P_1,P_2=N_A\setminus P_1\}$, where 
\begin{align}
P_1=\{\{1,3\}, \{1,4\},\{1,5\},\{2,3\},\{2,4\},\{2,5\} \},
\end{align}
yield the sum-capacity upper bound 14.5. With the intersecting refinement grouping,
\begin{align}
\Pcal_{\rm t} \wedge \Pcal_{\mathrm{fd}}&=\{ T_{\{1\}}\cap P_1,T_{\{1\}}\cap P_2,  \nonumber  \\
&\qquad \qquad T_{\{2,3,4,5\}}\cap P_1,T_{\{2,3,4,5\}}\cap P_2 \},
\end{align} 
a tighter upper bound of 14 on the sum-capacity is established, which matches the lower bound. For the latter, we use Theorem~\ref{thm:compcod} with $(\Pv,\Duv)$, where $P_i = N_A$, $i\in[n]$, and $\Duv$ is generated by Algorithm \ref{alg:natural:D} in Section \ref{sec:numerical}.
\end{example} 

Based on Proposition \ref{propo:refine}, we can establish the tightest grouping PM outer bound by using the ``finest'' server grouping $\Pcal^* = \{\{J\}: J \in N\setminus \{ \emptyset \} \}$ with $m = 2^n-1$, referred to as the \emph{single-server grouping}, which consists of all single nonempty servers and is a refinement of every other valid server grouping. 

We present the following corollary, namely, the \emph{single-server grouping outer bound}, without repeating \eqref{eq:group:rate}-\eqref{eq:axiom6}.
\begin{corollary}\label{cor:pm:finest} 
The capacity region of the distributed
index coding problem $(i|A_i)$, $i \in [n]$, with link capacity tuple $\Cv$ satisfies
\[
\Cr(\Cv) \subseteq \Rr^*, 
\]
where $\Rr^*$ 
denotes the outer bound given by the grouping PM outer bound with the single-server grouping $\Pcal^*$. 
\end{corollary}

\begin{remark}\label{rem:shannon}
In a similar fashion as in \cite{liu:vellambi:kim:sadeghi:itw18}, it can be shown that the single-server grouping outer bound region $\Rr^*$ is as tight as the apparently stronger outer bound in which all Shannon-type inequalities of the entropy function for the distributed index coding problem are used. 
\end{remark}

If all servers are active, the computational complexity of $\Rr^*$ is prohibitive even for small $n$ as the number of the intermediate variables $f(G,K),G\subseteq [m],K\subseteq [n]$ in Corollary \ref{cor:pm:finest} is $2^{|N\setminus \{ \emptyset \}|+n}=2^{2^n-1+n}$, which is doubly exponential to $n$. 

Finally, based on Proposition~\ref{propo:refine} we can establish the loosest grouping PM outer bound
by using the ``coarsest'' server grouping $\Pcal_* = \{N\}$ with $m = 1$, referred to as the \emph{all-server grouping}, which consists of a single all-server group and is an aggregation of every other valid server grouping. 
With $\Pcal_*$, the grouping PM outer bound in Theorem~\ref{thm:pm:group} simplifies to $\Rr_*$, namely, the \emph{all-server grouping outer bound}, shown as follows.

\begin{corollary}\label{cor:pm:coarsest} 
The capacity region of the distributed
index coding problem $(i|A_i)$, $i \in [n]$, with link capacity tuple $\Cv$ satisfies
\[
\Cr(\Cv) \subseteq \Rr_*,
\]
where $\Rr_*$ consists of all rate tuples $\Rv$ such that
\begin{align}
R_i \le g(B_i\cup \{i\})-g(B_i), \qquad i\in[n],     \label{eq:allserver:rate}
\end{align}
for some $g(K),K\subseteq [n]$, satisfying
\begin{align}
&g(\emptyset)=0,                          \label{eq:allserver:axiom1}                                                           \\
&g(K)\leq \sum_{J: J \in T_K} C_J,                                     \label{eq:allsever:axiom2}                                   \\
&g(K)\leq g(K'),           \enskip \qquad \qquad \qquad \qquad \text{if $K\subseteq K'$},            \label{eq:allserver:axiom3}         \\
&g(K\cap K') + g(K\cup K')\leq g(K) + g(K'),           \label{eq:allserver:axiom4}         \\
&g(B_i\cup \{i\})-g(B_i)= g(\{i\}),   \qquad \qquad i\in [n]. \label{eq:allserver:axiom:addi:de}
\end{align}
\end{corollary}

\begin{IEEEproof}
As $G\subseteq [1]$, $G$ can be either $\emptyset$ or $\{1\}$. Also, $f(\emptyset,K)=0$ for any $K\subseteq [n]$. Therefore, it suffices to use a single set function $g(K)=f(\{1\},K),K\subseteq [n]$ in the axioms and the rate constraint inequality in Corollary \ref{cor:pm:coarsest}. With all-server grouping $\Pcal_*=\{N\}, m=1$, we have $(P_G\cup P_{G'})\setminus (P_G \cap P_{G'})\subseteq T_{\ntouch{K}}$ only for $G=G'$ or $K=\emptyset$, and thus Axiom \eqref{eq:axiom1} in Theorem~\ref{thm:pm:group} becomes trivial. Also, there never exists two disjoint nonempty sets $K,K'\subseteq [n]$ such that $P_{\{1\}} = N \subseteq (N\setminus T_{K,K'})$. This means that Axiom \eqref{eq:axiom6} can only give trivial inequalities with the all-server grouping. In summary, in Corollary~\ref{cor:pm:coarsest}, there are no constraints corresponding to Axioms \eqref{eq:axiom1} and \eqref{eq:axiom6} in Theorem~\ref{thm:pm:group}.
\end{IEEEproof}

Even though the all-server grouping outer bound $\Rr_*$ on the capacity region is the loosest bound one can get from the grouping PM outer bound, it is already no looser than the (corrected) outer bound\footnote{In the original version of the outer bound in \cite{Sadeghi--Arbabjolfaei--Kim2016}, the rate constraint inequality was expressed as $R_i\leq f_L(B_i\cup \{i\})-f_L(B_i), i\in L$, which is not correct, as the set function $f_L(K)$ is only defined for $K\subseteq L$ and $B_i$ may not be a subset of $L$. We rectify this by taking the intersection of $L$ and the argument of $f_L(K)$  in \eqref{eq:itw:rate}.} proposed in \cite{Sadeghi--Arbabjolfaei--Kim2016}, which we refer to as $\Rr_{f_L}$ and restate it as follows.
\begin{proposition}[\hspace{-1.5mm}~\cite{Sadeghi--Arbabjolfaei--Kim2016}] \label{propo:itw}
The capacity region of the distributed
index coding problem $(i|A_i)$, $i \in [n]$, with link capacity tuple $\Cv$ satisfies
\[
\Cr(\Cv) \subseteq \Rr_{f_L},
\]
where $\Rr_{f_L}$ consists of all rate tuples $\Rv$ such that for any $L \subseteq [n]$
\begin{align}
R_i\leq f_L((B_i\cup \{i\})\cap L)-f_L(B_i\cap L), \qquad i\in L, \label{eq:itw:rate}
\end{align}
for some $f_L(K)$, $K\subseteq L$, such that 
\begin{align}
&f_L(\emptyset)=0,        \label{eq:itw:axiom1}                                                                               \\
&f_L(L)\leq \sum_{J: J \in T_L} C_J,    \label{eq:itw:axiom2}                                                                      \\
&f_L(K)\leq f_L(K'),        \qquad \qquad \qquad \text{if $K\subseteq K'\subseteq L$},   \label{eq:itw:axiom3}                      \\
&f_L(K\cap K')+f_L(K\cup K')\leq f_L(K) + f_L(K').     \label{eq:itw:axiom4}
\end{align}
\end{proposition}

\begin{proposition} \label{propo:allserver:itw}
For any problem $(i|A_i),i\in [n]$, with link capacity tuple $\Cv$, it holds that $\Rr_{*}\subseteq \Rr_{f_L}$.
\end{proposition}

The proof is presented in Appendix \ref{app:allserver:itw}.

The computational complexity of the all-server grouping outer bound $\Rr_{*}$ will be the lowest. And even performing FME to compute the outer bound on the entire capacity region for a general $\Cv$ is possible for small to moderate $n$ as the total number of variables is only $2^{n}+2^{n}-1+n$ in Corollary~\ref{cor:pm:coarsest}, accounting for $2^{n}$ $g(K),K\subseteq [n]$ variables, $2^{n}-1$ link capacity variables $C_J,J\in N\setminus \{ \emptyset \}$, and $n$ rate variables $R_i,i\in [n]$. 

\begin{example} \label{exm:pm:allserver}
We first revisit the problem $(1|-),(2|4),(3|4),(4|3)$ discussed in Example \ref{exm:uv}. A looser upper bound of 22 on the sum-capacity is given by the all-server grouping outer bound $\Rr_{*}$ in comparison to the tight bound established earlier. However, 
$\Rr_*$ can sometimes yield tight bounds. For example, consider the 
problem $(1|4),(2|1,4),(3|1,2,4),(4|1,2,3)$ with equal link capacities $C_J = 1, J \in N\setminus \{ \emptyset \}$. The all-server grouping outer bound $\Rr_*$ yields the tight upper bound of 22 on the sum-capacity, which matches the lower bound 
in Corollary~\ref{corr:allservers} with $P_i = N$ and $D_i = [n]\setminus A_i$, $i\in [n]$.
\end{example}

\subsection{Summary of the Outer Bounds}
Different server groupings that we presented in this section are summarized in Table~\ref{tab:compare}.

\begin{table*}[!h]
\begin{center}
\caption{Special cases of the grouping PM outer bound and their indicative computational complexities.}
\begin{tabular}{|c|c|c|c|}
\hline
Server groupings&Entropic variables&References&Total \# of variables\\
\hline
\opertypemed{The touch grouping  $\Pcal_{\mathrm{t}}=\{T_{L_1}, \ldots, T_{L_{m}}\}$ with $m \leq n$ groups: $f(G,K),G\subseteq [m], K\subseteq [n]$} & $\frac{1}{r}H(\Yv_{T_{L_G}}|\Xv_{K^c})$ &\opertypesmall{Cor. \ref{cor:pm:touch:aggregate}}&  $2^{m+n}+2^{n}-1+n$\\
\hline
\opertypemed{The individual touch grouping  $\Pcal_{\mathrm{t}}^*=\{T_{\{1\}}, \ldots, T_{\{n\}}\}$ with $m=n$ groups: $f(G,K), G, K\subseteq [n]$} & $\frac{1}{r}H(\Yv_{T_G}|\Xv_{K^c})$ &\opertypesmall{\cite[Thm. 4]{isit:2017}} &  $2^{2n}+2^{n}-1+n$\\
\hline
\opertypemed{The $\mathrm{fd}$ grouping $\Pcal_{\mathrm{fd}}$ constructed according to \eqref{eq:def:pcalfd:1}-\eqref{eq:def:pcalfd:3} with $m=2^k-k+1$ groups, where $k$ is the number of pairwisely disjoint cycles: $f(G,K),G\subseteq [m], K\subseteq [n]$} & $\frac{1}{r}H(\Yv_{P_G}|\Xv_{K^c})$ &\opertypesmall{Cor. \ref{cor:pm:fd:group}, Exm.  \ref{exm:fd:1}, \ref{example:3fd}} &  $2^{2^k-k+1+n}+2^{n}-1+n$\\
\hline
\opertypemed{The intersecting refined grouping $\Pcal \wedge \Qcal = \{P\cap Q: P \in \Pcal, Q \in \Qcal\}$ with $m\ell$ groups, where $m = |\Pcal|, \ell = |\Qcal|$: $f(G,K),G\subseteq [m\ell], K\subseteq [n]$} & $\frac{1}{r}H(\Yv_{(P\cap Q)_G}|\Xv_{K^c})$ &\opertypesmall{Def. \ref{def:refined}, \ref{def:refine:intersecting}, Prop. \ref{propo:refine}, Exm. \ref{exm:refine:touch:fd}}&  $2^{m\ell+n}+2^{n}-1+n$\\
\hline
\opertypemed{The single-server grouping $\Pcal^* = \{\{J\}: J \in N\setminus \{ \emptyset \} \}$ with $m = 2^n-1$: $f(G,K),G\subseteq [2^n-1], K\subseteq [n]$\\} &$\frac{1}{r}H(\Yv_{P_G}|\Xv_{K^c})$& Cor. \ref{cor:pm:finest} &$ 2^{2^n-1+n}+2^{n}-1+n$\\
\hline
\opertypemed{The all-server grouping $\Pcal_* = \{N\}$ with $m = 1$ server group: $g(K), K\subseteq [n]$\\} &\opertypesmalltwo{ $\frac{1}{r}H(\Yv_N|\Xv_{K^c})$ }& Cor. \ref{cor:pm:coarsest}, Prop. \ref{propo:allserver:itw}, Exm. \ref{exm:pm:allserver} &$2^{n}+2^{n}-1+n$\\
\hline
\end{tabular}\label{tab:compare}
\end{center}
\end{table*}


\section{Numerical Results}\label{sec:numerical}

We numerically evaluate lower and upper bounds on the sum-capacity for all 218 non-isomorphic four-message distributed index coding problems with equal link capacities, $C_J=1$, $J\in N\setminus \{ \emptyset \}$. For brevity, each problem in this section is represented with a problem number and the corresponding receiver side information is listed in Appendix \ref{app:numerical:graph4}. The upper bounds on the sum-capacity are computed using the special cases of the grouping PM outer bound proposed in Section \ref{sec:outer:bounds}. The lower bound are given by distributed composite coding, computed using a fixed decoding configuration in Theorem~\ref{thm:compcod}. For both upper and lower bounds, we use LP to maximize the sum-rate $R_1+R_2+R_3+R_4$. It turns out that the lower bounds match the upper bounds, thus establishing the sum-capacity, for all 218 problems. 

The results are summarized in Table \ref{tab:numerical}. On the right column, each tuple denotes a list of problem numbers, followed by their sum-capacity in bold face. For example, (16, 30, 60, 102; \textbf{19}) means that problems 16, $\cdots$, 102 have the same sum-capacity of 19. 
The left column shows outer bounds that are used to yield the tight upper bounds on the sum-capacity. 
It turns out that the all-server grouping outer bound $\Rr_*$ in Corollary \ref{cor:pm:coarsest} with $m=1$ group can solve 145 out of 218 problems with minimum computational complexity.
For the 63 problems shown in the middle row in Table \ref{tab:numerical}, tight upper bounds on the sum-capacity can be obtained by the touch grouping outer bound $\Rr_{\rm t}$ with $\Pcal_{\rm \Vv}$ defined in \eqref{eq:pcal:uv}. 
Notice that for these aforementioned $63+145=208$ problems, except for the $6$ problems  (149, 176, 179, 200, 203, 212; \textbf{26}) with sum-capacity of $26$, they are also solvable by Proposition \ref{propo:uv}. 
For the remaining 10 problems, the touch grouping outer bound $\Rr_{\rm t}$ gives loose results, and the $\rm fd$ grouping outer bound $\Rr_{\rm fd}$ is necessary to yield tight upper bounds. A subset of these problems were discussed in Example \ref{exm:fd:1} and shown in Figure \ref{fig:fd:1} and they all involve either isolated vertices or disjoint cycles in their side information graph.

For the lower bounds for all 218 problems, we used $(\Pv,\Dv)$ in Theorem \ref{thm:compcod} where $P_i = N$, $i\in [n]$, and $\Dv=\Duv$, generated according to \cite[Algorithm 2]{isit:2018}, 
which is repeated as follows. 

\begin{algorithm}[bth]\label{alg:natural:D}
    \SetKwInOut{Input}{Input}
    \SetKwInOut{Output}{Output}
    \Input{Index coding problem $(i| A_i)$, $i \in [n]$.}
    \Output{Decoding message set $\Duv=(\Du_i,i\in [n])$.}
Initialize $\Du_i = \{i\}$, $i\in [n]$.

As long as there exists $i,j \in [n]$ such that $A_j \subseteq A_i\cup \Du_i$ and $\Du_j \not \subseteq A_i\cup \Du_i$, update $\Du_i \leftarrow \Du_i \cup (\Du_j \setminus A_i)$. If no such $i,j$ exist, terminate the algorithm.
     \caption{Decoding message sets $\Duv$}
\end{algorithm}

\begin{table*}[!h]
\begin{center}
\caption{Sum-capacity for all 218 non-isomorphic $4$-message distributed index coding problems. Link capacities are $C_J = 1$, $J \in N\setminus \{  \emptyset \}$.}
\begin{tabular}{|c|c|}
\hline

Outer bounds&(Problem numbers; \textbf{sum-capacity})\\
\hline
\opertypemedminus{The all-server grouping outer bound $\Rr_{\rm t}$ in Corollary \ref{cor:pm:coarsest}} & \opertype{(1, 2, 3, 5, 6, 7, 8, 10, 11, 12, 13, 15, 17, 19, 20, 22, 25, 26, 33, 35, 38, 39, 40, 41, 49, 63, 65, 67, 69, 70, 100; \textbf{15}), (47; \textbf{18.6667}), (43, 78, 83, 85, 130, 132; \textbf{20}), (42, 44, 45, 71, 72, 73, 74, 75, 76, 77, 79, 80, 82, 84, 103, 104, 105, 106, 107, 108, 109, 110, 111, 113, 116, 117, 118, 120, 122, 123, 124, 125, 126, 127, 128, 131, 133, 142, 143, 144, 145, 147, 151, 152, 153, 154, 158, 159, 161, 162, 163, 164, 165, 166, 167, 168, 169, 174, 177, 182, 183, 184, 185, 186, 187, 201; \textbf{22}), (114, 121, 129, 146, 150, 155, 156, 157, 160, 170, 171, 175, 178, 180, 181, 188, 189, 190, 191, 192, 194, 195, 196, 197, 198, 202, 204, 206, 208, 210, 216; \textbf{24}), (207; \textbf{26}), (193, 205, 209, 211, 213, 214, 215, 217; \textbf{28}), (218; \textbf{32})} \\
\hline
\opertypemedminus{The touch grouping outer bound $\Rr_{\rm t}$ in Corollary \ref{cor:pm:touch:aggregate} with $\Pcal_{\rm \Vv}$ in \eqref{eq:pcal:uv}}& \opertype{(4, 9, 18, 21, 23, 24, 34, 36, 48, 55, 64, 66, 68, 86, 95, 99, 138; \textbf{19}), (14, 27, 28, 29, 31, 32, 37, 50, 51, 52, 53, 54, 56, 57, 58, 59, 61, 62, 87, 88, 89, 90, 91, 92, 94, 96, 97, 98, 101, 134, 136, 137, 139, 140, 141, 173; \textbf{21}), (93, 135, 172, 199; \textbf{25}), (149, 176, 179, 200, 203, 212; \textbf{26})} \\
\hline
\opertypemedminus{The $\rm{fd}$ grouping outer bound $\Rr_{\rm fd}$ in Corollary \ref{cor:pm:fd:group}} & \opertype{(16, 30, 60, 102; \textbf{19}), (46; \textbf{23.3333}), (81, 112, 115, 119, 148; \textbf{23.5})} \\
\hline
\end{tabular}\label{tab:numerical}
\end{center}
\end{table*}


\section{Conclusion}\label{sec:conclusion}

In this work, we studied the distributed index coding problem in a general model, where for $n$ messages in the system all possible $2^n-1$ servers, each containing a different nonempty subset of messages were taken into account. Due to the exponential size of the problem and its distributed nature, this problem is even more challenging than its centralized counterpart, which itself is an open problem. We showed that cooperative encoding among distributed servers combined with flexible utilization of degrees of freedom for decoding can achieve strictly larger rate regions compared to the existing work \cite{Sadeghi--Arbabjolfaei--Kim2016, isit:2017, Li--Ong--Johnson--ISIT--2017, Li--Ong--Johnson2017}. The new enhanced fractional rate splitting methods of Theorem \ref{thm:compcod:fractional} can have applications in other network information theory problems to improve upon standard convexification techniques that are based on time sharing. This work also inspired us to develop a more advanced three-layer composite coding scheme in \cite{liu2018three} for the centralized index coding problem that harvests more degrees of freedom in composite coding. For the outer bound on the capacity region of the distributed index coding problem, we
developed a new grouping PM outer bound and showed a general hierarchy that allows a full spectrum of tradeoffs between computational complexity and tightness of outer bounds. We demonstrated the utility of the achievable rate regions and performance bounds through several simple examples and extensive numerical results for a small number ($n = 4$) of messages. 

We conclude this paper with several open questions and intriguing research directions for the future. 

The computational complexity of the proposed distributed composite coding scheme can be of a practical concern, given the large number of composite index rate variables and decoding configurations. To address this issue, one may extend and further improve the simplification techniques in \cite{isit:2018,liu2018three} developed for the centralized scenario. Even for the centralized index coding problem the composite coding inner bound may not be tight (see \cite[Figure 6.15]{foundation}), which will carry over to the distributed case. It is a worthwhile research direction to better understand the limitations of composite coding and try to address them more systematically. 

Practical code design is an important open problem in distributed index coding. 
Recently in \cite{li2018multi,kim2019linear}, the celebrated minrank approach for the centralized \emph{linear} index coding was extended to the distributed index coding problem, providing non-trivial yet still suboptimal and computationally inefficient code design techniques. 
While composite coding is not practical due to random coding, one interesting observation is that after applying the simplification techniques proposed in \cite{isit:2018} to eliminate unnecessary composite indices, the remaining composite indices may guide which linear codes are efficient. For example, consider the centralized index coding problem $(1|4),(2|1,3),(3|1,2),(4|1,3)$ with equal message length $t_i=t$, $i\in [n]$, whose symmetric capacity of $1/2$ can be achieved using composite coding with $\Duv$ generated according to Algorithm \ref{alg:natural:D}.  
Among the $2^4-1=15$ composite indices $S_K$, $K\subseteq [n]$, $K\neq \emptyset$, only two indices, $S_{\{1,4\}}$ and $S_{\{1,2,3,4\}}$, remain after applying the heuristic reduction method in \cite{isit:2018}. 
Note that the symmetric capacity can also be achieved via sending the scalar linear codeword $(x_1\oplus x_4,x_1\oplus x_2\oplus x_3\oplus x_4)$ of length $2t$. Note the correspondence between the remaining composite indices and the optimal linear codeword, which can also be observed in many other problems. However, a precise theory behind such a correspondence remains unknown at the moment and whether such a relation holds for a general class of centralized or distributed index coding problems is an open problem.

As shown in \cite{sun2015index}, Shannon-type inequalities of the entropy function are not sufficient to obtain tight performance bounds even for the centralized index coding problem. Specific centralized problems where non-Shannon-type inequalities are needed for tight performance bounds were identified in \cite{sun2015index,baber2013multiple,isit:2019:arxiv}. It follows automatically that non-Shannon-type inequalities are also needed for some distributed index coding problems. Many questions remain, however, such as for which problems non-Shannon type inequalities are needed and how much would they improve upon Shannon-type inequalities. 


\appendices

\section{Proof of Theorem \ref{thm:compcod}} \label{app:error:proof}

Analysis of error for the first-step decoding is as follows. We partition the error event according to the collection $M \subseteq \Gamma_*(P_i) \setminus 2^{A_i}$ for erroneous composite indices. That is, $\wh_K \neq w_K$ iff $K \in M$. Therefore, by the union bound, we have
\begin{align}
P_e &= \mathrm{P}\{\text{$y_J = y_J(\wh_K, K \in 2^J)$ for all $J \in P_i$ for some}  \nonumber  \\
&\qquad \text{$(\wh_K, K \in \Gamma_*(P_i)) \neq (w_K, K \in \Gamma_*(P_i))$}\}  \nonumber  \\
&\leq \sum_{M \subseteq \Gamma_*(P_i) \setminus2^{A_i}}\,\,\sum_{\substack{(\wh_K, K \in \Gamma_*(P_i)):\\\wh_K \neq w_K, K \in M,\\\wh_K = w_K, K \notin M} }  \nonumber  \\
&\qquad \mathrm{P}\left\{\bigcap_{\substack{J \in P_i, J \in \Gamma^*(M)}}\left\{y_J = y_J(\wh_K, K \in 2^J)\right\}\right\} \nonumber  \\
&< \sum_{M \subseteq \Gamma_*(P_i) \setminus2^{A_i}} 2^{\sum_{K \in M} s_K-\sum_{J \in \Gamma^*(M)\,\cap \, P_i}r_J}\label{eq:error1}\\
&< \sum_{M \subseteq \Gamma_*(P_i) \setminus2^{A_i}} 2^{\sum_{K \in M} (rS_K+1)-\sum_{J \in \Gamma^*(M)\,\cap \, P_i}(rC_J-1)}  \nonumber  \\
&= \sum_{M \subseteq \Gamma_*(P_i) \setminus2^{A_i}} \frac{2^{r\sum_{K\in M}S_K}\cdot 2^{|M|+|\Gamma^*(M)\,\cap \, P_i|}}{2^{r\sum_{J \in \Gamma^*(M)\,\cap \, P_i}C_J}},  \nonumber
\end{align}
where \eqref{eq:error1} holds since for each composite index collection $M$, the number of erroneous tuples is $\prod_{K\in M}(2^{s_K}-1) < 2^{\sum_{K \in M} s_K}$, and for each erroneous composite index tuple with $\wh_K \neq w_K$, iff $K \in M$, the probability that it is mapped to the same codeword $y_J$ as the correct composite index tuple for all $J \in  \Gamma^*(M)\,\cap \, P_i$ is $2^{-\sum_{J \in \Gamma^*(M)\,\cap \, P_i}r_J}.$ 
Note that only servers in  $\Gamma^*(M)$ can generate composite index (indices) in the collection $M$ and the intersection with $P_i$ is necessary due to receiver $i$'s choice of server group $P_i$. 

Therefore, the error probability $P_e$ tends to zero as $r \to \infty$, provided that
\begin{align*}
\sum_{K \in M} S_K < \sum_{J \in \Gamma^*(M)\,\cap \, P_i} C_J,  \enskip \forall M \subseteq \Gamma_*(P_i) \setminus2^{A_i}, i\in[n].
\end{align*}

Analysis of error for the second-step decoding is as follows. We partition the error event according to message index subsets $L \subseteq \Delta_i$. That is, $\xh_j \neq x_j$ iff $j \in L$. Therefore, by the union bound, we have
\begin{align}
P_e &= \mathrm{P}\{\text{$w_K(\xvh_K) = \wh_K$ for all $K \in  \Gamma_*(P_i)$, $K \subseteq \Delta_i \cup A_i$}  \nonumber  \\
&\qquad \qquad \text{for some $\xh_j \neq x_j$, $ j \in \Delta_i$}\}  \nonumber  \\
&\leq \sum_{L \subseteq \Delta_i}\,\,\sum_{\substack{\xvh_{\Delta_i}:\\\xh_j \neq x_j, j \in L\\\xh_j = x_j, j \notin L} }\mathrm{P}\left\{\bigcap_{\substack{K \subseteq \Delta_i \cup A_i,\\ K \in \Gamma_*(P_i),\\K\cap L \neq \emptyset}}\left\{w_K(\xvh_K)= \wh_K\right\}\right\}  \nonumber  \\&\leq \sum_{L \subseteq \Delta_i} 2^{\sum_{j \in L} t_j-\sum_{K \in \Kcal}s_K}  \label{eq:error2}  \\
&< \sum_{L \subseteq \Delta_i}2^{\sum_{j\in L}(rR_j+1)-\sum_{K\in \Kcal}rS_K}  \nonumber  \\
&= \sum_{L \subseteq \Delta_i}2^{r(\sum_{j\in L}R_j-\sum_{K\in \Kcal}S_K)+|L|},  \nonumber
\end{align}
where $\Kcal = \{K:K \subseteq \Delta_i \cup A_i, K \in \Gamma_*(P_i),K\cap L \neq \emptyset\}$ and \eqref{eq:error2} holds since for each message index subset $L$, the number of erroneous messages is $\prod_{j\in L}(2^{t_j}-1) < 2^{\sum_{j \in L} t_j}$ and for each erroneous message tuple with $\xh_j \neq x_j$, iff $j \in L$, the probability that it is mapped to the same composite index $w_K$ as the correct message tuple for all $K\in \Kcal $ is $2^{-\sum_{K \in \Kcal}s_K}$. 

Therefore, the error probability $P_e$ tends to zero as $r \to \infty$, provided that
\begin{align*}
\sum_{j \in L} R_j < \sum_{\substack{K \subseteq \Delta_i \cup A_i,\\ K \in \Gamma_*(P_i),\\ K \cap L \neq \emptyset}} S_K, \quad \quad \forall L \subseteq \Delta_i, i\in[n].
\end{align*}

Note that only composite indices that are both in $\Gamma_*(P_i)$ and $2^{\Delta_i \cup A_i}$ are useful for receiver $i$ decoding of $\xv_{\Delta_i}$.

\section{Proof of Proposition \ref{propo:compcod:equiv}}
\label{app:compcod:equiv}

We first prove the following lemma.
\begin{lemma}\label{lemma:PP'}
If $P' \subseteq P$, then 
\begin{align}\label{eq:lemma:Gamma}
 \Gamma_*(P') \setminus \Gamma_*(P\setminus P')= \Gamma_*(P) \setminus \Gamma_*(P\setminus P').\end{align}
\end{lemma}
\begin{IEEEproof}
As $P'\subseteq P$, we have that $\Gamma_*(P')\subseteq \Gamma_*(P)$, and therefore, $\Gamma_*(P') \setminus \Gamma_*(P\setminus P') \subseteq \Gamma_*(P) \setminus \Gamma_*(P\setminus P')$.

Now consider an arbitrary $J\in \Gamma_*(P) \setminus \Gamma_*(P\setminus P')$. As $J\in \Gamma_*(P)$, there must exist some $J_1\in P$ such that $J\subseteq J_1$. Since $J \not \in \Gamma_*(P\setminus P')$, we know that $J_1 \not \in P\setminus P'$. Hence, $J_1 \in P'$ and thus $J\in \Gamma_*(P')$. Therefore, we have $J \in \Gamma_*(P') \setminus \Gamma_*(P\setminus P')$ and thus $ \Gamma_*(P) \setminus \Gamma_*(P\setminus P') \subseteq \Gamma_*(P') \setminus \Gamma_*(P\setminus P')$.

In summary, we have $\Gamma_*(P') \setminus \Gamma_*(P\setminus P')= \Gamma_*(P) \setminus \Gamma_*(P\setminus P').$
\end{IEEEproof}

Now we prove Proposition \ref{propo:compcod:equiv} as follows. 

For easier reference, we repeat \eqref{eq:first:decode:fixed} here for a given $(P_i, i\in [n])$, 
\begin{align}
\sum_{K \in M} S_K &< \sum_{J \in \Gamma^*(M)\,\cap \, P_i} C_J,  \nonumber  \\
&\qquad \qquad \forall M \subseteq \Gamma_*(P_i) \setminus2^{A_i}, i \in [n].  \label{eq:first:decode:fixed:rep2}
\end{align}

According to Lemma \ref{lemma:PP'}, for any $Q \subseteq P_i$ we have  $$\Gamma_*(Q)\setminus \Gamma_*(P_i \setminus Q)\setminus 2^{A_i} = \Gamma_*(P_i)\setminus \Gamma_*(P_i \setminus Q)\setminus 2^{A_i}.$$ 

Therefore,  for the same $P_i$ as chosen above, \eqref{eq:first:decode:fixed:minli} can be written as
\begin{align}\label{eq:first:decode:fixed:minli:rep}
\sum_{K \in \Gamma_*(P_i)\setminus \Gamma_*(P_i \setminus Q)\setminus 2^{A_i}} S_K &< \sum_{J \in Q} C_J,  \nonumber  \\
&\qquad \qquad \forall Q \subseteq P_i,i\in[n].
\end{align}

We prove that for any given inequality from the system of inequalities \eqref{eq:first:decode:fixed:rep2} there exists an inequality in the system of inequalities \eqref{eq:first:decode:fixed:minli:rep} that is no looser and vice versa.

First, for any $M \subseteq \Gamma_*(P_i) \setminus2^{A_i}$ in \eqref{eq:first:decode:fixed:rep2} we construct $Q = \Gamma^*(M)\,\cap \, P_i$. Therefore, the RHS of \eqref{eq:first:decode:fixed:rep2} and \eqref{eq:first:decode:fixed:minli:rep} become identical. Our claim is that $M\subseteq \Gamma_*(P_i)\setminus \Gamma_*(P_i \setminus Q)\setminus 2^{A_i}$ or that if $K \in M$ then $K \in \Gamma_*(P_i)\setminus \Gamma_*(P_i \setminus Q)\setminus 2^{A_i}$. Note that we have $M \subseteq \Gamma_*(P_i) \setminus2^{A_i}$, therefore, if $K \in M$ it is automatic that $K \in \Gamma_*(P_i)\setminus 2^{A_i}$. So it remains to show that $K \notin \Gamma_*(P_i \setminus Q)$, which can be proven by contradiction as follows. For any $K\in M$, assume that $K\in \Gamma_*(P_i\setminus Q)=\Gamma_*(P_i\setminus \Gamma^*(M))$, which indicates that there exists some $J\in P_i\setminus \Gamma^*(M)$ such that $K\subseteq J$. However, as $K\in M$ and $K\subseteq J$, we have $J\in \Gamma^*(M)$, which contradicts with $J\in P_i\setminus \Gamma^*(M)$. Therefore, for any $K\in M$, we must have $K \notin \Gamma_*(P_i \setminus Q)$. 
In summary, for any given $M \subseteq \Gamma_*(P_i) \setminus2^{A_i}$ and the corresponding inequality from \eqref{eq:first:decode:fixed:rep2}, we have proved that there exists an inequality in \eqref{eq:first:decode:fixed:minli:rep} that is no looser.

To prove the other direction, for any $Q \subseteq P_i$ we construct $M = \Gamma_*(P_i)\setminus \Gamma_*(P_i \setminus Q)\setminus 2^{A_i}$. Therefore, the LHS of \eqref{eq:first:decode:fixed:rep2} and \eqref{eq:first:decode:fixed:minli:rep} become identical.  Our  claim is that if $J \in \Gamma^*(M)\,\cap \, P_i$ then $J \in Q$ or that $\Gamma^*(M)\,\cap \, P_i\subseteq  Q$. Before proving our claim, we show that with the choice of $M = \Gamma_*(P_i)\setminus \Gamma_*(P_i \setminus Q)\setminus 2^{A_i}$, we have 
\begin{align}\label{eq:prove:M}
M=\Gamma^*(M) \cap \Gamma_*(P_i).
\end{align}
The direction $M \subseteq \Gamma^*(M) \cap \Gamma_*(P_i)$ is easy, as $M \subseteq \Gamma^*(M)$ and $M \subseteq \Gamma_*(P_i)$ by construction. To show $ \Gamma^*(M) \cap \Gamma_*(P_i) \subseteq M$, for all $J \in\Gamma^*(M) \cap \Gamma_*(P_i)$, we have $J \in \Gamma^*(M)$ and $J \in \Gamma_*(P_i)$. One can show a contradiction in assuming $J \in  \Gamma_*(P_i \setminus Q)$ or $J \in 2^{A_i}$. Therefore, $J \in \Gamma_*(P_i)$, $J\notin\Gamma_*(P_i \setminus Q)$ and $J \notin 2^{A_i}$, hence $J \in M$, which completes the proof of \eqref{eq:prove:M}.

Now we go back to proving the claim $\Gamma^*(M)\,\cap \, P_i\subseteq  Q$ or equivalently, proving $P_i \setminus Q \subseteq P_i \setminus (\Gamma^*(M)\,\cap \, P_i)$. For all $J \in P_i \setminus Q$, we have $J \in P_i$, which means $J \in \Gamma_*(P_i)$. Also, $J \in P_i \setminus Q$ means $J \in \Gamma_*(P_i\setminus Q)$. Since $M$ was constructed as $M = \Gamma_*(P_i)\setminus \Gamma_*(P_i \setminus Q)\setminus 2^{A_i}$, then $J \in \Gamma_*(P_i\setminus Q)$ means that $J \notin M$. However, from $J \notin M$, $J \in \Gamma_*(P_i)$ and \eqref{eq:prove:M}, we conclude that $J \notin \Gamma^*(M)$, which means $J \notin \Gamma^*(M) \cap P_i$. Therefore, $J \in P_i \setminus (\Gamma^*(M) \cap P_i)$, which completes the proof of our claim. In summary, for any given $Q\subseteq P_i$ and the corresponding inequality from \eqref{eq:first:decode:fixed:minli:rep}, we have proved that there exists an inequality in \eqref{eq:first:decode:fixed:rep2} that is no looser.

In conclusion, we have proved that the system of inequalities \eqref{eq:first:decode:fixed:rep2} and \eqref{eq:first:decode:fixed:minli:rep} are identical for a given $(P_i, i\in [n])$.

\section{Example of Theorem \ref{thm:compcod} with selected active servers} \label{app:example}

\begin{example}\label{exmp:simple4}
Consider the distributed index coding problem $(1|4),(2|1,3),(3|1,2),(4|2,3)$ with two active servers $N_A = \{\{1,2,3\}, \{2,3,4\}\}$ with positive link capacities $C_J, J \in N_A$ and $C_J = 0$, $J \in N \setminus N_A$. Note that
\begin{align*}
&\Gamma_*(N_A)=\bigcup_{J \in N_A}2^J\\ 
&= \{\emptyset, \{1\}, \{2\}, \{1,2\}, \{3\}, \{1,3\}, \{2,3\}, \{1,2, 3\}\}  \\
&\qquad \cup\{\emptyset,\{2\}, \{3\}, \{2,3\}, \{4\}, \{2,4\}, \{3,4\}, \{2,3,4\}\}\\
&= \{\emptyset, \{1\}, \{2\}, \{1,2\}, \{3\}, \{1,3\}, \{2,3\}, \{1,2, 3\},   \\
&\qquad \{4\}, \{2,4\}, \{3,4\}, \{2,3,4\}\}.
\end{align*}
We choose $P_i = N_A$, $i \in [n]$, $D_1 = \{1\}$, $D_2 = \{2,4\}$, $D_3 = \{3,4\}$ and $D_4 = \{1,4\}$. Hence, $D_1\cup A_1 = \{1,4\}$, and $D_i\cup A_i = \{1,2,3,4\}$,  $i = 2, 3, 4$. Note that $\Delta_i = D_i$, $i \in [n]$. Active inequalities from \eqref{eq:second:decode:fixed} are given in \eqref{eq:exmp:simple4:first}.

Before detailing \eqref{eq:first:decode:fixed}, we note that with respect to $N_A$, for $M_1 = \{\{1\}, \{1,2\},\{1,3\},\{1,2,3\}\}$, we have $\Gamma^*(M_1) = \{\{1,2,3\}\}$, for $M_2 = \{\{4\}, \{2,4\},\{3,4\},\{2,3,4\}\}$, we have $\Gamma^*(M_2) = \{\{2,3,4\}\}$, and for $M_3 = \Gamma_*(N_A)$, we have $\Gamma^*(M_3) = N_A$. 
With this in mind, we write the active inequalities \eqref{eq:first:decode:fixed} as \eqref{eq:exmp:simple4:second}.

\begin{figure*}[t]
\begin{align}
\begin{matrix*}[l]
R_1 < S_{\{1\}}, & i = 1, L = \Delta_1,\\
R_2 < S_{\{2\}}+S_{\{1,2\}}+S_{\{2,3\}}+S_{\{1,2,3\}}+S_{\{2,4\}}+S_{\{2,3,4\}}, & i = 2, L = \{2\}\subset \Delta_2,\\
R_4 < S_{\{4\}}+S_{\{2,4\}}+S_{\{3,4\}}+S_{\{2,3,4\}}, & i = 2,3,4, L = \{4\}\subset \Delta_i,\\
R_2+R_4 < S_{\{2\}}+S_{\{1,2\}}+S_{\{1,3\}}+S_{\{2,3\}}+S_{\{1,2,3\}}+S_{\{4\}}+S_{\{2,4\}}+S_{\{2,3,4\}}, & i = 2, L = \Delta_2,\\
R_3 < S_{\{3\}}+S_{\{1,3\}}+S_{\{2,3\}}+S_{\{1,2,3\}}+S_{\{3,4\}}+S_{\{2,3,4\}}, & i = 3, L = \{3\}\subset \Delta_3,\\
R_3+R_4 < S_{\{3\}}+S_{\{1,3\}}+S_{\{2,3\}}+S_{\{1,2,3\}}+S_{\{4\}}+S_{\{2,4\}}+S_{\{3,4\}}+S_{\{2,3,4\}}, & i = 3, L= \Delta_3\\
R_1+R_4 < S_{\{1\}}+S_{\{1,2\}}+S_{\{1,3\}}+S_{\{1,2,3\}}+S_{\{4\}}+S_{\{2,4\}}+S_{\{3,4\}}+S_{\{2,3,4\}}, & i = 4, L= \Delta_4.
\end{matrix*}
\label{eq:exmp:simple4:first}
\end{align}
\begin{align}
\begin{matrix*}[l]
 S_{\{1\}}+S_{\{1,2\}}+S_{\{1,3\}}+S_{\{1,2,3\}} < C_{\{1,2,3\}}, & i = 1,4,M = M_1,\\
 S_{\{4\}}+S_{\{2,4\}}+S_{\{3,4\}}+S_{\{2,3,4\}} < C_{\{2,3,4\}}, & i = 2,3,4, M = M_2,\\
 \sum_{K \in N, K \neq \{4\}} S_K < C_{\{1,2,3\}}+C_{\{2,3,4\}}, & i = 1,M = M_3\setminus \{\{4\}\},\\
\sum_{K\in \{ \{2\},\{1,2\},\{2,3\},\{1,2,3\},\{4\},\{2,4\},\{3,4\},\{2,3,4\} \}}S_K < C_{\{1,2,3\}}+C_{\{2,3,4\}}, & i = 2,M = M_3\setminus 2^{\{1,3\}},\\
\sum_{K\in \{ \{3\},\{1,3\},\{2,3\},\{1,2,3\},\{4\},\{2,4\},\{3,4\},\{2,3,4\} \}}S_K <C_{\{1,2,3\}}+C_{\{2,3,4\}}, & i = 3,M = M_3\setminus 2^{\{1,2\}},\\
\sum_{K\in \{ \{1\},\{1,2\},\{1,3\},\{1,2,3\},\{4\},\{2,4\},\{3,4\},\{2,3,4\} \}}S_K < C_{\{1,2,3\}}+C_{\{2,3,4\}}, & i = 4,M = M_3\setminus 2^{\{2,3\}}.\\
\end{matrix*}
\label{eq:exmp:simple4:second}
\end{align}
\rule[0.6ex]{\textwidth}{0.3mm}
\end{figure*}

We apply FME to eliminate all present $S_K$ variables and find that the achievable rate-capacity tuple $(\Rv, \Cv)$ satisfies
\begin{align*}
 R_1 &< C_{\{1,2,3\}},\\
 R_4 &< C_{\{2,3,4\}},\\
 R_1+R_2 &<C_{\{1,2,3\}}+C_{\{2,3,4\}},\\
R_1+R_3 &< C_{\{1,2,3\}}+C_{\{2,3,4\}},\\
 R_2+R_4 &<C_{\{1,2,3\}}+C_{\{2,3,4\}},\\
R_3+R_4 &< C_{\{1,2,3\}}+C_{\{2,3,4\}}.
\end{align*}
It can be verified that the above rate region matches the all-server grouping outer bound $\Rr_{*}$ in Corollary \ref{cor:pm:coarsest}, and thus establishes the capacity region for the problem.

In the following we rewrite active first-step decoding inequalities in the form of \eqref{eq:first:decode:fixed:minli}. 
Recall that $P_i = N_A = \{ \{1,2,3\},\{2,3,4\} \}$, $i \in [n]$. 
For $Q_1 = \{\{1,2,3\}\}$, we have $\Gamma_*(Q_1)\setminus\Gamma_*(P_i\setminus Q_1) = \{\{1\},  \{1,2\}, \{1,3\}, \{1,2, 3\}\}$, 
for $Q_2= \{\{2,3,4\}\}$, we have $\Gamma_*(Q_2)\setminus\Gamma_*(P_i\setminus Q_2) = \{\{4\}, \{2,4\}, \{3,4\}, \{2, 3,4\}\}$, 
and for $Q_3 = P_i = N_A$, we have 
\begin{align*}
&\Gamma_*(Q_3)\setminus\Gamma_*(P_i\setminus Q_3) \\
&= \{\{1\}, \{2\}, \{1,2\}, \{3\}, \{1,3\}, \{2,3\}, \\
&\qquad \{1,2,3\}, \{4\}, \{2,4\}, \{3,4\}, \{2,3,4\}\}.
\end{align*}
Inequality \eqref{eq:first:decode:fixed:minli} (excluding inactive inequalities) gives \eqref{eq:exmp:simple4:second}, which showcases the equivalence of \eqref{eq:first:decode:fixed} and \eqref{eq:first:decode:fixed:minli} as claimed by Proposition~\ref{propo:compcod:equiv}. 
\end{example}

\section{Remainder of Proof of Theorem \ref{thm:pm:group}}  \label{app:outer:group}
First we present a lemma based on the encoding condition in \eqref{eq:encoding:condition} and the touch structure.
\begin{lemma} \label{lemma:JSbar}
For any set $K\subseteq [n]$, $H(T_{\ntouch{K}}|\Xv_{K^c})=0$.
\end{lemma}
\begin{IEEEproof}
For any set $K\subseteq [n]$, we have
\begin{align*}
H(T_{\ntouch{K}}|\Xv_{K^c})&=H(\Yv_{\{J:J \in N, J \cap K = \emptyset\}}|\Xv_{K^c})  \\
&=H(\Yv_{\{J:J\subseteq K^c\}}|\Xv_{K^c})  \\
&\leq \sum_{J\subseteq K^c}H(Y_J|\Xv_{K^c})=0,
\end{align*}
where the last equality follows from the encoding condition in \eqref{eq:encoding:condition}.
\end{IEEEproof}

Now we prove that the set function $f(G,K)$ defined in \eqref{eq:group:pm:proof:def:f} satisfies Axioms \eqref{eq:axiom1}-\eqref{eq:axiom6} of Theorem \ref{thm:pm:group}. 
Toward that end, we first show that for any $\e>0$, the set function $f_{\e}(G,K)$ defined in \eqref{eq:f-def2} satisfies the following conditions, which are counterparts of \eqref{eq:axiom1}-\eqref{eq:axiom5} and \eqref{eq:axiom6} (we deal with \eqref{eq:axiom:addi:de} later): 
\begin{align}
&f_{\e}(G,K) = f_{\e}(G',K),    \nonumber  \\
& \qquad \qquad \qquad \text{if $(P_G\cup P_{G'})\setminus (P_G \cap P_{G'})\subseteq T_{\ntouch{K}}$},   \label{eq:f:epsilon:axiom1}
\\
&f_{\e}(\emptyset,K)= f_{\e}(G,\emptyset) =0,      \label{eq:f:epsilon:axiom2}                                                                               \\
&f_{\e}(G,K)\leq \sum_{J: J \in P_G, J \in T_K} C_J,    \label{eq:f:epsilon:axiom3}                                                                      \\
&f_{\e}(G,K)\leq f_{\e}(G',K'),      \qquad   \quad  \text{if $K\subseteq K', G\subseteq G'$},   \label{eq:f:epsilon:axiom4}                      \\
&f_{\e}(G\cup G',K\cap K') + f_{\e}(G\cap G',K\cup K')   \nonumber  \\
&\qquad \qquad \leq f_{\e}(G,K) + f_{\e}(G',K'),    \label{eq:f:epsilon:axiom5}\\
&f_{\e}(G,K)+f_{\e}(G,K')=f_{\e}(G,K\cup K'),      \nonumber  \\
&\qquad \qquad \qquad \text{if $K\cap K'=\emptyset, P_G\subseteq (N\setminus T_{K,K'})$}.  \label{eq:f:epsilon:axiom6}
\end{align}
\begin{enumerate}
\item
We show that $f_{\e}(G,K)$ satisfies \eqref{eq:f:epsilon:axiom1} as follows. Note that if $((P_G\cup P_{G'})\setminus (P_G\cap P_{G'}))\subseteq T_{\ntouch{K}}$, according to Lemma \ref{lemma:JSbar}, we have
\begin{align}
0 &= H(T_{\ntouch{K}}|\Xv_{K^c})  \nonumber\\
& \geq H((P_G\cup P_{G'})\setminus (P_G\cap P_{G'})|\Xv_{K^c})  \nonumber  \\
&\geq H(P_G\setminus P_{G'}|\Xv_{K^c})  \nonumber  \\
&\geq H(P_G\setminus P_{G'}|\Xv_{K^c},P_{G'}), \label{eq:axiomA1:1:mid} \\
0 &= H(T_{\ntouch{K}}|\Xv_{K^c})\nonumber\\
&\geq H((P_G\cup P_{G'})\setminus (P_G\cap P_{G'})|\Xv_{K^c})  \nonumber  \\
&\geq H(P_{G'}\setminus P_G|\Xv_{K^c})  \nonumber  \\
&\geq H(P_{G'}\setminus P_G|\Xv_{K^c},P_G).
 \label{eq:axiomA1:2:mid}
\end{align}
Given \eqref{eq:axiomA1:1:mid} and \eqref{eq:axiomA1:2:mid} as well as the nonnegativity of entropy, we have
\begin{align}
H(P_G\setminus P_{G'}|\Xv_{K^c},P_{G'})&=0,  \label{eq:axiomA1:1}  \\
H(P_{G'}\setminus P_G|\Xv_{K^c},P_G)&=0.  \label{eq:axiomA1:2}
\end{align}
Therefore,
\begin{align*}
f_{\e}(G,K)&=\frac{1}{r}H(P_G|\Xv_{K^c}) \\
                       &=\frac{1}{r}H(P_G,P_{G'}\setminus P_G|\Xv_{K^c})  \\
                       &=\frac{1}{r}H(P_{G'},P_G\setminus P_{G'}|\Xv_{K^c})   \\
                       &=\frac{1}{r}H(P_{G'}|\Xv_{K^c})=f_{\e}(G',K),
\end{align*}
where the second equality follows from \eqref{eq:axiomA1:2} and the fourth equality follows from \eqref{eq:axiomA1:1}.
\item
We show that $f_{\e}(G,K)$ satisfies \eqref{eq:f:epsilon:axiom2} as follows. It is obvious that for any $G\subseteq [m]$, $K\subseteq [n]$,
\begin{align}
f_{\e}(\emptyset,K) &=\frac{1}{r}H(\emptyset|\Xv_{K^c})=0,\\
f_{\e}(G,\emptyset) &=\frac{1}{r}H(P_G|\Xv_{[n]})=0, \label{eq:axiomA2:1}
\end{align}
where the last equality of \eqref{eq:axiomA2:1} is due to the encoding condition in \eqref{eq:encoding:condition}. 
\item 
We show that $f_{\e}(G,K)$ satisfies \eqref{eq:f:epsilon:axiom3} as follows. According to Lemma \ref{lemma:JSbar} and the fact that conditioning cannot increase entropy, we have
\begin{align*}
f_{\e}(G,K)&=\frac{1}{r}H(P_G|\Xv_{K^c})  \\
&=\frac{1}{r}H(P_G\cup T_{\ntouch{K}}|\Xv_{K^c})  \\
&=\frac{1}{r}H(T_{\ntouch{K}},P_G\setminus T_{\ntouch{K}}|\Xv_{K^c})  \\
&=\frac{1}{r}H(P_G\setminus T_{\ntouch{K}}|\Xv_{K^c})     \\
&\leq \frac{1}{r}H(P_G\setminus T_{\ntouch{K}})  \\
&\leq \frac{1}{r}\sum_{J:J\in P_G,J\in T_K}r_J \leq \sum_{J:J\in P_G,J\in T_K}C_J.
\end{align*}
\item
We show that $f_{\e}(G,K)$ satisfies \eqref{eq:f:epsilon:axiom4} as follows. Note that if $G\subseteq G'\subseteq [m]$, $K\subseteq K' \subseteq [n]$, then $K'^c \subseteq K^c$. Therefore, we have
\begin{align*}
f_{\e}(G,K)&=\frac{1}{r}H(P_G|\Xv_{K^c})  \\
&\leq \frac{1}{r}H(P_G|\Xv_{K'^c})  \\
&\leq \frac{1}{r}H(P_{G'}|\Xv_{{K'}^c})=f_{\e}(G',K').
\end{align*}
\item
We show that $f_{\e}(G,K)$ satisfies \eqref{eq:f:epsilon:axiom5} as follows. Let us define $G_1=G\setminus G'$, $G_2=G'\setminus G$ and $G_0=G\cap G'$, so that $G\cup G'=G_0 \cup G_1 \cup G_2$ is the union of three disjoint sets and $G = G_0\cup G_1$ and $G' = G_0\cup G_2$. Similarly, define  $K_0 = K^c \cap K'^c$, $K_1 = K^c\setminus K'^c$ and $K_2 = K'^c\setminus K^c$ so that $K^c \cup K'^c = K_0 \cup K_1\cup K_2$, $K^c = K_0 \cup K_1$ and $K'^c= K_0 \cup K_2$. 

Set $$f_1=f_{\e}(G\cup G',K\cap K')+f_{\e}(G\cap G',K\cup K')+\frac{1}{r}H_1,$$ where $H_1 = H(\Xv_{K_0 \cup K_1\cup K_2})+H(\Xv_{K_0})$, and set $$f_2=f_{\e}(G,K)+f_{\e}(G',K')+\frac{1}{r}H_2,$$ where $H_2=H(\Xv_{K_0 \cup K_1})+H(\Xv_{K_0 \cup K_2})$. Due to the message independence in \eqref{eq:message:independence} and sets $K_0,K_1, K_2$ being disjoint, we have $H_1=H_2$.

We can verify that for any server grouping $\Pcal=\{P_1,P_2,\cdots,P_m\}$, $G\subseteq [m]$, we have
\begin{align}
P_{G\cup G'}&=\bigcup_{i\in G_0\cup G_1\cup G_2}P_i=P_G \cup P_{G'},  \label{eq:axiomA5:union}  \\
P_{G\cap G'}&=\bigcup_{i\in G_0}P_i \subseteq P_G \cap P_{G'}, \label{eq:axiomA5:intersect}
\end{align}
where \eqref{eq:axiomA5:intersect} is due to the possible overlapping between two different server groups, e.g., even for two disjoint sets $G_1,G_2\subseteq [m]$, $P_{G_1}\cap P_{G_2}$ may not be $\emptyset$.
If $P_1,\cdots,P_m$ happen to be disjoint server groups, we will have $P_{G\cap G'}=P_G \cap P_{G'}$.

Therefore, we can write
\begin{align*}
rf_1&=H(P_{G\cup G'}|\Xv_{K_0 \cup K_1\cup K_2})+H(P_{G\cap G'}|\Xv_{K_0})\\\nonumber
&\quad\quad+H_1\\
     &=H(P_{G\cup G'},\Xv_{K_0 \cup K_1\cup K_2})+H(P_{G\cap G'},\Xv_{K_0})\\
     &\leq H(P_G \cup P_{G'},\Xv_{K_0 \cup K_1\cup K_2})\\\nonumber
     &\quad\quad +H(P_G \cap P_{G'},\Xv_{K_0})\\
     &\leq H(P_G,\Xv_{K_0\cup K_1}) +H(P_{G'},\Xv_{K_0\cup K_2})\\
     &=H(P_G|\Xv_{K_0\cup K_1})+H(P_{G'}|\Xv_{K_0\cup K_2})+H_2\\\nonumber
     &=rf_2.
\end{align*}
where the first inequality is due to \eqref{eq:axiomA5:union} and \eqref{eq:axiomA5:intersect} and the second inequality is due to the submodularity of the entropy function. Finally, we have
\begin{align*}
f_{\e}&(G\cup G',K\cap K')+f_{\e}(G\cap G',K\cup K')  \\
&=f_1\!-\!\frac{1}{r}H_1\leq f_2\!-\!\frac{1}{r}H_2=f_{\e}(G,K)+f_{\e}(G',K').
\end{align*}
\item We show that $f_{\e}(G,K)$ satisfies \eqref{eq:f:epsilon:axiom6} as follows. For any $K,K'\subseteq [n]$ such that $K\cap K'=\emptyset$, set $L=[n]\setminus (K\cup K')$. For $P_G\subseteq (N \setminus T_{K,K'})$, according to Proposition \ref{propo:fdseparation} (to be presented in Appendix~\ref{sec:fd}), we have 
\begin{align}
0&=I(\Xv_K;\Xv_{K'}|P_G,\Xv_L)  \nonumber  \\
 &=H(\Xv_K|P_G,\Xv_L)+H(\Xv_{K'}|P_G,\Xv_L)  \nonumber  \\
 &\qquad -H(\Xv_{K},\Xv_{K'}|P_G,\Xv_L)  \nonumber  \\
 &=H(P_G,\Xv_{K\cup L})-H(P_G,\Xv_L)+H(P_G,\Xv_{K'\cup L})  \nonumber  \\
 &\qquad -H(P_G,\Xv_L)-H(P_G,\Xv_{[n]})+H(P_G,\Xv_L)  \nonumber  \\
 &=H(P_G,\Xv_{K\cup L})+H(P_G,\Xv_{K'\cup L})  \nonumber  \\
 &\qquad -H(P_G,\Xv_L)-H(\Xv_{[n]})  \label{eq:axiomA6:1}  \\
 &=H(P_G|\Xv_{K\cup L})+H(P_G|\Xv_{K'\cup L})  \nonumber  \\
 &\qquad -H(P_G|\Xv_L) \label{eq:axiomA6:2} \\
 &=rf_{\e}(G,K')+rf_{\e}(G,K)  \nonumber  \\
 &\qquad -rf_{\e}(G,K\cup K'), \label{eq:axiomA6:3}
\end{align}%
where \eqref{eq:axiomA6:1} follows from the encoding condition in \eqref{eq:encoding:condition}, and \eqref{eq:axiomA6:2} follows from the message independence in \eqref{eq:message:independence}. Obviously, given $r$ being positive, by \eqref{eq:axiomA6:3}, we have $f_{\e}(G,K')+f_{\e}(G,K)=f_{\e}(G,K\cup K')$.
\end{enumerate}

Now that we have shown that $f_{\e}(G,K)$ satisfies \eqref{eq:f:epsilon:axiom1}-\eqref{eq:f:epsilon:axiom6}, we use this result to show that $f(G,K)$ satisfies corresponding six axioms \eqref{eq:axiom1}-\eqref{eq:axiom5} and \eqref{eq:axiom6} in the following. 

\begin{enumerate}
\item For Axiom \eqref{eq:axiom1}, if $((P_G\cup P_{G'})\setminus (P_G\cap P_{G'}))\subseteq T_{\ntouch{K}}$, we have
\begin{align*}
f(G,K)&=\liminf_{\e \to 0}f_{\e}(G,K)  \nonumber  \\
&=\liminf_{\e \to 0}f_{\e}(G',K)=f(G',K),
\end{align*}
where the second equality follows from \eqref{eq:f:epsilon:axiom1}.
\item For Axiom \eqref{eq:axiom2}, due to \eqref{eq:f:epsilon:axiom2}, we have
\begin{align*}
f(\emptyset,K)&=\liminf_{\e \to 0}f_{\e}(\emptyset,K)=0,  \\
f(G,\emptyset)&=\liminf_{\e \to 0}f_{\e}(G,\emptyset)=0.
\end{align*}
\item For Axiom \eqref{eq:axiom3}, due to \eqref{eq:f:epsilon:axiom3}, we have
\begin{align*}
f(G,K)=\liminf_{\e \to 0}f_{\e}(G,K)\leq \sum_{J:J\in P_G,J\in T_K}C_J.
\end{align*}
\item For Axiom \eqref{eq:axiom4}, consider any $G\subseteq G'\subseteq [m]$, $K\subseteq K' \subseteq [n]$. By \eqref{eq:f:epsilon:axiom4}, we have
\begin{align*}
0&\le \liminf_{\e \to 0}( f_{\e}(G',K')-f_{\e}(G,K) )    \\
&=\liminf_{\e \to 0}f_{\e}(G',K')-\limsup_{\e \to 0}f(G,K)    \\
&\le \liminf_{\e \to 0}f_{\e}(G',K')-\liminf_{\e \to 0}f(G,K)    \\
&=f(G',K')-f(G,K).
\end{align*}
\item For Axiom \eqref{eq:axiom5}, consider any $G,G'\subseteq [m]$, $K,K' \subseteq [n]$. By \eqref{eq:f:epsilon:axiom5}, we have
\begin{align*}
0&\le \liminf_{\e \to 0}( f_{\e}(G',K')+f_{\e}(G,K)  \\
&\qquad -f_{\e}(G\cup G',K\cap K')-f_{\e}(G\cap G',K\cup K') )    \\
&=\liminf_{\e \to 0}f_{\e}(G',K')+\liminf_{\e \to 0}f_{\e}(G,K)   \\
&\qquad -\limsup_{\e \to 0}f(G\cup G',K\cap K')  \\
&\qquad -\limsup_{\e \to 0}f(G\cap G',K\cup K')    \\
&\le \liminf_{\e \to 0}f_{\e}(G',K')+\liminf_{\e \to 0}f_{\e}(G,K)   \\
&\qquad -\liminf_{\e \to 0}f(G\cup G',K\cap K')  \\
&\qquad -\liminf_{\e \to 0}f(G\cap G',K\cup K')    \\
&=f(G',K')+f(G,K)-f(G\cup G',K\cap K')  \\
&\qquad -f(G\cap G',K\cup K'). 
\end{align*}
\item For Axiom \eqref{eq:axiom6}, consider any $G\subseteq [m]$, $K,K' \subseteq [n]$ such that $K\cap K'=\emptyset$ and $P_G\subseteq (N \setminus T_{K,K'})$. By \eqref{eq:f:epsilon:axiom6}, we have
\begin{align*}
f(G,K)+f(G,K')&=\liminf_{\e \to 0}(f_{\e}(G,K)+f_{\e}(G,K'))  \\
&=\liminf_{\e \to 0}f_{\e}(G,K\cup K')  \\
&=f(G,K\cup K'). 
\end{align*}
\end{enumerate}
Finally for the last remaining axiom, Axiom \eqref{eq:axiom:addi:de}, we use an approach similar to the one used in the inequality leading up to \eqref{eq:rate:leading}. Consider any $\e>0$ and $i\in [n]$. We rearrange \eqref{eq:group:pm:proof:revision4:2} as 
\begin{align}
t_i\le \frac{H(\Yv_N|\Xv_{A_i})-H(\Yv_N|\Xv_{A_i\cup \{i\}})}{1-\d(\e)}. \label{eq:ti:epsilon:repeat}
\end{align}

We have
\begin{align}
rf_{\e}([m],\{i\})&=H(\Yv_N|\Xv_{\{i\}^c})-H(\Yv_N|\Xv_{[n]})   \label{eq:axiom:addi:de:1}        \\
                 &=I(X_i;\Yv_N|\Xv_{A_i\cup B_i})            \nonumber        \\
                 &=H(X_i|\Xv_{A_i\cup B_i})-H(X_i|\Yv_N,\Xv_{A_i\cup B_i})      \nonumber     \\
                 &\le H(X_i)  \nonumber  \\
                 &=t_i     \label{eq:axiom:addi:de:2}     \\
                 &\le \frac{H(\Yv_N|\Xv_{A_i})-H(\Yv_N|\Xv_{A_i\cup \{i\}})}{1-\d(\e)}   \label{eq:axiom:addi:de:3}               \\
                 &=\frac{rf_{\e}([m], B_i\cup \{i\})-rf_{\e}([m], B_i)}{1-\d(\e)}   \label{eq:axiom:addi:de:4}
\end{align}
where \eqref{eq:axiom:addi:de:1} follows from the encoding condition in \eqref{eq:encoding:condition}, 
\eqref{eq:axiom:addi:de:2} follows from the fact that the messages are uniformly distributed as specified in \eqref{eq:message:independence}, \eqref{eq:axiom:addi:de:3} follows from \eqref{eq:ti:epsilon:repeat}, and \eqref{eq:axiom:addi:de:4} follows from the definition of $f_{\e}(G,K)$. 
Dividing both sides of \eqref{eq:axiom:addi:de:4} by $r$ and then taking the limit infimum as $\e$ approaches zero, we have
\begin{align}
f([m],\{i\})&\leq\liminf_{\e \to 0}\frac{f_{\e}([m], B_i\cup \{i\})-f_{\e}([m], B_i)}{1-\d(\e)} \nonumber  \\
&=\liminf_{\e \to 0}( f_{\e}([m],B_i\! \cup\! \{i\}) - f_{\e}([m],B_i) )    \nonumber   \\
&=\liminf_{\e \to 0} f_{\e}([m],B_i \!\cup\! \{i\})\!-\! \limsup_{\e \to 0} f_{\e}([m],B_i )  \nonumber \\
&\le \liminf_{\e \to 0} f_{\e}([m],B_i\!\cup\! \{i\})\! -\! \liminf_{\e \to 0} f_{\e}([m],B_i ) \nonumber  \\
&=f([m],B_i\!\cup\! \{i\})-f([m],B_i).  \label{eq:axiom:addi:de:mid}
\end{align}
On the other hand, by Axiom \eqref{eq:axiom5} we have
\begin{align}
f([m],\{i\})\ge f([m],B_i\cup \{i\})-f([m],B_i).    \label{eq:axiom:addi:de:7}
\end{align}
Combining \eqref{eq:axiom:addi:de:mid} and \eqref{eq:axiom:addi:de:7} leads to Axiom \eqref{eq:axiom:addi:de}.

This concludes the proof of Theorem \ref{thm:pm:group}.

\section{Functional Dependence Graph and $\mathrm{fd}$-separation for Distributed Index Coding}\label{sec:fd}

We review the functional dependence graph (FDG), which was first introduced in \cite{kramer} and then further developed in \cite{satyajit}. We first restate the general definition of FDG and then specialize it to the distributed index coding FDG based on the distributed index coding problem setup.

Within this section, we use $\Gcal=(\Vcal,\Ecal)$ to denote a directed graph with vertex set $\Vcal=\{ V_1,V_2,\cdots \}$ and directed edge set $\Ecal=\{ e_1,e_2,\cdots \}$. 
We use $\mathrm{tail}(e)$ and $\mathrm{head}(e)$ to denote the tail and the head of the directed edge $e, \forall e\in \Ecal$, respectively.
For any $V_j,V_k\in \Vcal$, $j<k$, we say that vertices $V_j, V_k \in \Vcal$ are \emph{connected} if there exist vertices in $\Vcal$, $V_j,V_{j+1},\dots,V_k$, and edges in $\Ecal$, $e_j,\dots,e_{k-1}$, 
such that for any $i\in [j:k-1]$, we have either ${\rm tail}(e_i)=V_i$, ${\rm head}(e_i)=V_{i+1}$ or ${\rm tail}(e_i)=V_{i+1}$, ${\rm head}(e_i)=V_{i}$. 
We call such vertex sequence $V_j,V_{j+1},\dots,V_k$ a path between $V_j$ and $V_k$. Correspondingly, we say that $V_j$ and $V_k$ are \emph{disconnected} if such intermediate vertices and edges do not exist (i.e., there is no path between $V_j$ and $V_k$). Note that we ignore the direction of the edges when determining whether two vertices are connected or not.

\begin{definition}[Functional dependence graph] \label{FDG}
Let $\Vcal=\{ V_1,V_2,\dots \}$ be a set of random variables. A directed graph $\Gcal =(\Vcal,\Ecal)$ is called a functional dependence graph (FDG) for $\Vcal$ if and only if 
\begin{align}
H(V_i|V_j:(V_j,V_i)\in \Ecal)=0, \quad \forall V_i \in \Vcal.
\end{align}
\end{definition}

\begin{definition}[Distributed index coding FDG] \label{index:FDG}
For a given distributed index coding problem, its distributed index coding FDG is a directed graph $\Gcal=(\Vcal,\Ecal)$ defined as follows.
\begin{itemize}
\item The set of vertices $\Vcal=\Xv_{[n]}\cup \Yv_N$. 
\item For any $V,V'\in \Vcal$, $(V, V')\in \Ecal$ if and only if it satisfies one of the following conditions:
\begin{enumerate}
\item $V=X_i,V'=Y_J,J\in N, i\in J$, i.e., $(V,V')$ denotes message availability at server $J$;\label{vertex1}
\item $V=X_i,V'=X_j,j\in [n], i\in A_j$, i.e., $(V,V')$ denotes side information availability at receiver $j$;\label{vertex2}
\item $V=Y_J,V'=X_i,J\in N, i\in [n]$, i.e., $(V,V')$ denotes a broadcast link from server $J$ to receiver $i$;\label{vertex3}
\end{enumerate}
\end{itemize}
\end{definition}
Note that the encoding conditions $H(Y_J|\Xv_J)=0$ are captured in the distributed index coding FDG due to the existence of edges as defined in item \ref{vertex1} above. 
The decoding conditions $H(X_i|\Yv_N,\Xv_{A_i})=0$ are captured due to the existence of edges as defined in items \ref{vertex2} and \ref{vertex3}.\footnote{Note that for the distributed index coding FDG, we assume zero-error decoding conditions at receivers for simplicity. However, as the $\rm fd$-separation and Proposition \ref{propo:fdseparation} (to be defined shortly) depend only on the message independence and the encoding conditions at servers, they hold in the general case of vanishing decoding error probability.}
Hence, it can be verified that for any distributed index coding FDG $\Gcal=(\Vcal, \Ecal)$ we have
\begin{align*}
H(V_i|V_j:(V_j,V_i)\in \Ecal)=0, \quad \forall V_i \in \Vcal.
\end{align*}
Therefore, the distributed index coding FDG defined in Definition \ref{index:FDG} is indeed an FDG satisfying Definition \ref{FDG}.

\begin{figure*}[t]
\begin{center}
\includegraphics[scale=0.18]{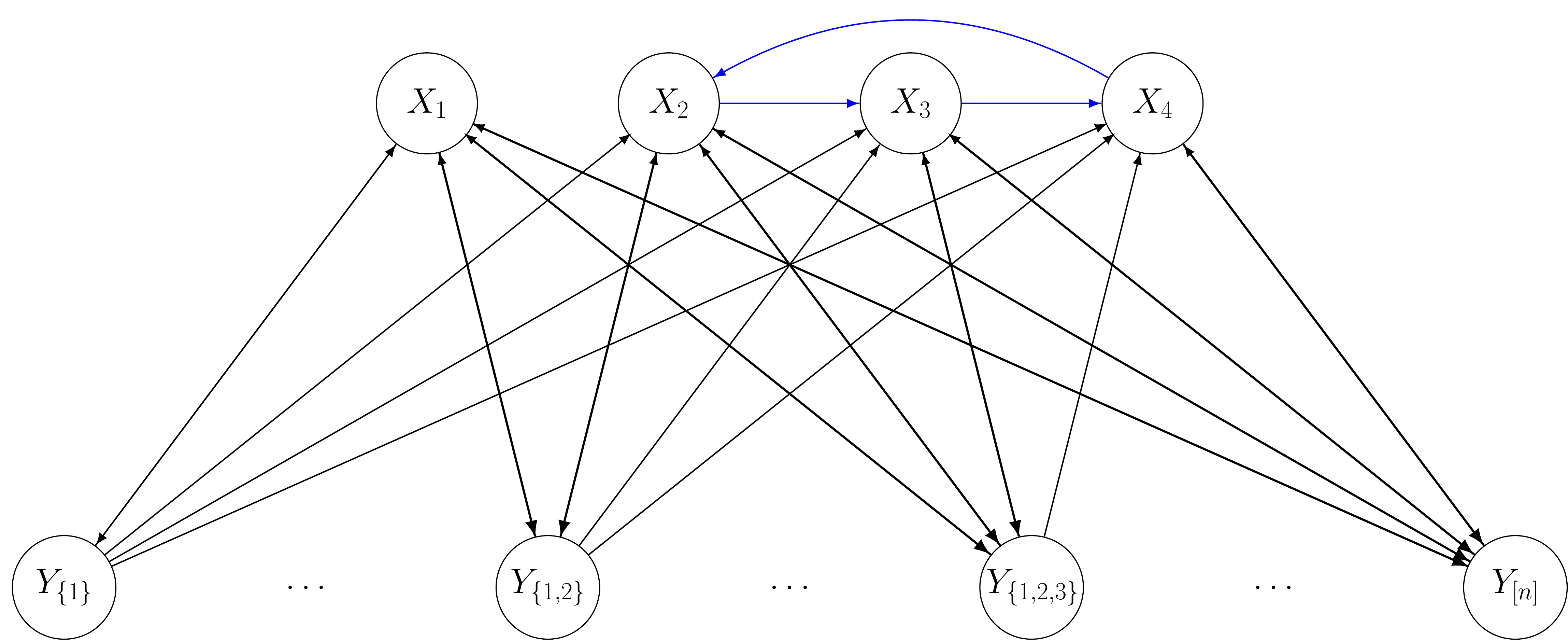}
\caption{The distributed index coding FDG for the $4$-message problem: $(1|-),(2|4),(3|2),(4|3)$, with all $2^4-1=15$ servers. For simplicity, only four output variables, $\Yv_{\{\{1\},\{1,2\},\{1,2,3\},\{1,2,3,4\}\}}$, and their corresponding links are shown. To avoid clutter, whenever there exist directed edges in both directions between any two vertices, we simply draw an edge with arrows at both ends between the vertices, instead of drawing two separate directed edges. 
Note that the edges defined in item \ref{vertex2} of Definition \ref{index:FDG} are shown as blue, while all the other edges defined in items \ref{vertex1} and \ref{vertex3} are shown in black.}
\label{fig:fdg:case16}
\end{center}
\end{figure*}

\begin{example}\label{exm:index:FDG}
See Figure \ref{fig:fdg:case16} for the distributed index coding FDG for the problem: $(1|-),(2|4),(3|2),(4|3)$. 
\end{example}

\begin{figure*}[t]
\begin{center}
\includegraphics[scale=0.16]{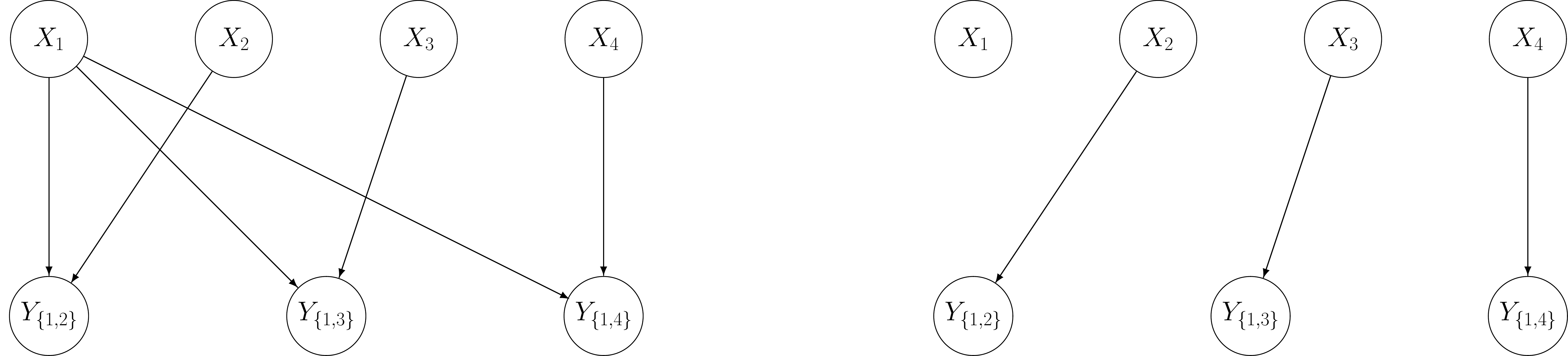}
\caption{The ancestral graph $\Gcal_{\rm{An}(\Ucal \cup \Wcal \cup \Zcal)}$ is shown on the left. And the remaining of $\Gcal_{\rm{An}(\Ucal \cup \Wcal \cup \Zcal)}$ after removing all edges outgoing from vertices in $\Ucal$ is shown on the right, where $X_2$ becomes disconnected from both $X_3$ and $X_4$.}
\label{fig:fd:separation}
\end{center}
\end{figure*}

Now we review the $\mathrm{fd}$-separation criterion, also from \cite{satyajit,kramer}, which leads to the conditional message independence utilized in Axiom \eqref{eq:axiom6} of Theorem  \ref{thm:pm:group} in Section \ref{sec:outer:bounds}. Similar to the distributed index coding FDG, the $\mathrm{fd}$-separation presented here has also been specialized to the distributed index coding scenario.

\begin{definition}[Ancestral graph] \label{def:ancestral}
Consider the distributed index coding FDG $\Gcal=(\Vcal,\Ecal)$ of a given distributed index coding problem. For any subset $\Acal \subseteq \Vcal$, let $\mathrm{An}(\Acal)$ be the set of all vertices in $\Vcal \setminus \Acal$ such that for every vertex $V \in \mathrm{An}(\Acal)$, there is a directed path from $V$ to some vertex $V'$ in $\Acal$ in the subgraph $\bar{\Gcal}=\Gcal \setminus \{ e\in \Ecal:\exists i\in [n],\mathrm{head}(e)=X_i \}$. The ancestral graph with respect to $\Acal$, denoted by $\Gcal_{\rm{An}(\Acal)}$, is a vertex-induced subgraph of $\Gcal$ consisting of vertices $(\Acal \cup \rm{An}(\Acal))$ and edges $e\in \Ecal$ such that $\rm{head}(e), \rm{tail}(e) \in \Acal \cup \rm{An}(\Acal)$. 

\end{definition}

\begin{definition}[$\mathrm{fd}$-separation]\label{def:fdseparation}
Let $\Gcal=(\Vcal,\Ecal)$ be the distributed index coding FDG of a given distributed index coding problem, and let $\Ucal,\Wcal,\Zcal$ be three nonempty disjoint subsets of $\Vcal$. Set $\Ucal$ $\mathrm{fd}$-separates sets $\Wcal$ and $\Zcal$ if every vertex in $\Wcal$ is disconnected from every vertex in $\Zcal$ in what remains of $\Gcal_{\rm{An}(\Ucal \cup \Wcal \cup \Zcal)}$ after removing all edges outgoing from vertices in $\Ucal$.
\end{definition}

\begin{example}\label{exm:fdseparation}
Recall the $4$-message problem whose distributed index coding FDG is shown in Figure \ref{fig:fdg:case16}. Set $\Ucal=\{ \Xv_{\{1\}},\Yv_{\{\{1,2\},\{1,3\},\{1,4\}\}} \},\Wcal=\{ \Xv_{\{2\}} \},\Zcal=\{ \Xv_{\{3,4\}} \}$. It can be verified that $\Ucal$ $\mathrm{fd}$-separates $\Wcal$ and $\Zcal$, as illustrated by Figure \ref{fig:fd:separation}. 

\end{example}

It can be verified that once the subset of vertices $\Ucal$ $\mathrm{fd}$-separates $\Zcal$ and $\Wcal$ in the distributed index coding FDG for any distributed index coding problem, then it also $\mathrm{fd}$-separates $\Zcal$ and $\Wcal$ in the corresponding network FDG \cite[Definition 11]{satyajit}. Therefore, according to Lemma 4 in \cite{satyajit}, we conclude that the random variables denoted by $\Zcal$ and $\Wcal$ are conditionally independent given the random variables denoted by $\Ucal$, i.e.,
\begin{align}
&I(\Zcal;\Wcal|\Ucal)=0,  \quad \text{if $\Ucal$ $\mathrm{fd}$-separates $\Zcal$ and $\Wcal$}  \nonumber  \\
&\qquad \qquad \text{in the distributed index coding FDG.} \label{eq:fd:lemma4}
\end{align}

Now we can state the following proposition.
\begin{proposition}\label{propo:fdseparation}
For any distributed index coding problem and two disjoint nonempty subsets $K,K'\subset [n]$, set $L=[n]\setminus (K\cup K')$. Then, we have
\begin{align} 
I(\Xv_{K};\Xv_{K'}|\Xv_L,\Yv_P) = 0,  \label{eq:fdg2:1} 
\end{align}
for any subset of servers $P\subseteq (N\setminus T_{K,K'})$.
\end{proposition}
\begin{IEEEproof}
Set $\Ucal=\Xv_L\cup \Yv_P$, $\Zcal=\Xv_{K}$, $\Wcal=\Xv_{K'}$.
Since $P\subseteq N\setminus T_{K,K'}= (T_{K,\ntouch{K'}} \cup T_{K',\ntouch{K}} \cup T_{\ntouch{K},\ntouch{K'}})$, according to the touch structure in Definition \ref{def:structure:J}, we know that in the ancestral graph $\Gcal_{\rm{An}(\Ucal \cup \Wcal \cup \Zcal)}$, for any $X_i\in \Zcal, X_j\in \Wcal$, vertices $X_i$ and $X_j$ are either disconnected, or connected with at least one vertex from $\Xv_L\subseteq \Ucal$ in the path between them.

After removing all edges outgoing from vertices in $\Ucal$, any connected vertices $X_i\in \Zcal$ and $X_j\in \Wcal$ become disconnected. 
Hence, we can conclude that the vertex set $\Ucal$ $\mathrm{fd}$-separates $\Zcal$ and $\Wcal$ in the distributed index coding FDG. Therefore, from (\ref{eq:fd:lemma4}), we conclude that $I(\Xv_{K};\Xv_{K'}|\Xv_L,\Yv_P)=I(\Zcal;\Wcal|\Ucal)=0$.
\end{IEEEproof}

\section{Proof of Proposition \ref{propo:uv}} \label{app:uv:proof}

For easier reference, we repeat \eqref{eq:uv:propo} here for a given problem $(i|A_i),i\in [n]$ with peripheral set $U$ and an augmentation group $\Vv=(V_1,\cdots,V_k)$, 
\begin{align}
\sum_{i\in [n]}R_i \le \sum_{J\in N}C_J+\sum_{\ell \in [k]} \sum_{J\in \Tcal_\ell}C_J,    \label{eq:uv:propo:repeat}
\end{align}
where $\Tcal_\ell=T_{V_{\ell}, (\bigcup_{j\in [\ell+1:k]}V_j)\cup W}$ and $W=[n]\setminus U\setminus (\bigcup_{j\in [k]}V_j)$. Note that $[k+1:k]$ simply means $\emptyset$. 

In the following, we prove the proposition by showing that \eqref{eq:uv:propo:repeat} is implied by the rate constraint inequality \eqref{eq:group:rate} as well as the Axioms \eqref{eq:axiom1}-\eqref{eq:axiom6} of Theorem \ref{thm:pm:group} with a specific server grouping $\Pcal_{\rm \Vv}=\{ T_{V_1},T_{V_2},\cdots,T_{V_k},T_{(\bigcup_{j\in [k]}V_j)^c} \}$. 

For any $i\in [n]$ and any $B\subseteq B_i$, according to \eqref{eq:group:rate} as well as Axiom \eqref{eq:axiom5}, we have
\begin{align}
R_i&\leq f([m],B_i \cup \{i\}) - f([m],B_i)  \nonumber  \\
&\leq f([m],B\cup \{i\})-f([m],B), \nonumber \\& \qquad \qquad \qquad \enskip \forall B\subseteq B_i,i\in [n].  \label{eq:uv:proof:rate}
\end{align}

Consider any two disjoint sets $L,K\subseteq [n]$. If $L$ is an augmentation set of $K$, then by Definition \ref{def:augmentation:set}, there exists an ordering $\{i_1,i_2,\cdots,i_{|L|}\}$ of the elements in $L$ such that $A_{i_j}\subseteq \{i_1,\cdots,i_{j-1}\}\cup K, j\in [|L|]$, which indicates that 
\begin{align}
(\{i_1,\cdots,i_{j-1}\}\cup K)^c \subseteq B_{i_j}, \qquad \forall j\in [|L|].  \label{eq:uv:proof:B}
\end{align}
Hence, according to \eqref{eq:uv:proof:rate} and \eqref{eq:uv:proof:B}, we have 
\begin{align*}
\sum_{i\in L}R_i&=\sum_{j\in [|L|]}R_{i_j}  \\
&\leq  f([m],K^c)-f([m],K^c\setminus \{i_1\})\\&\qquad+f([m],K^c\setminus \{i_1\})  -f([m],K^c\setminus \{i_1\} \setminus \{i_2\})\\&\qquad+f([m],K^c\setminus \{i_1\} \setminus \{i_2\})  \\
&\qquad -f([m],K^c\setminus \{i_1\} \setminus \{i_2\} \setminus \{i_3\})+\cdots \nonumber \\
&\qquad +f([m],K^c\setminus \{i_1\} \setminus \cdots \setminus \{i_{|L|-1}\})\\&\qquad-f([m],K^c\setminus L)  \\
&=f([m],K^c)-f([m],K^c\setminus L).
\end{align*}

By Definitions \ref{def:uv:u} and \ref{def:uv:v}, $U$ is an augmentation set of $\emptyset$, and $V_j$ for any $j\in [k]$ is an augmentation set of $V_j^c$, and $W$ is an augmentation set of $W^c$. Therefore, we have
\begin{align}
\sum_{i\in U}R_i&\leq f([m],\emptyset^c)-f([m],\emptyset^c\setminus U)  \nonumber  \\
&=f([m],[n])-f([m],[n]\setminus U), \label{eq:uv:proof:u}
\end{align}
and
\begin{align}
\sum_{i\in V_j}R_i&\leq f([m],(V_j^c)^c)-f([m],(V_j^c)^c\setminus V_j)  \nonumber  \\
&=f([m],V_j), \qquad \forall j\in [k], \label{eq:uv:proof:v}
\end{align}
and
\begin{align}
\sum_{i\in W}R_i&\leq f([m],(W^c)^c)-f([m],(W^c)^c\setminus W)  \nonumber  \\
&=f([m],W). \label{eq:uv:proof:w}
\end{align}

As $U,V_1,\cdots,V_k,W$ are disjoint to each other and $[n]=U\cup (\bigcup_{j\in [k]}V_j)\cup W$, combining \eqref{eq:uv:proof:u} and \eqref{eq:uv:proof:w}, as well as \eqref{eq:uv:proof:v} for every $j\in [k]$, we have
\begin{align*}
\sum_{i\in [n]}R_i&=\sum_{i\in U}R_i+\sum_{j\in [k]}\sum_{i\in V_j}R_i+\sum_{i\in W}R_i        \\
&\leq f([m],[n])-f([m],[n]\setminus U)  \\
&\qquad +\sum_{j\in [k]}f([m],V_j)+f([m],W)           \\
&\leq f([m],[n])+\sum_{j\in [k]}f(\{j\},V_j)  \\
&\qquad +f([m],W)-f([m],(\bigcup_{j\in [k]}V_j)\cup W),
\end{align*}
where the last inequality is due to Axiom \eqref{eq:axiom1} of Theorem \ref{thm:pm:group}. 
According to Axiom \eqref{eq:axiom3} of Theorem \ref{thm:pm:group}, we have 
\begin{align*}
&f([m],[n])+\sum_{\ell \in [k]}f(\{ \ell \},(\bigcup_{j\in [\ell+1:k]}V_j) \cup W)  \\
&\qquad \qquad \leq \sum_{J\in N}C_J+\sum_{\ell \in [k]} \sum_{J\in \Tcal_\ell}C_J.
\end{align*}
Therefore, to complete the proof of the proposition, it suffices to show that 
\begin{align}\nonumber
&f([m],[n])+\sum_{\ell \in [k]}f(\{ \ell \},(\bigcup_{j\in [\ell+1:k]}V_j) \cup W) \geq\\\nonumber
&\qquad f([m],[n])+\sum_{j\in [k]}f(\{j\},V_j)  \nonumber  \\
&\qquad +f([m],W)-f([m],(\bigcup_{j\in [k]}V_j)\cup W) \label{eq:uv:proof:goal}
\end{align}
is implied by the rate constraint inequality \eqref{eq:group:rate} as well as Axioms \eqref{eq:axiom1}-\eqref{eq:axiom6} of Theorem \ref{thm:pm:group} with $\Pcal_{\rm \Vv}$ from Definition \ref{def:pcal:uv}.

By Axioms \eqref{eq:axiom5} and \eqref{eq:axiom4}, for any $\ell\in [k]$, we have
\begin{align}
&f([m],(\bigcup_{j\in [\ell:k]}V_j)\cup W) + f(\{\ell\},(\bigcup_{j\in [\ell+1:k]}V_j) \cup W)  \nonumber  \\
&\qquad \geq f([m],(\bigcup_{j\in [\ell+1:k]}V_j)\cup W)+f(\{ \ell \},V_{\ell}). \label{eq:uv:proof:sub:mono}
\end{align}
Summing both sides of \eqref{eq:uv:proof:sub:mono} for every $\ell \in [k]$, we have
\begin{align}
&\sum_{\ell \in [k]}f([m],(\bigcup_{j\in [\ell:k]}V_j)\cup W)  \nonumber  \\
&\qquad +\sum_{\ell \in [k]}f(\{\ell\},(\bigcup_{j\in [\ell+1:k]}V_j) \cup W)    \nonumber   \\
&\ge \sum_{\ell \in [k]}f([m],(\bigcup_{j\in [\ell+1:k]}V_j)\cup W)+\sum_{\ell \in [k]}f(\{ \ell \},V_{\ell})  \nonumber  \\
&=\sum_{\ell \in [2:k+1]}f([m],(\bigcup_{j\in [\ell:k]}V_j)\cup W)  \nonumber  \\
&\qquad +\sum_{\ell \in [k]}f(\{ \ell \},V_{\ell}).  \label{eq:uv:proof:sub:mono:all}
\end{align}
Note that 
$$\sum_{\ell \in [1]}f([m],(\bigcup_{j\in [\ell:k]}V_j)\cup W)=f([m],(\bigcup_{j\in [k]}V_j)\cup W)$$ 
and 
$$\sum_{\ell \in [k+1:k+1]}f([m],(\bigcup_{j\in [\ell:k]}V_j)\cup W)=f([m],W).$$ 
Using the above relations, we have
\begin{align}
&f([m],(\bigcup_{j\in [k]}V_j)\cup W)+\sum_{\ell \in [k]}f(\{ \ell \},(\bigcup_{j\in [\ell+1:k]}V_j) \cup W)  \nonumber \\
&=\sum_{\ell \in [k]}f([m],(\bigcup_{j\in [\ell:k]}V_j)\cup W)  \nonumber  \\
&\qquad\qquad -\sum_{\ell \in [2:k]}f([m],(\bigcup_{j\in [\ell:k]}V_j)\cup W)  \nonumber  \\
&\qquad\qquad +\sum_{\ell \in [k]}f(\{ \ell \},(\bigcup_{j\in [\ell+1:k]}V_j) \cup W)    \nonumber     \\
&\ge \sum_{\ell \in [2:k+1]}f([m],(\bigcup_{j\in [\ell:k]}V_j)\cup W)       \nonumber  \\
&\qquad \qquad -\sum_{\ell \in [2:k]}f([m],(\!\bigcup_{j\in [\ell:k]}\!V_j)\cup W)+\sum_{\ell \in [k]}f(\{ \ell \},V_{\ell})  \nonumber  \\     
&=f([m],W)+\sum_{\ell \in [k]}f(\{ \ell \},V_{\ell}),  \nonumber  
\end{align}
where the inequality follows from \eqref{eq:uv:proof:sub:mono:all}. This completes the proof of \eqref{eq:uv:proof:goal} being implied by  the rate constraint inequality \eqref{eq:group:rate} as well as Axioms \eqref{eq:axiom1}-\eqref{eq:axiom6} of Theorem \ref{thm:pm:group} with $\Pcal_{\rm \Vv}$, and thus completes the proof of this proposition.

\section{Proof of Corollary \ref{cor:pm:touch:aggregate}} \label{app:pm:touch:aggregate}

The goal is to show that Axiom \eqref{eq:axiom1} in Theorem \ref{thm:pm:group} simplifies to Axiom \eqref{eq:touch:axiom1} in Corollary \ref{cor:pm:touch:aggregate}. For easier reference, we repeat Axiom \eqref{eq:axiom1}, with $\Pcal_{\mathrm{t}}=\{T_{L_1},T_{L_2},\ldots,T_{L_m}\}$, as follows.
\begin{align*}
f(G,K) &= f(G',K),  \\
&\qquad \text{if $((T_{L_G}\cup T_{L_{G'}})\setminus (T_{L_G} \cap T_{L_{G'}}))\subseteq T_{\ntouch{K}}$}.
\end{align*}
We also repeat Axiom \eqref{eq:touch:axiom1} as follows.
\begin{align*}
f(G,K) = f(G',K),           \qquad \text{if $K\subseteq (L_G\cap L_{G'})$}.
\end{align*}

For brevity, set $L=L_G$, $L'=L_{G'}$, and also set $T=(T_{L}\cup T_{L'})\setminus(T_{L}\cap T_{L'})=((T_{L_G}\cup T_{L_{G'}})\setminus (T_{L_G} \cap T_{L_{G'}}))$. 
For any $K\subseteq [n]$, $G,G'\subseteq [m]$, we are going to show that $K\subseteq  (L\cap L')$ is the sufficient and necessary condition for $T\subseteq T_{\ntouch{K}}$. 
If $G=G'$, then both \eqref{eq:axiom1} and \eqref{eq:touch:axiom1} becomes trivial. Hence we only consider the case when $G\ne G'$, and since $L_1,\cdots,L_m$ are disjoint to each other, we have $L\ne L'$. 

First, to show the sufficiency, we assume that $K\subseteq(L\cap L')$. Consider any $J\in T$, we know that $J$ touches either $L$ or $L'$, but not both. As $K\subseteq(L\cap L')$, we know that $J\cap K=\emptyset$, which means that $J\in T_{\ntouch{K}}$. Therefore, we have $T\subseteq T_{\ntouch{K}}$, which proves the sufficient condition.

Second, to show the necessity, we assume that $T\subseteq T_{\ntouch{K}}$. Since $L\neq L'$, without loss of generality, assume there exists some $j_1\in L\setminus L'$. Now we show that $K\subseteq (L\cap L')$ by contradiction.

Assume that $K\setminus L' \neq \emptyset$, then there exists some $j_2\in K\setminus L'$. Note that $j_1,j_2$ may be the same index. Now set $J=\{j_1\} \cup \{j_2\} \in N$. Then we have 
\begin{align}
J\cap L \neq \emptyset, \quad J\cap L' =\emptyset,\quad J\cap K \neq \emptyset.
\end{align}
Hence, we have $J\in T$ (since $J\in (T_{L}\cup T_{L'}),J\notin (T_{L}\cap T_{L'})$), and also $J\notin T_{\ntouch{K}}$. This contradicts with the assumption that $T\subseteq T_{\ntouch{K}}$. And therefore, we must have $K\setminus L' = \emptyset$, i.e., $K\subseteq L'$.

Now assume that $K\setminus L \neq \emptyset$, then there exists some $j_3\in K\setminus L$. Since $j_3\in K\setminus L$ and $K\subseteq L'$, for $\{j_3\} \in N$, we have 
\begin{align}
\{j_3\} \cap L =\emptyset,\quad \{j_3\} \cap L' \neq \emptyset,\quad \{j_3\} \cap K \neq \emptyset.
\end{align}
Hence, we have $\{j_3\} \in T$ (since $\{j_3\} \in (T_{L}\cup T_{L'}),\{j_3\} \notin (T_{L}\cap T_{L'})$), and also $\{j_3\} \notin T_{\ntouch{K}}$. This contradicts with the assumption that $T\subseteq T_{\ntouch{K}}$. And therefore, we must have $K\setminus L = \emptyset$, i.e., $K\subseteq L$.

Now we have both $K\subseteq L'$ and $K\subseteq L$, which means that $K\subseteq(L\cap L')$ and proves the necessary condition. Therefore, Axioms \eqref{eq:axiom1} and \eqref{eq:touch:axiom1} are the same under the touch grouping. This completes the proof.

\section{Proof of Proposition \ref{propo:refine}}
\label{app:propo:refine}

For a given distributed index coding problem, consider two valid server groupings $\Qcal=\{Q_1, \cdots, Q_{\ell}\}$ and $\Pcal=\{P_1,\cdots,P_m\}$ such that $\Pcal$ is a refinement of $\Qcal$. For any $E\subseteq [\ell]$, let $Q_{E}=\bigcup_{i\in E}Q_i$, and for any $G\subseteq [m]$, let $P_{G}=\bigcup_{i\in G}P_i$. Denote the outer bound on the capacity region given by the grouping PM outer bound with $\Qcal$ and $\Pcal$ as $\Rr_{\Qcal}$ and $\Rr_{\Pcal}$, respectively.

According to Definition \ref{def:refined}, for any $i\in [\ell]$, there exists some set $G\subseteq [m]$ such that $P_G=Q_i$. Define the mapping function $G$ that maps any set $E\subseteq [\ell]$ to a corresponding set $G(E)\subseteq [m]$ as follows.
\begin{align}
G(E)=\bigcup_{i\in E}\,\,\bigcup_{\substack{G\subseteq [m]:\\P_G=Q_i}}G.
\end{align}

Then for any $E\subseteq [\ell]$, we have
\begin{align}
P_{G(E)}=\bigcup_{i\in E}\,\,\bigcup_{\substack{G\subseteq [m]:\\P_G=Q_i}}P_G=\bigcup_{i\in E}Q_i=Q_E. \label{eq:refine:F}
\end{align}

It can also be verified that the mapping function $G$ has following properties.
\begin{align}
G(\emptyset)&= \emptyset. \label{eq:refine:F:empty}& \\
G(E)&\subseteq G(E'),&\quad \forall E\subseteq E'\subseteq [\ell]. \label{eq:refine:F:mono} \\
G(E\cup E')&=G(E)\cup G(E'), &\quad \forall E,E'\subseteq [\ell]. \label{eq:refine:F:sub1}  \\
G(E\cap E')&\subseteq G(E)\cap G(E'), &\quad \forall E,E'\subseteq [\ell].  \label{eq:refine:F:sub2}
\end{align}

Now assume that some rate tuple $\Rv=(R_i,i\in [n])$ is in $\Rr_{\Pcal}$. Then there exists $f(G,K), G\subseteq [m],K\subseteq [n]$ such that $\Rv$ and $f(G,K)$ satisfy Axioms \eqref{eq:axiom1}-\eqref{eq:axiom6}, as well as \eqref{eq:group:rate}, with server grouping $\Pcal$. 
Construct $f_Q(E,K)=f(G(E),K)$, $E\subseteq [\ell],K\subseteq [n]$. We now show that the rate tuple $\Rv$ is also in $\Rr_{\Qcal}$ by showing that $\Rv$ and $f_Q(E,K)$ satisfy Axioms \eqref{eq:axiom1}-\eqref{eq:axiom6}, as well as \eqref{eq:group:rate} with server grouping $\Qcal$.

For Axiom \eqref{eq:axiom1}, consider any $E,E'\subseteq [\ell],K\subseteq [n]$ such that $(Q_{E}\cup Q_{E'})\setminus (Q_{E}\cap Q_{E'})\subseteq T_{\ntouch{K}}$, we have
\begin{align*}
&(P_{G(E)}\cup P_{G(E')})\setminus (P_{G(E)}\cap P_{G(E')})  =
\\&\quad\quad (Q_{E}\cup Q_{E'})\setminus (Q_{E}\cap Q_{E'}) \subseteq T_{\ntouch{K}},
\end{align*}
where the first equality is due to \eqref{eq:refine:F}. 
As $f(G,K)$ satisfies Axiom \eqref{eq:axiom1} with server grouping $\Pcal$, we have $f(G(E),K)=f(G(E'),K)$. Therefore, by the construction of $f_Q(E,K)$, we have $f_Q(E,K)=f(G(E),K)=f(G(E'),K)=f_Q(E',K)$.

For Axiom \eqref{eq:axiom2}, it is clear that for any $E\subseteq [\ell],K\subseteq [n]$, due to \eqref{eq:refine:F:empty} as well as $f(G,K)$ satisfying Axiom \eqref{eq:axiom2}, we have $f_Q(\emptyset,K)=f(G(\emptyset),K)=f(\emptyset,K)=0$ and $f_Q(E,\emptyset)=f(G(E),\emptyset)=0$.

For Axiom \eqref{eq:axiom3}, for any $E\subseteq [\ell],K\subseteq [n]$, due to \eqref{eq:refine:F} as well as $f(G,K)$ satisfying Axiom \eqref{eq:axiom3}, we have
\begin{align*}
f_Q(E,K)=f(G(E),K)&\leq \sum_{J:J\in P_{G(E)},J\in T_{K}}C_J  \\
&=\sum_{J:J\in Q_{E},J\in T_{K}}C_J.
\end{align*}

For Axiom \eqref{eq:axiom4}, for any $E\subseteq E'\subseteq [\ell],K\subseteq K'\subseteq [n]$, we have
\begin{align*}
f_Q(E,K)=f(G(E),K)\leq f(G(E'),K')=f_Q(E',K'),
\end{align*}
where the inequality is due to \eqref{eq:refine:F:mono} and $f(G,K)$ satisfying Axiom \eqref{eq:axiom4}.

For Axiom \eqref{eq:axiom5}, for any $E,E'\subseteq [\ell],K,K'\subseteq [n]$, we have
\begin{align*}
&f_Q(E\cup E',K\cap K')+f_Q(E\cap E',K\cup K') \\
&\quad =f(G(E\cup E'),K\cap K')+f(G(E\cap E'),K\cup K') \\
                                                        &\quad \leq f(G(E)\cup G(E'),K\cap K')+f(G(E)\cap G(E'),K\cup K') \\
                                                        &\quad\leq f(G(E),K)+f(G(E'),K') \\
                                                        &\quad=f_Q(E,K)+f_Q(E',K'),
\end{align*}
where the first inequality is due to \eqref{eq:refine:F:sub1} and \eqref{eq:refine:F:sub2} and $f(G,K)$ satisfying Axiom \eqref{eq:axiom4}, and the second inequality is due to $f(G,K)$ satisfying Axiom \eqref{eq:axiom5}.

For Axiom \eqref{eq:axiom:addi:de}, for any $K\subseteq [n]$, according to \eqref{eq:refine:F}, we have 
\begin{align*}
&(P_{[m]}\cup P_{G([\ell])})\setminus (P_{[m]}\cap P_{G([\ell])})  \\
&\quad=(P_{[m]}\cup Q_{[\ell]})\setminus (P_{[m]}\cap Q_{[\ell]})  \\
&\quad=(N\cup N)\setminus (N\cap N)=\emptyset \subseteq T_{\ntouch K}.  
\end{align*}
Therefore, for any $i\in [n]$, as $f(G,K)$ satisfies Axioms \eqref{eq:axiom1} and \eqref{eq:axiom:addi:de}, we have
\begin{align}
&f_Q([\ell],B_i\cup \{i\})-f_Q([\ell],B_i)  \nonumber  \\
&\quad=f(G([\ell]),B_i\cup \{i\})-f(G([\ell]),B_i)  \nonumber  \\
                                          &\quad=f([m],B_i\cup \{i\})-f([m],B_i)                 \label{eq:refine:for:rate}                 \\
&\quad=f([m],\{i\})=f(G([\ell]),\{i\})=f_Q([\ell],\{i\}).  \nonumber
\end{align}

For Axiom \eqref{eq:axiom6}, for any $E\subseteq [\ell],K,K'\subseteq [n]$ such that $Q_{E}\subseteq (N\setminus T_{K,K'})$, according to \eqref{eq:refine:F}, we have $P_{G(E)}=Q_{E}\subseteq (N\setminus T_{K,K'})$. 
As $f(G,K)$ satisfies Axiom \eqref{eq:axiom6} with server grouping $\Pcal$, we have 
\begin{align*}
f_Q(E,K)+f_Q(E,K')  &=f(G(E),K)+f(G(E),K')  \\
&=f(G(E),K\cup K')\\&=f_Q(E,K\cup K').
\end{align*}

Finally, given \eqref{eq:refine:for:rate} and that $f(G,K)$ and $\Rv$ jointly satisfy \eqref{eq:group:rate}, for any $i\in [n]$, we have
\begin{align*}
&f_Q([\ell],B_i\cup \{i\})-f_Q([\ell],B_i)  \\
&\quad =f([m],B_i\cup \{i\})-f([m],B_i)\geq R_i,
\end{align*}
which finishes the proof that $f_Q(E,K)$ and $\Rv$ jointly satisfy \eqref{eq:group:rate}.

So far we have shown that for any $\Rv$ in rate region $\Rr_{\Pcal}$, it must be also in $\Rr_{\Qcal}$. Therefore, $\Rr_{\Pcal}\subseteq \Rr_{\Qcal}$.

\section{Proof of Proposition \ref{propo:allserver:itw}}
\label{app:allserver:itw}

We show that $\Rr_* \subseteq \Rr_{f_L}$ as follows. 

Consider any rate tuple $\Rv \in \Rr_*$. Then, there exists some set function $g(K),K\subseteq [n]$ such that $g(K)$ satisfies Axioms \eqref{eq:allserver:axiom1}-\eqref{eq:allserver:axiom:addi:de} and $\Rv$ and $g(K)$ jointly satisfy \eqref{eq:allserver:rate}. Construct $f_L(S),S\subseteq L\subseteq [n]$ as $f_L(S)=g(S)$. 

It can be verified with relative ease that $f_L(S)$ satisfies all the axioms for proposition \ref{propo:itw} (Axioms \eqref{eq:itw:axiom1}-\eqref{eq:itw:axiom4}). So it remains to show that $\Rv$ and $f_L(S)$ jointly satisfy \eqref{eq:itw:rate} as follows.

For any $L \subseteq [n]$ and $i\in L$, we have
\begin{align*}
R_i & \leq g(B_i\cup \{i\})-g(B_i) \\
     & \leq g((B_i\cup \{i\})\cap L)-g(B_i\cap L) \\
     & = f_L((B_i\cup \{i\})\cap L)-f_L(B_i\cap L).
\end{align*}
where the first inequality is due to \eqref{eq:allserver:rate}, and the second inequality is due to the fact that $B_i\cup ((B_i\cup \{i\})\cap L)=B_i\cup \{i\},B_i\cap ((B_i\cup \{i\})\cap L)=B_i\cap L$ and that $g(K)$ satisfies the submodularity axiom, Axiom \eqref{eq:allserver:axiom4}.

Therefore, we can conclude that $\Rv \in \Rr_{f_L}$ and $\Rr_* \subseteq \Rr_{f_L}$.

\section{List of all Non-isomorphic Index Coding Problems with $n=4$ Messages}
\label{app:numerical:graph4}
See Table~\ref{tab:numerical:graph4}.

\begin{table*}[t]
\tiny
\begin{center}
\caption{List of all 218 Non-isomorphic Index Coding Problems with $n=4$ Messages}
\begin{tabular}{|c|c|c|c|c|c|}
\hline
Problem No. &Side Information Sequence&Problem No. &Side Information Sequence&Problem No.&Side Information Sequence\\
\hline
1 &  $(1|-),(2|-),(3|-),(4|-)$ & 74 &  $(1|4),(2|4),(3|2),(4|1,2)$ & 147 & $(1|4),(2|3),(3|1,4),(4|1,2,3)$\\
\hline
2 &  $(1|2),(2|-),(3|-),(4|-)$ & 75 & $(1|4),(2|4),(3|2,4),(4|3)$ & 148 & $(1|4),(2|3),(3|1,2),(4|1,2,3)$\\
\hline
3 &  $(1|2,3),(2|-),(3|-),(4|-)$ & 76 & $(1|4),(2|4),(3|2,4),(4|2)$ & 149 & $(1|4),(2|3,4),(3|2,4),(4|2,3)$\\
\hline
4 &  $(1|-),(2|-),(3|4),(4|3)$ & 77 & $(1|4),(2|4),(3|2,4),(4|1)$ & 150 & $(1|4),(2|3,4),(3|2,4),(4|1,3)$\\
\hline
5 &  $(1|-),(2|-),(3|4),(4|2)$ & 78 & $(1|4),(2|4),(3|1,2),(4|3)$ & 151 & $(1|4),(2|3,4),(3|1),(4|1,2,3)$\\
\hline
6 &  $(1|-),(2|-),(3|2),(4|2)$ & 79 & $(1|4),(2|4),(3|1,2),(4|2)$ & 152 & $(1|4),(2|3,4),(3|1,4),(4|2,3)$\\
\hline
7 &  $(1|-),(2|-),(3|2),(4|1)$ & 80 & $(1|4),(2|3),(3|2),(4|2,3)$ & 153 & $(1|4),(2|3,4),(3|1,4),(4|1,3)$\\
\hline
8 &  $(1|2,3,4),(2|-),(3|-),(4|-)$ & 81 & $(1|4),(2|3),(3|2),(4|1,3)$ & 154 & $(1|4),(2|3,4),(3|1,4),(4|1,2)$\\
\hline
9 &  $(1|-),(2|-),(3|4),(4|2,3)$ & 82 & $(1|4),(2|3),(3|2,4),(4|2)$ & 155 & $(1|4),(2|3,4),(3|1,2),(4|2,3)$\\
\hline
10 &  $(1|-),(2|-),(3|4),(4|1,2)$ & 83 & $(1|4),(2|3),(3|1),(4|2,3)$ & 156 & $(1|4),(2|3,4),(3|1,2),(4|1,3)$\\
\hline
11 &  $(1|-),(2|-),(3|2),(4|2,3)$ & 84 & $(1|4),(2|3),(3|1),(4|1,2)$ & 157 & $(1|4),(2|3,4),(3|1,2),(4|1,2)$\\
\hline
12 &  $(1|-),(2|-),(3|2),(4|1,3)$ & 85 & $(1|4),(2|3,4),(3|1),(4|3)$ & 158 & $(1|4),(2|3,4),(3|1,2,4),(4|3)$\\
\hline
13 &  $(1|-),(2|-),(3|2),(4|1,2)$ & 86 & $(1|2,3,4),(2|1,3,4),(3|-),(4|-)$ & 159 & $(1|4),(2|3,4),(3|1,2,4),(4|2)$\\
\hline
14 &  $(1|-),(2|4),(3|4),(4|3)$ & 87 & $(1|-),(2|4),(3|2,4),(4|1,2,3)$ & 160 & $(1|4),(2|3,4),(3|1,2,4),(4|1)$\\
\hline
15 &  $(1|-),(2|4),(3|4),(4|1)$ & 88 & $(1|-),(2|4),(3|1,4),(4|1,2,3)$ & 161 & $(1|4),(2|1),(3|1,2),(4|1,2,3)$\\
\hline
16 &  $(1|-),(2|4),(3|2),(4|3)$ & 89 & $(1|-),(2|4),(3|1,2),(4|1,2,3)$ & 162 & $(1|4),(2|1),(3|1,2,4),(4|2,3)$\\
\hline
17 &  $(1|-),(2|4),(3|2),(4|1)$ & 90 & $(1|-),(2|4),(3|1,2,4),(4|2,3)$ & 163 & $(1|4),(2|1),(3|1,2,4),(4|1,2)$\\
\hline
18 &  $(1|-),(2|4),(3|1),(4|2)$ & 91 & $(1|-),(2|4),(3|1,2,4),(4|1,3)$ & 164 & $(1|4),(2|1,4),(3|1,4),(4|2,3)$\\
\hline
19 &  $(1|-),(2|4),(3|1),(4|1)$ & 92 & $(1|-),(2|4),(3|1,2,4),(4|1,2)$ & 165 & $(1|4),(2|1,4),(3|1,4),(4|1,3)$\\
\hline
20 &  $(1|-),(2|1),(3|1),(4|1)$ & 93 & $(1|-),(2|3,4),(3|2,4),(4|2,3)$ & 166 & $(1|4),(2|1,4),(3|1,2),(4|2,3)$\\
\hline
21 &  $(1|2,3,4),(2|1),(3|-),(4|-)$ & 94 & $(1|-),(2|3,4),(3|2,4),(4|1,3)$ & 167 & $(1|4),(2|1,4),(3|1,2),(4|1,3)$\\
\hline
22 &  $(1|-),(2|-),(3|2),(4|1,2,3)$ & 95 & $(1|-),(2|3,4),(3|1),(4|1,2,3)$ & 168 & $(1|4),(2|1,4),(3|1,2),(4|1,2)$\\
\hline
23 &  $(1|-),(2|-),(3|2,4),(4|2,3)$ & 96 & $(1|-),(2|3,4),(3|1,4),(4|1,3)$ & 169 & $(1|4),(2|1,4),(3|1,2,4),(4|1)$\\
\hline
24 &  $(1|-),(2|-),(3|2,4),(4|1,3)$ & 97 & $(1|-),(2|3,4),(3|1,4),(4|1,2)$ & 170 & $(1|4),(2|1,3),(3|1,2),(4|2,3)$\\
\hline
25 &  $(1|-),(2|-),(3|2,4),(4|1,2)$ & 98 & $(1|-),(2|3,4),(3|1,2),(4|1,2)$ & 171 & $(1|4),(2|1,3),(3|1,2),(4|1,3)$\\
\hline
26 &  $(1|-),(2|-),(3|1,2),(4|1,2)$ & 99 & $(1|-),(2|1),(3|1,4),(4|1,2,3)$ & 172 & $(1|2,3,4),(2|1,3,4),(3|1,2),(4|-)$\\
\hline
27 &  $(1|-),(2|4),(3|4),(4|2,3)$ & 100 & $(1|-),(2|1),(3|1,2),(4|1,2,3)$ & 173 & $(1|-),(2|1,4),(3|1,2,4),(4|1,2,3)$\\
\hline
28 &  $(1|-),(2|4),(3|4),(4|1,3)$ & 101 & $(1|-),(2|1,4),(3|1,4),(4|1,3)$ & 174 & $(1|4),(2|4),(3|1,2,4),(4|1,2,3)$\\
\hline
29 &  $(1|-),(2|4),(3|2),(4|2,3)$ & 102 & $(1|-),(2|1,4),(3|1,2),(4|1,3)$ & 175 & $(1|4),(2|3),(3|1,2,4),(4|1,2,3)$\\
\hline
30 &  $(1|-),(2|4),(3|2),(4|1,3)$ & 103 & $(1|4),(2|4),(3|4),(4|1,2,3)$ & 176 & $(1|4),(2|3,4),(3|2,4),(4|1,2,3)$\\
\hline
31 &  $(1|-),(2|4),(3|2),(4|1,2)$ & 104 & $(1|4),(2|4),(3|2),(4|1,2,3)$ & 177 & $(1|4),(2|3,4),(3|1,4),(4|1,2,3)$\\
\hline
32 &  $(1|-),(2|4),(3|2,4),(4|2)$ & 105 & $(1|4),(2|4),(3|2,4),(4|2,3)$ & 178 & $(1|4),(2|3,4),(3|1,2),(4|1,2,3)$\\
\hline
33 &  $(1|-),(2|4),(3|2,4),(4|1)$ & 106 & $(1|4),(2|4),(3|2,4),(4|1,3)$ & 179 & $(1|4),(2|3,4),(3|1,2,4),(4|2,3)$\\
\hline
34 &  $(1|-),(2|4),(3|1),(4|2,3)$ & 107 & $(1|4),(2|4),(3|2,4),(4|1,2)$ & 180 & $(1|4),(2|3,4),(3|1,2,4),(4|1,3)$\\
\hline
35 &  $(1|-),(2|4),(3|1),(4|1,3)$ & 108 & $(1|4),(2|4),(3|1,2),(4|2,3)$ & 181 & $(1|4),(2|3,4),(3|1,2,4),(4|1,2)$\\
\hline
36 &  $(1|-),(2|4),(3|1),(4|1,2)$ & 109 & $(1|4),(2|4),(3|1,2),(4|1,2)$ & 182 & $(1|4),(2|1),(3|1,2,4),(4|1,2,3)$\\
\hline
37 &  $(1|-),(2|4),(3|1,4),(4|2)$ & 110 & $(1|4),(2|4),(3|1,2,4),(4|3)$ & 183 & $(1|4),(2|1,4),(3|1,4),(4|1,2,3)$\\
\hline
38 &  $(1|-),(2|4),(3|1,4),(4|1)$ & 111 & $(1|4),(2|4),(3|1,2,4),(4|2)$ & 184 & $(1|4),(2|1,4),(3|1,2),(4|1,2,3)$\\
\hline
39 &  $(1|-),(2|4),(3|1,2),(4|1)$ & 112 & $(1|4),(2|3),(3|2),(4|1,2,3)$ & 185 & $(1|4),(2|1,4),(3|1,2,4),(4|2,3)$\\
\hline
40 &  $(1|-),(2|3,4),(3|1),(4|1)$ & 113 & $(1|4),(2|3),(3|2,4),(4|2,3)$ & 186 & $(1|4),(2|1,4),(3|1,2,4),(4|1,3)$\\
\hline
41 &  $(1|-),(2|1),(3|1),(4|1,3)$ & 114 & $(1|4),(2|3),(3|2,4),(4|1,3)$ & 187 & $(1|4),(2|1,4),(3|1,2,4),(4|1,2)$\\
\hline
42 &  $(1|4),(2|4),(3|4),(4|3)$ & 115 & $(1|4),(2|3),(3|2,4),(4|1,2)$ & 188 & $(1|4),(2|1,3),(3|1,2),(4|1,2,3)$\\
\hline
43 &  $(1|4),(2|4),(3|2),(4|3)$ & 116 & $(1|4),(2|3),(3|1),(4|1,2,3)$ & 189 & $(1|4),(2|1,3),(3|1,2,4),(4|2,3)$\\
\hline
44 &  $(1|4),(2|4),(3|2),(4|2)$ & 117 & $(1|4),(2|3),(3|1,4),(4|2,3)$ & 190 & $(1|4),(2|1,3),(3|1,2,4),(4|1,3)$\\
\hline
45 &  $(1|4),(2|4),(3|2),(4|1)$ & 118 & $(1|4),(2|3),(3|1,4),(4|1,2)$ & 191 & $(1|4),(2|1,3),(3|1,2,4),(4|1,2)$\\
\hline
46 &  $(1|4),(2|3),(3|2),(4|1)$ & 119 & $(1|4),(2|3),(3|1,2),(4|1,2)$ & 192 & $(1|4),(2|1,3,4),(3|1,2,4),(4|1)$\\
\hline
47 &  $(1|4),(2|3),(3|1),(4|2)$ & 120 & $(1|4),(2|3,4),(3|2,4),(4|3)$ & 193 & $(1|3,4),(2|3,4),(3|2,4),(4|2,3)$\\
\hline
48 &  $(1|2,3,4),(2|1,3),(3|-),(4|-)$ & 121 & $(1|4),(2|3,4),(3|2,4),(4|1)$ & 194 & $(1|3,4),(2|3,4),(3|2,4),(4|1,3)$\\
\hline
49 &  $(1|-),(2|-),(3|1,2),(4|1,2,3)$ & 122 & $(1|4),(2|3,4),(3|1),(4|2,3)$ & 195 & $(1|3,4),(2|3,4),(3|2,4),(4|1,2)$\\
\hline
50 &  $(1|-),(2|4),(3|4),(4|1,2,3)$ & 123 & $(1|4),(2|3,4),(3|1),(4|1,3)$ & 196 & $(1|3,4),(2|3,4),(3|1,2),(4|1,2)$\\
\hline
51 &  $(1|-),(2|4),(3|2),(4|1,2,3)$ & 124 & $(1|4),(2|3,4),(3|1),(4|1,2)$ & 197 & $(1|3,4),(2|1,4),(3|2,4),(4|2,3)$\\
\hline
52 &  $(1|-),(2|4),(3|2,4),(4|2,3)$ & 125 & $(1|4),(2|3,4),(3|1,4),(4|3)$ & 198 & $(1|3,4),(2|1,4),(3|1,2),(4|2,3)$\\
\hline
53 &  $(1|-),(2|4),(3|2,4),(4|1,3)$ & 126 & $(1|4),(2|3,4),(3|1,4),(4|2)$ & 199 & $(1|2,3,4),(2|1,3,4),(3|1,2,4),(4|-)$\\
\hline
54 &  $(1|-),(2|4),(3|2,4),(4|1,2)$ & 127 & $(1|4),(2|3,4),(3|1,4),(4|1)$ & 200 & $(1|4),(2|3,4),(3|1,2,4),(4|1,2,3)$\\
\hline
55 &  $(1|-),(2|4),(3|1),(4|1,2,3)$ & 128 & $(1|4),(2|3,4),(3|1,2),(4|3)$ & 201 & $(1|4),(2|1,4),(3|1,2,4),(4|1,2,3)$\\
\hline
56 &  $(1|-),(2|4),(3|1,4),(4|2,3)$ & 129 & $(1|4),(2|3,4),(3|1,2),(4|1)$ & 202 & $(1|4),(2|1,3),(3|1,2,4),(4|1,2,3)$\\
\hline
57 &  $(1|-),(2|4),(3|1,4),(4|1,3)$ & 130 & $(1|4),(2|1),(3|1,2),(4|2,3)$ & 203 & $(1|4),(2|1,3,4),(3|1,2,4),(4|2,3)$\\
\hline
58 &  $(1|-),(2|4),(3|1,4),(4|1,2)$ & 131 & $(1|4),(2|1),(3|1,2),(4|1,2)$ & 204 & $(1|4),(2|1,3,4),(3|1,2,4),(4|1,3)$\\
\hline
59 &  $(1|-),(2|4),(3|1,2),(4|2,3)$ & 132 & $(1|4),(2|1),(3|1,2,4),(4|2)$ & 205 & $(1|3,4),(2|3,4),(3|2,4),(4|1,2,3)$\\
\hline
60 &  $(1|-),(2|4),(3|1,2),(4|1,3)$ & 133 & $(1|4),(2|1,4),(3|1,4),(4|1)$ & 206 & $(1|3,4),(2|3,4),(3|1,2),(4|1,2,3)$\\
\hline
61 &  $(1|-),(2|4),(3|1,2),(4|1,2)$ & 134 & $(1|2,3,4),(2|1,3,4),(3|1),(4|-)$ & 207 & $(1|3,4),(2|1,4),(3|2,4),(4|1,2,3)$\\
\hline
62 &  $(1|-),(2|4),(3|1,2,4),(4|2)$ & 135 & $(1|-),(2|3,4),(3|2,4),(4|1,2,3)$ & 208 & $(1|3,4),(2|1,4),(3|1,2),(4|1,2,3)$\\
\hline
63 &  $(1|-),(2|4),(3|1,2,4),(4|1)$ & 136 & $(1|-),(2|3,4),(3|1,4),(4|1,2,3)$ & 209 & $(1|3,4),(2|1,4),(3|1,2,4),(4|1,3)$\\
\hline
64 &  $(1|-),(2|3,4),(3|2,4),(4|1)$ & 137 & $(1|-),(2|3,4),(3|1,2),(4|1,2,3)$ & 210 & $(1|3,4),(2|1,4),(3|1,2,4),(4|1,2)$\\
\hline
65 &  $(1|-),(2|3,4),(3|1),(4|1,3)$ & 138 & $(1|-),(2|1),(3|1,2,4),(4|1,2,3)$ & 211 & $(1|3,4),(2|1,3,4),(3|1,4),(4|1,3)$\\
\hline
66 &  $(1|-),(2|3,4),(3|1),(4|1,2)$ & 139 & $(1|-),(2|1,4),(3|1,4),(4|1,2,3)$ & 212 & $(1|2,3,4),(2|1,3,4),(3|1,2,4),(4|1)$\\
\hline
67 &  $(1|-),(2|1),(3|1),(4|1,2,3)$ & 140 & $(1|-),(2|1,4),(3|1,2),(4|1,2,3)$ & 213 & $(1|3,4),(2|3,4),(3|1,2,4),(4|1,2,3)$\\
\hline
68 &  $(1|-),(2|1),(3|1,4),(4|1,3)$ & 141 & $(1|-),(2|1,4),(3|1,2,4),(4|1,2)$ & 214 & $(1|3,4),(2|1,4),(3|1,2,4),(4|1,2,3)$\\
\hline
69 &  $(1|-),(2|1),(3|1,4),(4|1,2)$ & 142 & $(1|4),(2|4),(3|2,4),(4|1,2,3)$ & 215 & $(1|3,4),(2|1,3,4),(3|1,4),(4|1,2,3)$\\
\hline
70 &  $(1|-),(2|1),(3|1,2),(4|1,2)$ & 143 & $(1|4),(2|4),(3|1,2),(4|1,2,3)$ & 216 & $(1|3,4),(2|1,3,4),(3|1,2),(4|1,2,3)$\\
\hline
71 &  $(1|4),(2|4),(3|4),(4|2,3)$ & 144 & $(1|4),(2|4),(3|1,2,4),(4|2,3)$ & 217 & $(1|2,3,4),(2|1,3,4),(3|1,2,4),(4|1,2)$\\
\hline
72 &  $(1|4),(2|4),(3|2),(4|2,3)$ & 145 & $(1|4),(2|4),(3|1,2,4),(4|1,2)$ & 218 & $(1|2,3,4),(2|1,3,4),(3|1,2,4),(4|1,2,3)$\\
\hline
73 &  $(1|4),(2|4),(3|2),(4|1,3)$ & 146 & $(1|4),(2|3),(3|2,4),(4|1,2,3)$ &  & \\
\hline
\end{tabular}\label{tab:numerical:graph4}
\end{center}
\end{table*}

\newcommand{\noopsort}[1]{}


\begin{thebibliography}{10}
\providecommand{\url}[1]{#1}
\csname url@samestyle\endcsname
\providecommand{\newblock}{\relax}
\providecommand{\bibinfo}[2]{#2}
\providecommand{\BIBentrySTDinterwordspacing}{\spaceskip=0pt\relax}
\providecommand{\BIBentryALTinterwordstretchfactor}{4}
\providecommand{\BIBentryALTinterwordspacing}{\spaceskip=\fontdimen2\font plus
\BIBentryALTinterwordstretchfactor\fontdimen3\font minus
  \fontdimen4\font\relax}
\providecommand{\BIBforeignlanguage}[2]{{%
\expandafter\ifx\csname l@#1\endcsname\relax
\typeout{** WARNING: IEEEtran.bst: No hyphenation pattern has been}%
\typeout{** loaded for the language `#1'. Using the pattern for}%
\typeout{** the default language instead.}%
\else
\language=\csname l@#1\endcsname
\fi
#2}}
\providecommand{\BIBdecl}{\relax}
\BIBdecl

\bibitem{isit:2017}
Y.~Liu, P.~Sadeghi, F.~Arbabjolfaei, and Y.-H. Kim, ``{On the Capacity for
  Distributed Index Coding},'' in \emph{{Proc. {IEEE} Int. Symp. on Information
  Theory (ISIT)}}, Aachen, Germany, Jun. 2017, pp. 3055--3059.

\bibitem{Sadeghi--Arbabjolfaei--Kim2016}
P.~Sadeghi, F.~Arbabjolfaei, and Y.-H. Kim, ``Distributed index coding,'' in
  \emph{Proc. {IEEE} Information Theory Workshop (ITW)}, Cambridge, UK, Sep.
  2016, pp. 330--334.

\bibitem{Birk--Kol1998}
Y.~Birk and T.~Kol, ``Informed-source coding-on-demand ({ISCOD}) over broadcast
  channels,'' in \emph{Proc. IEEE Int. Conf. on Computer Communications
  (INFOCOM)}, Mar. 1998, pp. 1257--1264.

\bibitem{Birk--Kol2006}
------, ``Coding on demand by an informed source ({ISCOD}) for efficient
  broadcast of different supplemental data to caching clients,'' \emph{{IEEE}
  Trans. Inf. Theory}, vol.~52, no.~6, pp. 2825--2830, Jun. 2006.

\bibitem{bar2011index}
Z.~Bar-Yossef, Y.~Birk, T.~Jayram, and T.~Kol, ``Index coding with side
  information,'' \emph{{IEEE} Trans. Inf. Theory}, vol.~57, no.~3, pp.
  1479--1494, Jun. 2011.

\bibitem{lubetzky2009nonlinear}
E.~Lubetzky and U.~Stav, ``Nonlinear index coding outperforming the linear
  optimum,'' \emph{{IEEE} Trans. Inf. Theory}, vol.~55, no.~8, pp. 3544--3551,
  Aug. 2009.

\bibitem{tehrani2012bipartite}
A.~S. Tehrani, A.~G. Dimakis, and M.~J. Neely, ``Bipartite index coding,'' in
  \emph{{Proc. {IEEE} Int. Symp. on Information Theory (ISIT)}}, Boston, MA,
  Jul. 2012, pp. 2246--2250.

\bibitem{blasiak2011lexicographic}
A.~Blasiak, R.~Kleinberg, and E.~Lubetzky, ``Lexicographic products and the
  power of non-linear network coding,'' in \emph{Foundations of Computer
  Science (FOCS)}, Palm Springs, CA, Oct. 2011, pp. 609--618.

\bibitem{shanmugam2013local}
K.~Shanmugam, A.~G. Dimakis, and M.~Langberg, ``Local graph coloring and index
  coding,'' in \emph{{Proc. {IEEE} Int. Symp. on Information Theory (ISIT)}},
  Istanbul, Turkey, Jul. 2013, pp. 1152--1156.

\bibitem{arbabjolfaei2014local}
F.~Arbabjolfaei and Y.-H. Kim, ``Local time sharing for index coding,'' in
  \emph{{Proc. {IEEE} Int. Symp. on Information Theory (ISIT)}}, Honolulu, HI,
  Jun.--Jul. 2014, pp. 286--290.

\bibitem{Arbabjolfaei--Kim2015a}
------, ``Structural properties of index coding capacity using fractional graph
  theory,'' in \emph{{Proc. {IEEE} Int. Symp. on Information Theory (ISIT)}},
  Hong Kong, Jun. 2015, pp. 1034--1038.

\bibitem{el2010index}
S.~El~Rouayheb, A.~Sprintson, and C.~Georghiades, ``On the index coding problem
  and its relation to network coding and matroid theory,'' \emph{{IEEE} Trans.
  Inf. Theory}, vol.~56, no.~7, pp. 3187--3195, Jul. 2010.

\bibitem{effros2015equivalence}
M.~Effros, S.~El~Rouayheb, and M.~Langberg, ``An equivalence between network
  coding and index coding,'' \emph{{IEEE} Trans. Inf. Theory}, vol.~61, no.~5,
  pp. 2478--2487, May 2015.

\bibitem{Arbabjolfaei--Bandemer--Kim--Sasoglu--Wang2013}
F.~Arbabjolfaei, B.~Bandemer, Y.-H. Kim, E.~Sasoglu, and L.~Wang, ``On the
  capacity region for index coding,'' in \emph{{Proc. {IEEE} Int. Symp. on
  Information Theory (ISIT)}}, Istanbul, Turkey, Jul. 2013, pp. 962--966.

\bibitem{jafar2013topological}
S.~A. Jafar, ``Topological interference management through index coding,''
  \emph{{IEEE} Trans. Inf. Theory}, vol.~60, no.~1, pp. 529--568, 2013.

\bibitem{maleki2014index}
H.~Maleki, V.~R. Cadambe, and S.~A. Jafar, ``Index coding: {An} interference
  alignment perspective,'' \emph{{IEEE} Trans. Inf. Theory}, vol.~60, no.~9,
  pp. 5402--5432, Sep. 2014.

\bibitem{sun2015index}
H.~Sun and S.~A. Jafar, ``Index coding capacity: {How} far can one go with only
  {Shannon} inequalities?'' \emph{{IEEE} Trans. Inf. Theory}, vol.~61, no.~6,
  pp. 3041--3055, Jun. 2015.

\bibitem{foundation}
F.~Arbabjolfaei and Y.-H. Kim, \emph{Fundamentals of Index Coding}.\hskip 1em
  plus 0.5em minus 0.4em\relax Foundations and Trends in Communications and
  Information Theory, 2018.

\bibitem{isit:2018}
Y.~Liu, P.~Sadeghi, F.~Arbabjolfaei, and Y.-H. Kim, ``Simplified composite
  coding for index coding,'' in \emph{{Proc. {IEEE} Int. Symp. on Information
  Theory (ISIT)}}, Vail, CO, Jun. 2018, pp. 456--460.

\bibitem{Ong--Ho--Lim2016}
L.~Ong, C.~K. Ho, and F.~Lim, ``The single-uniprior index-coding problem: The
  single-sender case and the multi-sender extension,'' \emph{{IEEE} Trans. Inf.
  Theory}, vol.~62, no.~6, pp. 3165--3182, Jun. 2016.

\bibitem{Thapa--Ong--Johnson2016}
C.~Thapa, L.~Ong, and S.~J. Johnson, ``Graph-theoretic approaches to two-sender
  index coding,'' in \emph{{Proc. {IEEE Global Communications Conf.
  (GLOBECOM)}}}, Washington, DC, Dec. 2016, pp. 1--6.

\bibitem{Li--Ong--Johnson--ISIT--2017}
M.~Li, L.~Ong, and S.~J. Johnson, ``Improved bounds for multi-sender index
  coding,'' in \emph{{Proc. {IEEE} Int. Symp. on Information Theory (ISIT)}},
  Aachen, Germany, Jun. 2017, pp. 3060--3064.

\bibitem{Li--Ong--Johnson2017}
------, ``Cooperative multi-sender index coding,'' \emph{{IEEE} Trans. Inf.
  Theory}, vol.~65, no.~3, pp. 1725--1739, Mar. 2019.

\bibitem{kramer}
G.~Kramer, ``Directed information for channels with feedback,'' Ph.D.
  dissertation, ETH Series in Information Processing, 1998.

\bibitem{satyajit}
S.~Thakor, A.~Grant, and T.~Chan, ``Cut-set bounds on network information
  flow,'' \emph{{IEEE} Trans. Inf. Theory}, vol.~62, no.~4, pp. 1850--1865,
  Apr. 2016.

\bibitem{Langberg--Effros2011}
M.~Langberg and M.~Effros, ``Network coding: {I}s zero error always possible?''
  in \emph{Proc. 49th Ann. Allerton Conf. Comm. Control Comput.}, Sep. 2011,
  pp. 1478--1485.

\bibitem{tahmasbi2015critical}
M.~Tahmasbi, A.~Shahrasbi, and A.~Gohari, ``Critical graphs in index coding,''
  \emph{IEEE Journal on Selected Areas in Communications}, vol.~33, no.~2, pp.
  225--235, 2015.

\bibitem{elgamal_yhk}
A.~{El Gamal} and Y.-H. Kim, \emph{Network Information Theory}.\hskip 1em plus
  0.5em minus 0.4em\relax Cambridge: Cambridge University Press, 2011.

\bibitem{fme:permuter}
\BIBentryALTinterwordspacing
I.~B. Gattegno. {FME-IT package for MATLAB}. [Online]. Available:
  \url{http://www.ee.bgu.ac.il/~fmeit/download.html}
\BIBentrySTDinterwordspacing

\bibitem{liu:vellambi:kim:sadeghi:itw18}
\BIBentryALTinterwordspacing
Y.~Liu, Y.-H. Kim, B.~Vellambi, and P.~Sadeghi, ``On the capacity region for
  secure index coding,'' in \emph{Proc. {IEEE} Information Theory Workshop
  (ITW)}, Guanzhou, China, Nov. 2018. [Online]. Available:
  \url{http://arxiv.org/abs/1809.03615}
\BIBentrySTDinterwordspacing

\bibitem{tangliu2018itw}
T.~Liu and D.~Tuninetti, ``An information theoretic converse for the
  consecutive complete--s picod problem,'' in \emph{Proc. {IEEE}
  Information Theory Workshop (ITW)}.\hskip 1em plus 0.5em minus 0.4em\relax
  IEEE, 2018, pp. 1--5.

\bibitem{neely2013dynamic}
M.~J. Neely, A.~S. Tehrani, and Z.~Zhang, ``Dynamic index coding for wireless
  broadcast networks,'' \emph{{IEEE} Trans. Inf. Theory}, vol.~59, no.~11, pp.
  7525--7540, 2013.

\bibitem{liu2018three}
Y.~Liu, P.~Sadeghi, and Y.-H. Kim, ``Three-layer composite coding for index
  coding,'' in \emph{Proc. IEEE Information Theory Workshop}, 2018.

\bibitem{li2018multi}
M.~Li, L.~Ong, and S.~J. Johnson, ``Multi-sender index coding for collaborative
  broadcasting: A rank-minimization approach,'' \emph{{IEEE} Trans. Commun.},
  vol.~67, no.~2, pp. 1452--1466, Feb. 2019.

\bibitem{kim2019linear}
J.-W. Kim and J.-S. No, ``Linear index coding with multiple senders and
  extension to a cellular network,'' \emph{{IEEE} Trans. Commun.}, vol.~67,
  no.~12, pp. 8666--8677, Dec. 2019.

\bibitem{baber2013multiple}
R.~Baber, D.~Christofides, A.~N. Dang, S.~Riis, and E.~R. Vaughan, ``Multiple
  unicasts, graph guessing games, and {non-Shannon} inequalities,'' in
  \emph{Proc. Int. Symp. on Network Coding (NetCod)}, 2013, pp. 1--6.

\bibitem{isit:2019:arxiv}
Y.~Liu and P.~Sadeghi, ``Generalized alignment chain: Improved converse results
  for index coding,'' \emph{arXiv preprint arXiv:1901.09183}, 2019.

\end{thebibliography}
\end{document}